\documentclass[graybox]{svmult}

\usepackage{mathptmx}       
\usepackage{helvet}         
\usepackage{courier}        
\usepackage{type1cm}        
\usepackage{makeidx}         
\usepackage{graphicx}        

\usepackage{multicol}        
\usepackage[bottom]{footmisc}
\usepackage{hyperref,cite} 
\usepackage{amssymb}
\usepackage{amsmath}
\usepackage{amsfonts}
\usepackage{upgreek}
\usepackage{latexsym}
\usepackage{dsfont}

\definecolor{MyBlue}{rgb}{0.15,0.15,0.70}

\hypersetup{
colorlinks=true,
citecolor=MyBlue,
linkcolor=MyBlue,
urlcolor=MyBlue
}

\newcommand{\DP}{\Delta_4}
\newcommand*{\Scale}[2][4]{\scalebox{#1}{$#2$}}
\newcommand{\Boxbar}{\Scale[0.84]{\overline{\square}}}

\newcommand{\red}{\textcolor{red}} 

\newcommand{\nn}{\nonumber}

\newcommand{\iBox}{\Box^{-1}}
\newcommand{\Stu}{St\"uckelberg }

\newcommand{\Fmn}{F_{\mu\nu}}
\newcommand{\FMN}{F^{\mu\nu}}
\newcommand{\Am}{A_{\mu}}

\newcommand{\AMU}{A^{\mu}}
\newcommand{\AN}{A^{\nu}}

\renewcommand\({\left(}
\renewcommand\){\right)}
\renewcommand\[{\left[}
\renewcommand\]{\right]}

\newcommand\n{{\mbox {\boldmath $\nabla$}}}
\newcommand{\ra}{\rightarrow}

\def\lsim{\raise 0.4ex\hbox{$<$}\kern -0.8em\lower 0.62
ex\hbox{$\sim$}}

\def\gsim{\raise 0.4ex\hbox{$>$}\kern -0.7em\lower 0.62
ex\hbox{$\sim$}}

\def\lbar{{\hbox{$\lambda$}\kern -0.7em\raise 0.6ex
\hbox{$-$}}}

\newcommand\eq[1]{eq.~(\ref{#1})}
\newcommand\eqs[2]{eqs.~(\ref{#1}) and (\ref{#2})}
\newcommand\Eq[1]{Equation~(\ref{#1})}
\newcommand\Eqs[2]{Equations~(\ref{#1}) and (\ref{#2})}

\newcommand\eqst[2]{eqs.~(\ref{#1})--(\ref{#2})}
\newcommand\Eqst[2]{Eqs.~(\ref{#1})--(\ref{#2})}
\newcommand\pa{\partial}
\newcommand\p{\partial}

\newcommand\ee{\end{equation}}
\newcommand\be{\begin{equation}}
\def\bea{\begin{array}}
\def\eea{\end{array}}\def\ea{\end{array}}
\newcommand\ees{\end{eqnarray}}
\newcommand\bees{\begin{eqnarray}}
\def\nn{\nonumber}

\def\a{\alpha}

\def\s{\sigma}
\def\g{\gamma}

\def\d{\delta}

\def\eps{\epsilon}

\def\dslash{\hspace{-1mm}\not{\hbox{\kern-2pt $\partial$}}}
\def\Dslash{\not{\hbox{\kern-2pt $D$}}}
\def\pslash{\not{\hbox{\kern-2.1pt $p$}}}
\def\kslash{\not{\hbox{\kern-2.3pt $k$}}}
\def\qslash{\not{\hbox{\kern-2.3pt $q$}}}

\newcommand{\vac}{|0\rangle}
\newcommand{\cav}{\langle 0|}

\newcommand{\vk}{{\bf k}}
\newcommand{\vx}{{\bf x}}

\def\p1{{\bf p}_1}
\def\p2{{\bf p}_2}
\def\k1{{\bf k}_1}
\def\k2{{\bf k}_2}

\newcommand{\emn}{\eta_{\mu\nu}}

\newcommand{\eMN}{\eta^{\mu\nu}}
\newcommand{\eRS}{\eta^{\rho\sigma}}
\newcommand{\eMR}{\eta^{\mu\rho}}
\newcommand{\eNS}{\eta^{\nu\sigma}}
\newcommand{\eMS}{\eta^{\mu\sigma}}
\newcommand{\eNR}{\eta^{\nu\rho}}

\newcommand{\gmn}{g_{\mu\nu}}

\newcommand{\gMN}{g^{\mu\nu}}

\newcommand{\gbmn}{\bar{g}_{\mu\nu}}

\newcommand{\hmn}{h_{\mu\nu}}
\newcommand{\hrs}{h_{\rho\sigma}}
\newcommand{\hmr}{h_{\mu\rho}}

\newcommand{\hnr}{h_{\nu\rho}}

\newcommand{\hMN}{h^{\mu\nu}}
\newcommand{\hRS}{h^{\rho\sigma}}

\newcommand{\bhmn}{\bar{h}_{\mu\nu}}

\newcommand{\xim}{\xi_{\mu}}
\newcommand{\xin}{\xi_{\nu}}

\newcommand{\pam}{\pa_{\mu}}

\newcommand{\pan}{\pa_{\nu}}
\newcommand{\parho}{\pa_{\rho}}

\newcommand{\paM}{\pa^{\mu}}
\newcommand{\paN}{\pa^{\nu}}
\newcommand{\paR}{\pa^{\rho}}
\newcommand{\paS}{\pa^{\sigma}}

\newcommand{\Rmn}{R_{\mu\nu}}
\newcommand{\Gmn}{G_{\mu\nu}}
\newcommand{\RMN}{R^{\mu\nu}}

\newcommand{\Rmnrs}{R_{\mu\nu\rho\sigma}}

\newcommand{\Tmn}{T_{\mu\nu}}
\newcommand{\Smn}{S_{\mu\nu}}

\newcommand{\TMN}{T^{\mu\nu}}

\newcommand{\dddM}{\kern 0.2em \raise 1.9ex\hbox{$...$}\kern -1.0em \hbox{$M$}}
\newcommand{\dddQ}{\kern 0.2em \raise 1.9ex\hbox{$...$}\kern -1.0em \hbox{$Q$}}
\newcommand{\dddI}{\kern 0.2em \raise 1.9ex\hbox{$...$}\kern -1.0em\hbox{$I$}}
\newcommand{\dddJ}{\kern 0.2em \raise 1.9ex\hbox{$...$}\kern-1.0em
\hbox{$J$}}
\newcommand{\dddcalJ}{\kern 0.2em \raise 1.9ex\hbox{$...$}\kern-1.0em
\hbox{${\cal J}$}}

\newcommand{\dddO}{\kern 0.2em \raise 1.9ex\hbox{$...$}\kern -1.0em
\hbox{${\cal O}$}}
\def\dddz{\raise 1.5ex\hbox{$...$}\kern -0.8em \hbox{$z$}}
\def\dddd{\raise 1.8ex\hbox{$...$}\kern -0.8em \hbox{$d$}}
\def\dddbd{\raise 1.8ex\hbox{$...$}\kern -0.8em \hbox{${\bf d}$}}
\def\ddbd{\raise 1.8ex\hbox{$..$}\kern -0.8em \hbox{${\bf d}$}}
\def\dddx{\raise 1.6ex\hbox{$...$}\kern -0.8em \hbox{$x$}}

\newcommand{\msun}{M_{\odot}}

\newcommand{\Sch}{Schwarzschild }

\newcommand{\mplr}{m_{\rm Pl}}

\newcommand{\ode}{\Omega_{\rm DE}}
\newcommand{\oma}{\Omega_{M}}

\newcommand{\ola}{\Omega_{\Lambda}}

\newcommand{\rde}{\rho_{\rm DE}}
\newcommand{\rvac}{\rho_{\rm vac}}

\graphicspath{ {./Graphs/} }

\makeindex             

\begin{document}

\title*{Nonlocal Infrared Modifications of Gravity.\\ A Review}
\author{Michele Maggiore}
\institute{Michele Maggiore \at D\'epartement de Physique Th\'eorique and Center for Astroparticle Physics,  
Universit\'e de Gen\`eve, 24 quai Ansermet, CH--1211 Gen\`eve 4, Switzerland}
%
%
\maketitle


\vspace*{-25mm}

\abstract{We review  an approach developed in the last few years by our  group in which GR is modified in the infrared, at an effective level, by nonlocal terms associated to a mass scale. We  begin by recalling the notion of quantum effective action  and its associated nonlocalities, illustrating some of their features with the anomaly-induced effective actions in $D=2$ and $D=4$. We examine conceptual issues of nonlocal theories such as causality, degrees of freedoms and ghosts,   stressing the importance of the fact that these  nonlocalities only emerge at the effective level.
We discuss  a particular class of nonlocal theories where the nonlocal operator is associated to a mass scale, and we show that they perform very well in the comparison with  cosmological observations, to the extent that they fit CMB, supernovae, BAO and structure formation data at a level fully competitive with $\Lambda$CDM, with the same number of free parameters. We explore some extensions of these `minimal' models, and   we finally discuss some directions of investigation for deriving the required effective nonlocality from a fundamental local QFT.}

\section{Introduction}\label{sec:1}

I am very glad to contribute to this Volume in honor of prof.~Padmanabhan (Paddy, to his friends), on the occasion of his 60th birthday. I will take this opportunity to give a self-contained account of the work done in the last few years by our group in Geneva, on nonlocal modifications of gravity.

Our motivation comes from cosmology. In  particular, the observation of the  accelerated expansion of the Universe~\cite{Riess:1998cb,Perlmutter:1998np} has revealed the existence of dark energy (DE). The simplest explanation for dark energy is provided by a cosmological constant. Indeed, $\Lambda$CDM has gradually established itself as the cosmological paradigm, since it accurately fits all cosmological data, with a limited set of parameters. From a theoretical point of view, however,  the model is not fully satisfying, because a cosmological constant  is not technically natural from the point of view of the stability under radiative corrections. Independently of such theoretical `prejudices', the really crucial fact is that, with the present and forthcoming cosmological data, alternatives to $\Lambda$CDM are testable, and it is therefore worthwhile to explore them.

At the fundamental level QFT is  local, and in our approach we will not depart from this basic principle. However, both in classical and in quantum field theory, at an effective level nonlocal terms are unavoidably generated. Classically, this happens when one integrates out some degree of freedom to obtain an effective dynamics for the remaining degrees of freedom. Consider for instance a system with 
two degrees of freedom $\phi$ and $\psi$, described classically by two coupled equations of the generic form
$\Box\phi=j(\psi)$ and $\Box\psi=f(\phi)$. The first equation is solved by  $\phi=\Box^{-1} j(\psi)$. This solutions can then be re-injected in the equation for the remaining degree of freedom $\psi$, leading to a nonlocal  equations involving only $\psi$. In QFT, nonlocalities appear in the quantum effective action, as we will review below. The appearance of nonlocal terms involving inverse powers of the d'Alembertian is potentially interesting from a cosmological point of view, since we expect that the $\iBox$ operator becomes relevant in the infrared (IR).

This review is organized as follows. In Sect.~\ref{sect:Qeffact} we recall the notion of quantum effective action, in particular in gravity, and we discuss the associated nonlocalities.
In Sect.~\ref{sect:anominid} we examine two particularly important nonlocal quantum effective actions, the anomaly-induced effective actions in $D=2$ (i.e. the Polyakov quantum effective action) and in $D=4$. 
In Sect.~\ref{sect:nonlocmass} we introduce  a class of nonlocal theories in which the nonlocality  is associated to a  mass scale. In Sects.~\ref{sect:Hownot}, building also on the experience gained in Sect.~~\ref{sect:anominid} with the anomaly-induced effective actions,  we discuss conceptual issues of nonlocal theories, such as causality and degrees of freedom, emphasizing the importance of dealing with them as quantum effective  actions derived from a fundamental local QFT. In Sect.~\ref{sect:loc} we discuss how nonlocal  theories can be formally put in a local form, and we examine the conceptual subtleties associated to the localization procedure concerning the actual  propagating degrees of freedom of the theory.

The cosmological consequences of these nonlocal models are studied in Sect.~\ref{sect:backg} at the level of background evolution, while in Sect.~\ref{sect:cosmpert} we study the cosmological perturbations and in Sect.~\ref{sect:Baye} we present the results of a full Bayesian parameter estimation and the comparison with observational data and with $\Lambda$CDM. In Sect.~\ref{sect:exten} we discuss further possible extensions of the `minimal models', and their phenomenology. 

As we will see, these nonlocal models turn out to be phenomenologically very successful. The next  step will then be understanding how these  nonlocalities emerge. Possible directions of investigations for deriving the required nonlocality from a fundamental theory are briefly explored in 
Sect.~\ref{sect:toward}, although this part is still largely work in progress.

We use  units $\hbar=c=1$, and MTW conventions~\cite{MTW} for the curvature and signature, so in particular  $\emn=(-,+,+,+)$.

\section{Nonlocality and quantum effective actions}\label{sect:Qeffact}

At the quantum level nonlocalities are generated when massless or light particles run into quantum loops. The effect of loop corrections can be summarized into a quantum effective action which,  used at tree level,  takes into account the effect of quantum loops. The quantum effective action is a nonlocal object. 
For instance in QED, if we are interested  in amplitudes where only photons appear in the external legs, 
we can integrate out the electron. The corresponding quantum effective action $\Gamma_{\rm QED}$ is given by
\bees
e^{i\Gamma_{\rm QED}[\Am]}&=&\int {\cal D}\psi{\cal D}\overline{\psi}\, 
\exp\left\{i\int d^4x\, \[ -\frac{1}{4e^2}\Fmn\FMN+\overline{\psi}(i\Dslash -m_e+i\eps)\psi\]\right\}\nn\\
&=&e^{-\frac{i}{4e^2}\int d^4x\, \Fmn\FMN}\, {\rm det}(i\Dslash -m_e+i\eps)\, .
\ees
To quadratic order in the electromagnetic field  this gives
\be\label{qed}
\Gamma_{\rm QED}[\Am]=-\frac{1}{4}\int d^4x\, \[F_{\mu\nu}\frac{1}{e^2(\Box)}F^{\mu\nu} 
+ {\cal O}(F^4)\]\, ,
\ee
where, to one-loop order and in the $\overline{\rm MS}$ scheme~\cite{Dalvit:1994gf},
\be\label{e2MSbar}
\frac{1}{e^2(\Box)}=\frac{1}{e^2(\mu)}-\frac{1}{8\pi^2}\int_0^1dt\, (1-t^2)
\log\[\frac{m_e^2-\frac{1}{4}(1-t^2)\Box }{\mu^2}\]\, .
\ee
Here $\mu$ is the renormalization scale and $e(\mu)$ is the renormalized charge at the scale $\mu$.
In the limit $|\Box/m_e^2|\gg1$,   i.e. when the electron is light with respect to the relevant energy scale, the form factor $1/e^2(\Box)$ becomes 
\be\label{runninge}
\frac{1}{e^2(\Box)}\simeq\frac{1}{e^2(\mu)}-\beta_0\log\(\frac{-\Box}{\mu^2}\)\, ,
\ee
where  $\beta_0=1/(12\pi^2)$. The logarithm of the d'Alembertian is a nonlocal operator defined by
\be
\log\(\frac{-\Box}{\mu^2}\)=\int_0^{\infty}dm^2\, \[\frac{1}{m^2+\mu^2}-
\frac{1}{m^2-\Box}\]\, .
\ee
Thus,  in this case the nonlocality of the effective action is just the running of the coupling constant, expressed in coordinate space. In the opposite limit $|\Box/m_e^2|\ll 1$ the form factor (\ref{e2MSbar})
becomes local,
\be\label{runningelow}
\frac{1}{e^2(\Box)}\simeq\frac{1}{e^2(\mu)}-\beta_0\log\(\frac{m_e^2}{\mu^2}\)\, .
\ee
Observe that  the corresponding beta function, which is obtained by taking the derivative with respect to $\log \mu$, is independent of the fermion mass, so in particular in a theory with several fermions even the heavy fermions would contributes to the beta function, and would  not decouple. Actually, this is a pathology of the $\overline{\rm MS}$ subtraction scheme, and is related to the fact that, when $m_e^2$ is large, \eq{runningelow} develops large   logarithms $\log m_e^2/\mu^2$, so in this scheme perturbation theory breaks down for particles heavy with respect to the relevant energy scales. To study the limit $|\Box/m_e^2|\ll 1$ it can be more convenient to use 
a mass-dependent subtraction scheme, such as subtracting from a divergent graph its value at an Euclidean momentum $p^2=-\mu^2$. Then,  in the limit $|\Box/m_e^2|\ll 1$,
\be
\frac{1}{e^2(\Box)}\simeq\frac{1}{e^2(\mu)}
+\frac{4}{15\, (4\pi)^2}\, \frac{\Box}{m_e^2}\, ,
\ee
so the contribution of a fermion with mass $m_e$ to the beta function is suppressed by a factor $|\Box/m_e^2|$, so the decoupling of heavy particles is explicit~\cite{Manohar:1996cq}.\footnote{Alternatively, in a theory with $N$ fermion fields, one can still use the  $\overline{\rm MS}$ scheme. However, if $m_f$ is the mass of the heaviest among the $N$ fermions, 
at energies $E<m_f$, one must use the theory without the heavy fermion of mass $m_f$, and impose appropriate matching conditions at $E=m_f$ between the theory with $N$ fermions at $E>m_f$ and the theory with $N-1$ fermions at $E<m_f$. One proceeds similarly whenever, lowering the energy, we reach the mass of any of the other fermions. This is the standard way of treating weak interactions at low energies, 
`integrating out' the heavy quarks, see sects. 6 and 7 of \cite{Manohar:1996cq}.} 
Thus, using a mass-dependent subtraction scheme, the effect of a heavy fermion with mass $m_e$, at quadratic order in the fields, is to produce the local higher-derivative operator $\Fmn\Box\Fmn$, suppressed by a factor $1/m_e^2$. Adding to this also the terms 
of order $\Fmn^4$ gives the well-known local  Euler-Heisenberg effective action (see e.g. \cite{Dobado:1998mr} for the explicit computation),  valid in the limit $|\Box/m_e^2|\ll 1$,
\bees\label{Euler}
\Gamma_{\rm QED}[\Am]&\simeq&\int d^4x\, \bigg[-\frac{1}{4e^2(\mu)}F_{\mu\nu}F^{\mu\nu} 
-\frac{1}{15\, (4\pi)^2}\, \frac{1}{m_e^2}\, \Fmn\Box\FMN\nn\\
&&\hspace*{11mm}+\frac{e^2(\mu)}{90 (4\pi)^2}\, \frac{1}{m_e^4}\, \(  (\FMN\Fmn)^2+\frac{7}{4}(\FMN\tilde{F}_{\mu\nu})^2\)
\bigg]\, .
\ees
To sum up, nonlocalities emerge in the quantum effective action when we integrate out a particle which is light compared to the relevant energy scale. In contrast, heavy particles give local contributions which, if computed in a mass-dependent subtraction scheme, are encoded in higher-dimension local operators suppressed by inverse powers of  the particle mass.

The quantum effective action is a particularly useful tool in gravity, where the integration over matter fields gives the quantum effective action for the metric (see e.g. \cite{Birrell:1982ix,Buchbinder:1992rb,Mukhanov:2007zz,Shapiro:2008sf} for  pedagogical introductions).
Let us denote collectively all  matter fields as $\phi$, and the fundamental matter action by 
$S_m[\gmn,\phi]$. Then the quantum effective action $\Gamma$ is given by 
\be\label{Gamma}
e^{i\Gamma[\gmn]}=e^{iS_{\rm EH}[\gmn]}\, \int{\cal D}\phi\, e^{iS_m[\gmn,\phi]}\, ,
\ee
where $S_{\rm EH}$ is the Einstein-Hilbert action.\footnote{Depending on the conventions, $\Gamma$ can be defined so that it includes $S_{\rm EH}$, or just as the term to be added to $S_{\rm EH}$.}
The effective quantum action $\Gamma$ determines the  dynamics of the metric, including the backreaction from quantum loops of matter fields. Even if the fundamental action $S_m[\gmn,\phi]$ is local, again the quantum effective action for gravity is unavoidably nonlocal. Its nonlocal part  describes the  running of coupling constants, as in \eq{qed}, and other effects such as particle production in the external gravitational field.

The matter energy-momentum tensor $\TMN$ is given by the variation of the fundamental action, according to the standard GR expression
$\TMN=(2/\sqrt{-g})\d S_m/\d\gmn$. In contrast, the variation of the effective quantum action gives the {\em vacuum expectation value} of the energy-momentum tensor,
\be\label{Tmnvac}
\cav\TMN\vac=\frac{2}{\sqrt{-g}}\, \frac{\d\Gamma}{\d\gmn}\, .
\ee
More precisely, the in-out expectation value $\langle 0_{\rm out}|\TMN|0_{\rm in}\rangle$ is obtained when the path-integral in \eq{Gamma} is the standard Feynman path-integral, while using
the Schwinger-Keldish path integral gives 
the in-in
expectation value $\langle 0_{\rm in}|\TMN|0_{\rm in}\rangle$. This point will be important for the discussion of the causality of the effective nonlocal theory, and we will get back to it in Sect.~\ref{sect:causality}.

In principle, in \eq{Gamma} one could expand $\gmn=\emn+\hmn$ and compute perturbatively in $\hmn$. A much more powerful and explicitly covariant computational method is based on the heat-kernel technique (see e.g. \cite{Mukhanov:2007zz} for review), combined with an expansion in powers of the curvature. In this way
Barvinsky and Vilkovisky~\cite{Barvinsky:1985an,Barvinsky:1987uw} have developed a formalisms that allows one to  compute, in a covariant manner, the  gravitational effective action as an expansion in powers of the curvature, including the nonlocal terms. The resulting quantum effective action, up to terms quadratic in the curvature,  has the form 
\be\label{formfact}
\Gamma=\frac{\mplr^2}{2}\int d^4x \sqrt{-g}\,R
+\frac{1}{2(4\pi)^2}\,\int d^4x  \sqrt{-g}\, 
\[  R \,k_R(\Box) R +\frac{1}{2}C_{\mu\nu\rho\sigma}k_W(\Box)C^{\mu\nu\rho\sigma}\]\, ,
\ee
where $\mplr$ is the reduced Planck mass, $\mplr^2=1/(8\pi G)$, $C_{\mu\nu\rho\sigma}$ is the Weyl tensor, and we used as a basis for the quadratic term $R^2$, $C_{\mu\nu\rho\sigma}C^{\mu\nu\rho\sigma}$ and the Gauss-Bonnet term, that we have not written explicitly. Just as in \eq{runninge},
in the case of loops of
massless particles the  form factors $k_R(\Box)$ and $k_W(\Box)$ only contain logarithmic terms plus finite parts, i.e.  $k_{R,W}(\Box)=c_{R,W}\log (\Box/\mu^2)$,
where now $\Box$ is the  generally-covariant d'Alembertian, $\mu$ is the renormalization point, and $c_R ,c_W$  are known coefficients that depend on the number of matter species and on their spin. The form factors generated by loops of a massive particles  are more complicated. For instance, for a massive scalar field  with mass $m_s$ and  action
\be\label{scalaract}
S_s=-\frac{1}{2} \int d^4x\, \sqrt{-g}\, \(\gMN\pam\phi\pan\phi+m_s^2\phi^2+\xi R\phi^2\)\, ,
\ee
the form factors $k_R(-\Box/m_s^2)$ and $k_W(-\Box/m_s^2)$ in \eq{formfact} were  computed in \cite{Gorbar:2002pw,Gorbar:2003yt}
in closed form, for $(\Box/m_s^2)$ generic, in a mass-dependent subtraction scheme where the decoupling of heavy particles is explicit. After subtracting the divergent part, the result is 
\bees
k_W(-\Box/m_s^2) &=& \frac{8A}{15\,a^4}
\,+\,\frac{2}{45\,a^2}\,+\,\frac{1}{150}+\frac{1}{60}\log\frac{\mu^2}{m_s^2}\,, \label{kW}\\
k_R(-\Box/m_s^2) &=&
\bar{\xi}^2A
+ \(\frac{2A}{3a^2}-\frac{A}{6}+ \frac{1}{18}\)\bar{\xi}
+ A\( \frac{1}{9a^4}- \frac{1}{18a^2}
+ \frac{1}{144} \)\nn\\
&&+ \frac{1}{108\,a^2} -\frac{7}{2160}+\frac{1}{2}\bar{\xi}^2\log\frac{\mu^2}{m_s^2}
\,, \label{kR}
\ees
where $\bar{\xi}=\xi -(1/6)$, and
\be
A\,=\,1-\frac{1}{a}\log\,\Big(\frac{2+a}{2-a}\Big)\,, \qquad a^2 =
\frac{4\Box}{\Box - 4m_s^2}\, .
\ee
In the limit  $|\Box/m_s^2|\gg 1$ (i.e. in the limit in which the particle is very light compared to the typical energy or curvature scales), \eq{kR} has the expansion
\be\label{expan}
k_R\(\frac{-\Box}{m_s^2}\)=\alpha\log\(\frac{-\Box}{m_s^2}\)+\beta \frac{m_s^2}{\Box}
+\gamma \,\frac{m_s^2}{\Box}\log\(\frac{-\Box}{m_s^2}\)+\delta \, \frac{m_s^4}{\Box^2}+\ldots \, ,
\ee
and similarly for $k_W$. This result has also been re-obtained with effective field theory 
techniques~\cite{Donoghue:1994dn,Donoghue:2014yha,Codello:2015mba}. Similar results can also be obtained for different spins, so in the end the coefficients  $\a,\beta,\gamma,\delta$  depend on the number and type of massive particles.

The result further simplifies for a massless conformally-invariant scalar field.   Taking the limit $m_s\ra 0$, $\xi\ra 1/6$ in \eq{formfact} one finds that the terms involving $\log m_s^2$ cancel and  the form factor $k_R(\Box)$ becomes local, $k_R=-1/1080$, while $k_W(\Box)\ra -(1/60)\log (-\Box/\mu^2)$. Similar results, with different coefficients, are obtained from massless vectors and spinor fields. So, for conformal matter,  the one-loop effective action has the form
\be\label{formfactconfmat}
\Gamma_{\rm conf.\,  matter}=\int d^{4}x \sqrt{-g}\, \bigg[\frac{\mplr^2}{2}R +c_1R^2 +c_2C_{\mu\nu\rho\sigma}\log (-\Box/\mu^2)C^{\mu\nu\rho\sigma}+{\cal O}(\Rmnrs^3)\bigg]\, ,
\ee 
where $c_1,c_2$ are known coefficients that depends on the number and type of conformal matter fields, and we have stressed that the computation leading to \eq{formfactconfmat} has been performed only up to terms quadratic  in the curvature.

In contrast, when the particle is heavy compared to the relevant energy or curvature scales, i.e. in the 
limit $-\Box/m_s^2\ll 1$, the form factors in \eqs{kW}{kR} become local,
\be\label{dec}
k_W(-\Box/m_s^2),\, k_R(-\Box/m_s^2)={\cal O}(\Box/m_s^2)\, .
\ee
Again, this expresses  the fact that particles which are massive compared to the relevant energy scale decouple,  leaving a local contribution to the effective action proportional to higher derivatives, and  suppressed by inverse powers of the mass. This decoupling is explicit in the mass-dependent subtraction scheme used in refs.~\cite{Gorbar:2002pw,Gorbar:2003yt}.

\section{The anomaly-induced effective action}\label{sect:anominid}

In a theory with massless, conformally-coupled matter fields, in $D=2$ space-time dimensions, the quantum effective action can be computed {\em exactly}, at all perturbative orders, by integrating the conformal anomaly. In $D=4$ one can obtain in this way, again exactly, the part of the quantum effective  action that depends on the conformal mode of the metric. 

These examples of quantum effective actions for the gravitational field will be relevant for us when we discuss how the nonlocal models that we will propose can emerge from a fundamental local theory. They also provide an explicit example  of the fact that effective quantum actions must be treated differently from fundamental QFT, otherwise one might be fooled into believing that they contain, e.g., ghost-like degrees of freedom, when in fact the fundamental theories from which they are derived are perfectly healthy. We will then devote this section to recalling basic facts on the anomaly-induced effective action, both in $D=2$ and in $D=4$ (see e.g. ~\cite{Birrell:1982ix,Antoniadis:1991fa,Buchbinder:1992rb,Antoniadis:2006wq,Mukhanov:2007zz,Shapiro:2008sf} for reviews).

\subsection{The anomaly-induced effective action in $D=2$} 

Consider 2D gravity coupled to $N_s$ conformally-coupled massless scalars [i.e. $m_s=0$ and $\xi=1/6$ in \eq{scalaract}]
and $N_f$ massless Dirac fermions. We take these fields to be free, apart from their interaction with gravity.  For conformal matter fields, classically the  trace $T^{a}_{a}$ of the energy-momentum tensor  vanishes [in $D=2$ we use $a=0,1$ as Lorentz indices, and signature $\eta_{ab}=(-,+)$]. However, at the quantum level the vacuum expectation value of $T^{a}_{a}$ is non-zero, and is given by
\be\label{traceT}
\cav T^{a}_{a}\vac=\frac{N}{24\pi} R\, ,
\ee
where $N=N_s+N_f$. \Eq{traceT} is the trace anomaly. The crucial point about this result is that, even if it can be obtained with a one-loop computation, it is actually {\em exact}.\footnote{For the trace anomaly (\ref{traceT}), this can be shown using the Seeley-DeWitt expansion of the heat kernel, see sect.~14.3 of \cite{Mukhanov:2007zz}.}
No contribution to the trace anomaly comes from higher loops. We can now find the effective action that reproduces the trace anomaly, by integrating \eq{Tmnvac}. We write 
\be \label{confgauge}
g_{ab}=e^{2\sigma}\bar{g}_{ab}\, , 
\ee
where $\bar{g}_{ab}$ is a fixed reference metric. 
The corresponding Ricci scalar is 
\be\label{RRbar2D} 
R= e^{-2\sigma}(\overline{R}-2\Boxbar\sigma)\, , 
\ee
where the overbars denotes the quantities computed with the metric $\bar{g}_{ab}$. In $D=2$, \eq{Tmnvac} gives
\be
\d\Gamma=\frac{1}{2}\int d^2x\, \sqrt{-g}\,\, \cav T^{ab} \vac \d g_{ab} 
=\int d^2x\, \sqrt{-g}\,\, \cav T^{ab}\vac g_{ab} \d\sigma\, .
\ee
Therefore
\be\label{dGdsD2}
\frac{\d\Gamma}{\d\sigma}=2g_{ab} \frac{\d\Gamma}{\d g_{ab} }=
\sqrt{-g}\,  \cav T^{a}_{a}\vac\, ,
\ee
where $T^{a}_{a}=g_{ab} T^{ab}$. 
In $D=2$, without loss of generality, locally we can always write the metric as $g_{ab} =e^{2\sigma}\eta_{ab}$, i.e. we can chose $\bar{g}_{ab}=\eta_{ab}$. In this case, from \eq{RRbar2D}, 
\be\label{Rsigma2D}
R=-2e^{-2\sigma}\Box_{\eta}\sigma\, ,
\ee 
where $\Box_{\eta}$ is the flat-space d'Alembertian, $\Box_{\eta}=\eta^{ab}\pa_a\pa_b$.
Then, inserting \eq{traceT} into \eq{dGdsD2} and using  $\sqrt{-g}= e^{2\sigma}$, we get
\be\label{dGdsBoxs}
\frac{\d\Gamma}{\d\sigma}=-\frac{N}{12\pi}\, \Box\sigma\, .
\ee
This can be integrated to obtain
\be
\Gamma[\sigma]-\Gamma[0]=-\frac{N}{24\pi}\, \int d^2x\, \sigma\Box_{\eta}\sigma\, .
\ee
We see that, in general, the trace anomaly determines the effective action only modulo a term 
$\Gamma[0]$ independent of the conformal mode. However, in the special case $D=2$, when $\sigma=0$ we can choose the coordinates so that, locally, $g_{ab}=\eta_{ab}$. Thus, all curvature invariants vanish when $\sigma=0$, and therefore $\Gamma[0]=0$. Therefore, in $D=2$ the trace anomaly determines {\em exactly} the quantum effective action, at all perturbative orders!
Finally, we can rewrite this effective action in a generally-covariant but non-local form observing that $\Box_g=e^{-2\sigma}\Box_{\eta}$, where $\Box_g$ is the d'Alembertian computed with the full metric $g_{ab}=e^{2\sigma}\eta_{ab}$. Then, from \eq{Rsigma2D}, $R=-2\Box_g\sigma$, which can be inverted to give 
$\sigma=-(1/2) \Box_g^{-1}R$, so that
\bees\label{SPolyakov}
\Gamma[\gmn]&=&-\frac{N}{24\pi}\, \int d^2x\,e^{2\sigma} \sigma\Box_g\sigma\nn\\
&=&-\frac{N}{96\pi}\, \int d^2x\, \sqrt{-g}\, R\Box_g^{-1}R\, .
\ees
This is  the Polyakov quantum effective action. The remarkable fact about this effective quantum action is that, even if it has been obtained from the one-loop computation of the trace anomaly, it is the {\em exact} quantum effective action, to all perturbative orders.

In the above derivation we have studied matter fields in a fixed gravitational background. We now add the dynamics for the metric itself, i.e. we consider 
2D gravity, including also a cosmological constant $\lambda$, coupled to $N$ massless matter fields,
\be\label{Skappalambda}
S=\int d^2x\, \sqrt{-g } (\kappa R-\lambda) +S_m\, ,
\ee
where $S_m$  is the the action describing $N=N_S+N_F$  conformally-coupled massless scalar and massless Dirac fermion fields.
In 2D the Einstein-Hilbert term is a topological invariant and, once we integrate out the massless matter field, all the gravitational dynamics comes from the anomaly-induced effective action. The contribution of the $N$ matter fields is given by the Polyakov effective action (\ref{SPolyakov}). Diff invariance  fixes locally $g_{ab}=e^{2\sigma}\bar{g}_{ab}$, where $\bar{g}_{ab}$ is a reference metric. In  a theory with dynamical gravity, where in the path integral we also integrate over $g_{ab}$, this is now a gauge fixing condition, and the corresponding
reparametrization ghosts give a contribution $-26$ to be added to $N$, while  the conformal factor  $\sigma$ gives a contribution $+1$~\cite{Knizhnik:1988ak,David:1988hj,Distler:1988jt}.
Then, after dropping the topologically-invariant Einstein-Hilbert term,  the exact quantum effective action of $2D$ gravity reads 
\be\label{SPolyakov26}
\Gamma=-\frac{N-25}{96\pi}\int d^2x\sqrt{-g}\, R\frac{1}{\Box}R -\lambda\int d^2x\, \sqrt{-g } \, ,
\ee
with an overall factor in the nonlocal term  proportional to $(N-25)$.\footnote{In bosonic string theory
$\lambda=0$ and, beside diff invariance, one also has  Weyl invariance on the world-sheet. This  allows one to eliminate also $\sigma$, so  one only has the contribution $-26$ from the reparametrization ghosts, together with the contribution from the   $N=D$ matter fields  $X^{\mu}(\sigma_1,\sigma_2)$ living in the world-sheet, where $\mu=0,\ldots ,D-1$ and $D$ is the number of spacetime dimensions of the target space. Then the coefficient in the anomaly-induced effective  action is proportional to $D-26$, leading to the condition $D=26$ for the anomaly cancellation, necessary for the elimination of the ghost-like $X^0$ field. } 
Using \eq{RRbar2D} and dropping a $\sigma$-independent term 
$\sqrt{-\bar{g}}\, \overline{R}\Boxbar^{-1}\overline{R}$ we see that, in terms of the conformal mode,
\eq{SPolyakov26} becomes local,
\be\label{SPolyakov26sigmatot}
\Gamma=\int d^2x\, \sqrt{-\bar{g}}\[ \frac{N-25}{24\pi}\bar{g}^{ab}\pa_a\sigma\pa_b\sigma +\frac{N-25}{24\pi}\overline{R}\sigma -\lambda e^{2\sigma}\]\, ,
\ee
which is the action of Liouville field theory. 

\Eq{SPolyakov26sigmatot} also allows us to illustrate an issue that will emerge later, in the context of the nonlocal model that we will propose. 
If we try to read  the spectrum of the quantum theory from \eq{SPolyakov26sigmatot}, treating it as if it were the fundamental action of a QFT,  we would conclude that, for $N\neq 25$, there is one dynamical degree of freedom, $\sigma$. Recalling that our signature is $\eta_{ab}=(-,+)$, we would also conclude that for $N>25$ this degree of freedom is a ghost and for $N<25$ it has a normal kinetic term. 

However, this conclusion is wrong.
\Eq{SPolyakov26sigmatot} is the quantum effective action of a fundamental theory which is just 2D gravity coupled to $N$ healthy fields, in which there is no ghost in the spectrum of the fundamental  theory. If we perform the quantization of the fundamental theory in the conformal gauge (\ref{confgauge}), the fields involved are the matter fields, the reparametrization ghosts, and the only surviving component of the metric once we have fixed the conformal gauge, i.e. the conformal factor $\sigma$. Each of them has its own creation and annihilation operators, which  generate the full Hilbert space of the theory. However, as always in theories with a local invariance (in this case diff invariance) the physical Hilbert space is a subset of the full Hilbert space. The condition on physical states can be obtained  requiring that the amplitude $\langle f | i \rangle$ between an initial state $|i\rangle$ and a final state $|f\rangle$ is invariant under a change of gauge fixing (see e.g. chap.~4 of \cite{Polchinski:1998rq} for a discussion in the context of bosonic string theory). From this it follows that two states $|s\rangle$ and $|s'\rangle$ are physical if and only if
\be\label{physstateTab}
\langle s'| T^{ab}_{\rm tot} |s\rangle=0\, ,
\ee
where $T^{ab}_{\rm tot}$ is the sum of the energy-momentum tensors of matter, ghosts and $\sigma$. This condition (or, more, precisely,  the condition that physical states must by BRST invariant) eliminates from the physical spectrum both the states associated with the reparametrization ghosts, and the states generated by the creation operators of the conformal mode, as explicitly proven in \cite{Polchinski:1989fn}. Of course, the physical-state condition (\ref{physstateTab}) is the analogous of the physical-state condition
\be\label{physstateAmu}
\langle s'| \pam A^{\mu} |s\rangle=0\, 
\ee
in the Gupta-Bleuler quantization of electrodynamics, which again eliminates from the physical spectrum the would-be ghost states associated to $A_0$.

What we learn from this example is that, if we start from a theory such as (\ref{SPolyakov26sigmatot}), e.g. to explore its cosmological consequences, 
there is a huge difference between the situation in which we take it to be  a fundamental QFT, and the situation in which
we consider  it as  the quantum effective action of some underlying fundamental theory.
In the former case, in the theory  (\ref{SPolyakov26sigmatot}) we would treat $\sigma$ as a scalar field living in 2D, and the theory would have one degree of freedom, which is a ghost for $N>25$ and a healthy scalar for $N<25$, while for $N=25$ there would be no dynamics at all. In contrast, when \eq{SPolyakov26sigmatot} is treated as the effective quantum action derived from the fundamental QFT theory  (\ref{Skappalambda}), the interpretation is completely different. The field $\sigma$ is not just a scalar field living in 2D, but the component of the 2D metric that remains after gauge fixing. The physical spectrum of the fundamental theory is given by the quanta of the $N$ healthy matter fields, which are no longer visible in 
(\ref{SPolyakov26sigmatot}) because they have been integrated out. There is no ghost, independently of the value of $N$, and there are no physical quanta associated to $\sigma$, because they are removed by the physical-state condition associated to the diff invariance of the underlying fundamental theory.

As a final  remark, observe that the fact that no physical quanta are associated to $\sigma$ does not mean that the field $\sigma$ itself has no physical effects. The situation is again the same  as in electrodynamics, where there are no physical quanta associated to $A_0$, but still the interaction mediated by $A_0$ generates the Coulomb potential between static charges. In other words, the quanta associated to $\sigma$ (or to $A_0$ in QED) cannot appear in the external lines of Feynman diagram, since there are no physical states associated to them, but do appear in the internal lines.

\subsection{The anomaly-induced effective action in $D=4$}\label{sect:anomalyD4}

Let us now follow the same strategy in $D=4$ space-time dimensions, again for massless conformally-coupled matter fields. As we will see, in this case we will not be able to compute the  quantum effective action exactly, but still we will be able to obtain valuable non-perturbative information from the  trace anomaly.
In $D=4$ the trace anomaly is
\be\label{traceT4d}
\cav T^{\mu}_{\mu}\vac=b_1 C^2+ b_2\(E-\frac{2}{3}\Box R\)+b_3\Box R\, ,
\ee
where $C^2$ is the square of the Weyl tensor, $E$ the Gauss-Bonnet term,   and it is convenient to use as independent  combinations $[E-(2/3)\Box R]$ and $\Box R$, rather than $E$ and $\Box R$. The coefficients $b_1,b_2,b_3$ are known constants that depend on the number of massless conformally-coupled scalars, massless fermions and massless vector fields. Once again, the anomaly receives contribution only at one loop order, so \eq{traceT4d} is {\em exact}.
Let us now write again 
\be\label{defsig}
\gmn=e^{2\sigma}\gbmn\, .
\ee
A crucial difference compared to the $2D$ case is that in $D=4$ diff invariance no longer allows us to set $\gbmn=\emn$. \Eq{dGdsD2} still holds, so the anomaly-induced effective action satisfies
\be\label{dGdsD4}
\frac{\d\Gamma_{\rm anom}}{\d\sigma}=\sqrt{-g}\[ b_1 C^2+ b_2\(E-\frac{2}{3}\Box R\)+b_3\Box R\]\, .
\ee
We have added the subscript `anom' to stress that this is the part of the effective action which is obtained from the anomaly. The total quantum effective action is obtained adding $\Gamma_{\rm anom}$ to the classical Einstein-Hilbert term.

To integrate \eq{dGdsD4} we first of all observe that the $\Box R$ term can be obtained from the variation of a local $R^2$ term,
\be
\gmn\, \frac{\d}{\d\gmn}\int d^4x\, \sqrt{-g}\, R^2=-6\sqrt{-g}\,\, \Box R\, .
\ee
To integrate the other terms we observe that
\bees
\sqrt{-g} \,C^2&=&\sqrt{-\bar{g}}\, \bar{C}^2\, ,\\
\sqrt{-g}\,\(E-\frac{2}{3}\Box R\)&=&\sqrt{-\bar{g}}\, \(\bar{E}-\frac{2}{3}\Boxbar \,\bar{R}+4\bar{\Delta}_4\sigma \)\, ,
\ees
where the overbars denotes the quantities computed with the metric $\gbmn$, and $\Delta_4$ is the Paneitz operator
\be\label{defDP}
\Delta_4\equiv\Box^2+2\RMN\n_{\mu}\n_{\nu}-\frac{2}{3}R\Box+\frac{1}{3}\gMN\n_{\mu} R\n_{\nu}\, .
\ee
Thus, we get 
\bees
\Gamma_{\rm anom}[\gmn]&=&\Gamma_{\rm anom}[\gbmn]-\frac{b_3}{12}\int d^4x\, \sqrt{-g}\, R^2\nn\\
&&+\int d^4x\, \sqrt{-\bar{g}}\, \[ b_1 \sigma \bar{C}^2 +
b_2 \sigma   \(\bar{E}-\frac{2}{3}\Boxbar\, \bar{R}\) +2b_2\sigma\bar{\Delta}_4\sigma \]\, ,
\label{GammagmnRs}
\ees
where $\Gamma_{\rm anom}[\gbmn]$ is an undetermined integration `constant', i.e. a term independent of $\sigma$, equal to $\Gamma_{\rm anom}[\gmn]$ evaluated at $\sigma=0$.
We will discuss below the possible covariantizations of the term in the second line. First, we can rewrite everything in terms of $\sigma$ and $\gbmn$ using
\be
R=e^{-2\sigma}
\[ \bar{R}-6\Boxbar\sigma-6\overline{\n}_{\mu}\sigma\overline{\n}^{\mu}\sigma\]\, .
\ee
Then
\bees
\Gamma_{\rm anom}[\gmn]&=&
\Gamma_{\rm anom}[\gbmn]-\frac{b_3}{12}\int d^4x\, \sqrt{-\bar{g}}\, 
\[ \bar{R}-6\Boxbar\sigma-6\overline{\n}_{\mu}\sigma\overline{\n}^{\mu}\sigma\]^2
\nn\\
&&+\int d^4x\, \sqrt{-\bar{g}}\, \[ b_1 \sigma \bar{C}^2 +
b_2 \sigma   \(\bar{E}-\frac{2}{3}\Boxbar\, \bar{R}\) +2b_2\sigma\bar{\Delta}_4\sigma \]\, .
\label{anominducSeffD4}
\ees
Once again, the trace anomaly allowed us to determine {\em exactly} the dependence of the action on the conformal mode $\sigma$. However, we cannot determine in this way the $\sigma$-independent part of the effective action, $\Gamma_{\rm anom}[\gbmn]$. This is an important difference compared to the $D=2$ case, where we could show that $\Gamma_{\rm anom}[\bar{g}_{ab}]=0$ using the fact that locally we can always choose $g_{ab}=\eta_{ab}$.
In the end, the effective action must be a function of $\gbmn$ and $\sigma$ only in the combination $\gmn=e^{2\sigma}\gbmn$, so the 
$\sigma$-independent term $\Gamma_{\rm anom}[\gbmn]$ is just the conformally-invariant part of the effective action, $\Gamma_{\rm c}[\gmn]$, which by definition satisfies
\be
\Gamma_c[e^{2\sigma}\gbmn]=\Gamma_c[\gbmn]\, .
\ee
It is  interesting to compare the anomaly-induced effective action (\ref{anominducSeffD4}) with the conformal limit of the explicit one-loop computation given in \eq{formfactconfmat}. First of all,  the anomaly-induced effective action has a local $R^2$ term, coming  both from the explicit $b_3R^2$ term and from the term $(-2/3)b_2\sigma \Box \bar{R}$, corresponding to the two terms proportional to $\Box R$ in \eq{dGdsD4}.
The value of its overall  coefficient $-[b_3-(2/3)b_2]/12$,  obtained from the trace anomaly as a function of the number of conformal massless scalar, massless spinor and massless vector fields, agrees with the coefficient $c_1$ obtained from the one-loop computation, as it should. 
Consider now the Weyl-square term in \eq{formfactconfmat}. Recall  that \eq{formfactconfmat} is valid only up to second order in the curvature. Thus, strictly speaking, in the term $C_{\mu\nu\rho\sigma}\log (-\Box/\mu^2)C^{\mu\nu\rho\sigma}$, the $\Box$ operator is the flat-space d'Alembertian. If one would  compute  to higher orders in the curvature, this term should  naturally become  a covariant d'Alembertian acting on a tensor such as $C^{\mu\nu\rho\sigma}$. The covariantization of  the $\log(\Box)$ operator  acting on such a tensor is a non-trivial problem, see the discussion in \cite{Deser:1999zv,Donoghue:2015nba}. In any case we expect that, at least in the simple case of $\gmn=e^{2\sigma}\gbmn$ with $\sigma$ constant, we
will have $\Box_g=e^{-2\sigma}\Box_{\bar{g}}$, just as for the scalar d'Alembertian. Then,
\be
C_{\mu\nu\rho\sigma}\log (-\Box/\mu^2)C^{\mu\nu\rho\sigma}=-2\sigma C^2 +
C_{\mu\nu\rho\sigma}\log (-\Boxbar/\mu^2)C^{\mu\nu\rho\sigma}\, .
\ee
The second term on the right-hand side, once multiplied by $\sqrt{-g}$, is independent of $\sigma$ and therefore belongs to $\Gamma_c[\gbmn]$. On the other hand, the term proportional to $\sqrt{-g} \, \sigma C^2=\sqrt{-\bar{g}}\, \sigma \bar{C}^2$ is just the term proportional to $b_1$ in \eq{anominducSeffD4}. Once again, 
one  can check that the numerical value of the coefficient from the explicit one-loop computation and from the trace anomaly agree. We see that the anomaly-induced effective action and the explicit one-loop computation give complementary information. The anomaly-induced effective action misses all terms independent of $\sigma$, such as the term proportional to $C_{\mu\nu\rho\sigma}\log (-\Boxbar/\mu^2)C^{\mu\nu\rho\sigma}$  that gives the logarithmic running of the coupling constant associated to $C^2$. However, the terms that depend on the conformal mode are obtained {\em exactly}, without any restriction to quadratic order in the curvature.

One can now look for a covariantization of \eq{GammagmnRs}, in which everything is written in terms of $\gmn=e^{2\sigma}\gbmn$.  In general, the covariantization of an expression is not unique. A possible covariantization is given by the Riegert action~\cite{Riegert:1984kt}
\bees
\Gamma_{\rm anom}[\gmn]&=&\Gamma_c[\gmn]-\frac{b_3}{12}\int d^4x\, \sqrt{-g}\, R^2\label{GammaRiegert}\\
&&+\frac{1}{8}\int d^4x\sqrt{-g}\(E-\frac{2}{3}\Box R\) \Delta_4^{-1}
\[ b_2 \(E-\frac{2}{3}\Box R\)+2b_1 C^2\]\, .\nn
\ees
Just as for the Polyakov action, even if the anomaly-induced action is local when written in terms of the conformal factor, it becomes nonlocal when written in terms of curvature tensors. 
In this covariantization, as we have seen, the $\log\Box$ form factor in \eq{formfactconfmat} is not really visible since the term $C_{\mu\nu\rho\sigma}\log (-\Boxbar/\mu^2)C^{\mu\nu\rho\sigma}$ is hidden in $\Gamma_{\rm conf}[\gmn]$. Alternative ways of covariantizing the $\log\Box$ operator are discussed in~\cite{Deser:1999zv,Donoghue:2015nba}. In any case, in the approximation in which  one is interested only  in the dynamics of the conformal mode one can use the effective action in the form (\ref{anominducSeffD4}), simply dropping the $\sigma$-independent term $\Gamma[\gbmn]$, independently of the covariantization chosen.

Once again, if one  uses \eq{anominducSeffD4} as if it were a fundamental QFT, one would reach the conclusion that this theory contains a ghost. This would be an unavoidable consequence of the presence of the four-derivative term $\sigma\bar{\Delta}_4\sigma$ in
\eq{anominducSeffD4} which, expanding over flat space and after integrations by parts, is simply $(\Box\sigma)^2$. As a fundamental QFT, the theory defined by \eq{anominducSeffD4} would then be hopelessly sick. In contrast, we have seen that \eq{anominducSeffD4} is 
the quantum effective action derived from a fundamental and healthy quantum theory, with no ghost. One could still wander whether the appearance a four-derivative term $\sigma\bar{\Delta}_4\sigma$ signals the fact that a new ghost-like state emerges in the theory because of quantum fluctuations. To answer this question one can quantize the theory (\ref{anominducSeffD4}), and see which states survive the physical-state condition, analogous to \eq{physstateTab} in $D=2$, which reflects the diff-invariance of the underlying fundamental theory. This analysis has been carried out in \cite{Antoniadis:1995dy} and it was found that,
once one imposes the physical state condition, there is no local propagating degree of freedom associated to $\sigma$. Rather, we remain with an infinite  tower of {\em discrete} states, one for each level, all with positive norm. In the limit $Q^2/(4\pi)^2\equiv-2b_2\ra\infty$, these states have the form
$\int d^4x\sqrt{-g}\, R^n\vac$.

\section{Nonlocality and mass terms}\label{sect:nonlocmass}

In this section we introduce a class of nonlocal theories where the nonlocality is associated to a mass term.
In Sect.~\ref{sect:Hownot}, using also the experience gained with the study of the anomaly-induced effective action, we will discuss some conceptual issues (such as causality and ghosts) in these theories.
A different class of nonlocal models, which do not feature an explicit mass scale, has been introduced in
\cite{Deser:2007jk,Deser:2013uya}, and reviewed in \cite{Woodard:2014iga}. In this review we will rather focus on the nonlocal models where the nonlocal terms are associated to a mass scale.

\vspace*{-5mm}

\subsection{Nonlocal terms and massive gauge theories}


A simple and instructive example of how a nonlocal term can appear in the description of a massive gauge theory is  given by  massive electrodynamics. Consider the the Proca action with an external conserved current $j^{\mu}$
\be\label{1Lemmass}
S=\int d^4x\[ -\frac{1}{4}F_{\mu\nu}F^{\mu\nu}-\frac{1}{2}m_{\g}^2\, A_{\mu}A^{\mu}
-j_{\mu}A^{\mu}\] \, .
\ee
The equations of motion 
obtained from (\ref{1Lemmass}) are
\be\label{1FAj}
\pam F^{\mu\nu}-m_{\g}^2A^{\nu}=j^{\nu}\, .
\ee
Acting with $\pan$ on both sides and using  $\pan j^{\nu}=0$, \eq{1FAj} gives
\be\label{1mgAmu}
m_{\g}^2\, \pan A^{\nu}=0\, .
\ee
Thus, if $m_{\g}\neq 0$, we get the condition
$\pan A^{\nu}=0$ dynamically, as a consequence of the equation of motion, and we have eliminated one degree of
freedom. Making use of \eq{1mgAmu}, \eq{1FAj} becomes
\be\label{BoxmA0}
(\Box -m_{\g}^2) A^{\mu}=j^{\mu}\, .
\ee
\Eqs{1mgAmu}{BoxmA0} together describe the three degrees of freedom of a massive photon.
In this formulation locality is manifest, while the  $U(1)$ gauge invariance
of the massless theory  is lost, because of the non gauge-invariant term $m_{\g}^2\,A_{\mu}A^{\mu}$ in the Lagrangian. However,
as shown in \cite{Dvali:2006su}, this theory can be rewritten in a gauge-invariant but nonlocal form. Consider in fact the equation of motion
\be\label{eqnonlocFMN}
\(1-\frac{m_{\g}^2}{\Box}\)\pam\FMN=j^{\nu}\, ,
\ee
or, rewriting it in terms of $\Am$, 
\be\label{eqnonlocAN}
(\Box-m_{\g}^2)\AN=\(1-\frac{m_{\g}^2}{\Box}\)\paN\pam\AMU+j^{\nu}\, .
\ee
\Eq{eqnonlocFMN} is clearly gauge invariant. We can therefore chose the gauge $\pam A^{\mu}=0$. As we see more easily from \eq{eqnonlocAN}, in this gauge the nonlocal term vanishes, and \eq{eqnonlocAN} reduces to   the local equation $(\Box -m_{\g}^2) A^{\nu}=j^{\nu}$. Thus, we end up with the same equations  as in  Proca theory,
$(\Box -m_{\g}^2) A^{\mu}=j^{\mu}$ and  $ \pam A^{\mu}=0$.
Note however that  they were obtained in  a different way:
in the Proca theory there is no gauge invariance to be fixed, but \eq{1mgAmu} comes out dynamically, as a consequence of the equations of motion, while in the theory (\ref{eqnonlocFMN}) there is a gauge invariance and $\pam A^{\mu}=0$  can be imposed as a gauge condition. In any case, since the equations of motions are finally the same, we see that the theory defined by 
(\ref{eqnonlocFMN}) is  classical equivalent to the theory defined by \eq{1Lemmass}. Observe also that \eq{eqnonlocFMN} can be formally obtained by taking the variation of the nonlocal action
\be\label{Lnonloc}
S=-\frac{1}{4}\int d^4x\,\[ \Fmn \(1-\frac{m_{\g}^2}{\Box}\)\FMN
-j_{\mu}A^{\mu}\]\, ,
\ee
(apart from a subtlety in the variation of $\iBox$, that we will discuss in Sect.~\ref{sect:causality}).\footnote{The equivalence of the two theories can also be directly proved
using  the ``\Stu trick":    
one introduces a scalar  field $\varphi$ and  
replaces $\Am\ra\Am+(1/m_{\g})\pam\varphi$ in the action. The equation of motion of this new action 
$S[\Am,\varphi]$, obtained performing the variation with respect to $\varphi$, is 
$\Box\varphi+m_{\g}\pam\AMU=0$, which can be formally solved by
$\varphi(x)=-m_{\g}\Box^{-1}(\pam\AMU)$. Inserting this expression for $\varphi$ into 
$S[\Am,\varphi]$ one gets \eq{Lnonloc}, see  \cite{Dvali:2006su}.} Thus, \eq{Lnonloc} provides an alternative description of a massive photon which is explicitly gauge invariant, at the price of nonlocality. In this case, however, the nonlocality is only apparent, since we see from \eq{eqnonlocAN} that the nonlocal term can be removed with a suitable gauge choice. In the following we will study similar theories, in which however the nonlocality cannot be simply gauged away.

An interesting aspect of the nonlocal reformulation of massive electrodynamics is that it also  allows us  to generate the mass term dynamically, through a non-vanishing gauge-invariant condensate $\langle \Fmn\Box^{-1}\FMN\rangle\neq 0$. In the $U(1)$ theory we do not expect non-perturbative effects described by vacuum condensates. However, these considerations can be generalized to non-abelian gauge theories. Indeed, in pure Yang-Mills theory the introduction in the action of a nonlocal term
\be\label{Fmn2}
\frac{m^2}{2} {\rm Tr}\, \int d^4x\,   F_{\mu\nu} \frac{1}{D^2}F^{\mu\nu}\, ,
\ee
(where  $D_{\mu}^{ab}=\d^{ab}\pam-gf^{abc}A_{\mu}^c$ is the covariant derivative and $m$ is a mass scale) correctly reproduces the results on the non-perturbative gluon propagator in the IR, obtained from operator product expansions and lattice QCD~\cite{Boucaud:2001st,Capri:2005dy,Dudal:2008sp}. In this case this term is generated in the IR dynamically by the strong interactions. In other words, because of non-perturbative effects in the IR, at large distances  we have
\be
\langle  {\rm Tr}\, [ F_{\mu\nu}D^{-2}F^{\mu\nu}]\rangle\neq 0\, ,
\ee
which amounts to dynamically generating  a mass term for the gluons.

\subsection{Effective nonlocal modifications of GR}\label{sect:effmodGR}

We next  apply a similar strategy to GR. We will begin with a purely phenomenological approach, trying to construct potentially interesting IR modifications of GR by playing with nonlocal operators such as $m^2/\Box$, and exploring different possibilities. 

When one tries to construct an infrared  modification of GR, usually the aims that one has   in mind is  the construction of a fundamental QFT (possibly valid up to a UV cutoff, beyond which it needs a  suitable UV  completion). In that case a crucial requirement is the absence of ghosts, at least up to  the cutoff of the UV completion, as in the  dRGT theory of massive gravity~\cite{deRham:2010ik,deRham:2010kj,Hassan:2011hr}, or in ghost-free bigravity~\cite{Hassan:2011zd}. In the following we will instead take a different path, and present these models as {\em effective} nonlocal modification of GR, such as a quantum effective action. 
This change of perspective, from a fundamental action to an effective quantum action, is important since (as we already saw for the anomaly-induced effective action, and as we will see in Sect.~\ref{sect:loc} for the nonlocal theories that we will propose) the presence of an apparent ghost in the effective quantum action does not imply that a ghost is truly present in the physical spectrum of the theory. Similarly, we will see in Sect.~\ref{sect:causality} that the issue of causality is different for a  nonlocal fundamental QFT and a nonlocal
quantum effective action.

\vspace{2mm}\noindent
{\em  A nonlinear completion of  the degravitation model}. As a first example we consider the theory defined by the effective nonlocal equation of motion 
\be\label{degrav}
\(1-\frac{m^2}{\Box}\)\Gmn=8\pi G\,\Tmn\, ,
\ee
where $\Box$ is the fully covariant d'Alembertian. \Eq{degrav} is the most straightforward generalization of \eq{eqnonlocFMN} to GR. 
This model was  proposed in~\cite{ArkaniHamed:2002fu} to introduce the 
degravitation idea. Indeed, at least performing naively the inversion of the nonlocal operator, \eq{degrav} can be rewritten as $\Gmn =8\pi G\, [\Box/(\Box-m^2)]\Tmn$. Therefore the low-momentum modes of $\Tmn$, with $|k^2|\ll m^2$, are filtered out and in particular a constant term in $\Tmn$, such as that due to a cosmological constant, does not contribute.\footnote{Observe however that the inversion of the nonlocal operator is more subtle. Indeed, by definition, $\iBox$ is such that, on any differentiable  function $f(x)$, $\Box\iBox f=f$, i.e.  $\Box\iBox=\mathds{1}$. In contrast, from 
$\iBox\Box f=g$ it does not follows $f=g$. Rather, applying $\Box$ to both sides and using
$\Box\iBox=\mathds{1}$ we get $\Box (f-g)=0$, so $f=g+h$ where $h$ is any function such that $\Box h=0$. Therefore, $\iBox\Box\neq \mathds{1}$. The same holds for the inversion of $(\Box-m^2)$. Thus,
more precisely, the  inversion of \eq{degrav} is $\Gmn =8\pi G\,(\Box-m^2)^{-1} \Box \Tmn +\Smn$, where $\Smn$ is any tensor that satisfies $(\Box-m^2)\Smn=0$. In any case, a constant vacuum energy term $\Tmn=-\rho_{\rm vac}\emn$ does not contribute, because of the $\Box$ operator acting on $\Tmn$, while $\Smn$ only has modes with $k^2=-m^2$, so it cannot contribute to a constant vacuum energy.
}

The degravitation idea is very interesting, but 
\eq{degrav} has the problem that the energy-momentum tensor is no longer automatically conserved, since in curved space the covariant derivatives $\n_{\mu}$ do not commute, so $[\n_{\mu},\Box]\neq 0$ and therefore also  $[\n_{\mu},\iBox]\neq 0$. Therefore the Bianchi identity $\n^{\mu}\Gmn=0$ no longer ensures $\n^{\mu}\Tmn=0$. In \cite{Jaccard:2013gla} it was however observed that it is possible to cure this problem, by making use of the fact that 
any symmetric tensor $\Smn$ can be decomposed as 
\be\label{splitSmn}
S_{\mu\nu}=S_{\mu\nu}^{\rm T}+\frac{1}{2}(\n_{\mu}S_{\nu}+\n_{\nu}S_{\mu})\, , 
\ee
where $S_{\mu\nu}^{\rm T}$ is the transverse part of $\Smn$, i.e. it  satisfies
$\n^{\mu}S_{\mu\nu}^{\rm T}=0$. Such a  decomposition can be performed in a generic curved space-time~\cite{Deser:1967zzb,York:1974}. The extraction of the transverse part of a tensor is itself a nonlocal operation, which is the reason why it never appears  in the equations of motions of a local field theory.\footnote{In flat space $\n_{\mu}\ra\pam$ and, applying to both sides of \eq{splitSmn} $\paM$ and  $\paM\paN$ we find that
\be\label{extractSmnT}
S_{\mu\nu}^{\rm T}=\Smn
-\iBox (\pam\paR S_{\rho\nu}+\pan\paR S_{\rho\mu})
+\Box^{-2}\pam\pan\paR\paS S_{\rho\sigma}\, .
\ee
In a generic curved spacetime there is no such a simple formula, because $[\n_{\mu},\n_{\nu}]\neq 0$, but we will see in Sect.~\ref{sect:loc} how to deal, in practice, with the extraction of the transverse part.}  Here however we are already admitting nonlocalities, so we can make use of this operation. Then, in \cite{Jaccard:2013gla} (following a similar treatment in the context of nonlocal massive gravity in~\cite{Porrati:2002cp}) it was proposed to modify \eq{degrav} into
\be\label{GmnT}
\Gmn -m^2\(\iBox\Gmn\)^{\rm T}=8\pi G\,\Tmn\, ,
\ee
so that energy-momentum conservation $\n^{\mu}\Tmn=0$ is automatically ensured. This model can be considered as a nonlinear completion of the original degravitation idea. Furthermore, \eq{GmnT} still admits
a degravitating solution~\cite{Jaccard:2013gla}. Indeed, consider a modification of \eq{GmnT} of the form
\be\label{final3summary}
\Gmn -m^2\[ (\Box-\mu^2)^{-1}\, \Gmn\]^{\rm T}=8\pi G\,\Tmn\, ,
\ee
with $\mu$ is a regularization parameter to be eventually sent to zero.
If  we set $\Tmn=-\rho_{\rm vac}\gmn$,  
\eq{final3summary} admits a de~Sitter solution $\Gmn=-\Lambda\gmn$ with
$\Lambda=8\pi G\, [\mu^2/(m^2+\mu^2)]\,  \rho_{\rm vac}$. In the limit $\mu\ra 0$
we get $\Lambda\ra 0$, so the vacuum energy has been completely degravitated.
However, the  cosmological evolution of this model induced by the remaining cosmological fluid, such as radiation or non-relativistic matter, turns out to be unstable, already at the background level~\cite{Maggiore:2013mea,Foffa:2013vma}. We will see in Sect.~\ref{sect:cosmo} how such an instability emerges.  In any case, this means that the model (\ref{GmnT}) is not phenomenologically viable. 

\vspace{2mm}\noindent
{\em  The RT and RR models}.
The first phenomenologically successful nonlocal model of this type was then proposed in \cite{Maggiore:2013mea}, where it was noticed   that the instability is specific to the form of the $\iBox$ operator on a tensor such as $\Rmn$ or $\Gmn$, and does not appear when $\iBox$ is applied to a scalar, such as the Ricci scalar $R$. Thus, in  \cite{Maggiore:2013mea} it was proposed a model based on the nonlocal equation
\be\label{RT}
\Gmn -\frac{m^2}{3}\(\gmn\iBox R\)^{\rm T}=8\pi G\,\Tmn\, ,
\ee
where the factor $1/3$ is a useful normalization for the mass parameter $m$. We will discuss its phenomenological consequences in Sect.~\ref{sect:cosmo}. We will denote it as
the ``RT" model, where R stands for the Ricci scalar and T for the extraction of the transverse part.
A closed form for the action corresponding to \eq{RT} is currently not known. This model is however  closely related to another nonlocal model,  proposed in \cite{Maggiore:2014sia}, and defined by the effective action
\be\label{RR}
\Gamma_{\rm RR}=\frac{\mplr^2}{2}\int d^{4}x \sqrt{-g}\, 
\[R-\frac{m^2}{6} R\frac{1}{\Box^2} R\]\, .
\ee
Again, we will see that this model is phenomenologically viable, and  we will refer to it as the RR model.
The RT and RR models are related by the fact that,
if we compute the equations of motion  from \eq{RR} and we linearize them over Minkowski space, we  
find the same equations of motion  obtained by linearizing \eq{RT}. However, at the full nonlinear level, or linearizing over a background different from Minkowski, the two models are different. 

We have seen above that nonlocal terms of this sort may be related to a mass  for some degree of freedom. One might then ask whether this is the case also for the RR and RT models. In fact, the answer is quite interesting: the nonlocal terms in eqs.~(\ref{RT}) or (\ref{RR}) correspond to a mass term for the conformal mode of the metric~\cite{Maggiore:2015rma,Maggiore:2016fbn}.
Indeed, consider the conformal mode $\sigma(x)$, 
defined choosing a fixed fiducial metric $\gbmn$ and writing
$\gmn(x)=e^{2\sigma(x)}\gbmn(x)$. Let us restrict the dynamics to the conformal mode, and choose
for simplicity a flat fiducial metric $\gbmn=\emn$. The Ricci scalar computed from the metric $\gmn=e^{2\sigma(x)}\emn$ is then
\be
R=-6 e^{-2\sigma}\( \Box\sigma +\pam\sigma\paM\sigma\)\, .
\ee
Therefore, to linear order in $\sigma$,
$R=-6\Box\sigma +{\cal O}(\sigma^2)$ and (upon integration by parts)
\be\label{m2s2}
 R\frac{1}{\Box^2} R=36 \sigma^2 +{\cal O}(\sigma^3)\, .
\ee
Thus, the  $R\Box^{-2}R$ terms gives a nonlocal but diff-invariant mass term for the conformal mode, plus  higher-order interaction terms (which are nonlocal even in $\sigma$) which are required to reconstruct a diff-invariant quantity. The same is true for the nonlocal term in the RT model, since the RR and RT models coincide when linearized over Minkowski space.

\section{How not to deal with effective nonlocal theories}\label{sect:Hownot}

In this section we  discuss some conceptual aspects of general nonlocal theories, that involve some subtleties. 
The bottomline is that quantum field theory must be played according to its rules  and, as we have already seen in Sect.~\ref{sect:anominid} with the explicit example  of the anomaly-induced effective action,  the rules for  quantum  effective actions are different from the rules for the fundamental action of a  QFT. 
\vspace*{-6mm}
\subsection{Causality}\label{sect:causality}

\vspace*{-2mm}
We begin by examining causality in nonlocal theories (we  follow the discussion in app.~A of \cite{Cusin:2016nzi}; see also \cite{Tsamis:1997rk,Deser:2007jk,Barvinsky:2011rk,Deser:2013uya,Ferreira:2013tqn,Foffa:2013sma,Woodard:2014iga} for related discussions).
In a fundamental QFT with a  nonlocal action, the standard variational principle  produces acausal equations of motion.   Consider for instance a nonlocal term $\int dx \,\phi\iBox\phi$ in the action of a scalar field $\phi$, where  $\iBox$ is defined with respect to some Green's function $G(x;x')$. Then
\bees
&&\frac{\d}{\d\phi(x)}\int dx' \phi(x') (\iBox\phi )(x')=
\frac{\d}{\d\phi(x)} \int dx' dx'' \phi(x') G(x';x'') \phi(x'')\nn\\
&&=\int dx' [G(x;x')+G(x';x)] \phi(x')\, . \label{symGreen}
\ees
Thus, the variation   symmetrizes the Green's
function. However, the retarded Green's function is not symmetric; rather, 
$G_{\rm ret}(x';x)=G_{\rm adv}(x;x')$, and therefore it cannot be obtained from such a variation.
In a fundamental action, nonlocality implies the loss of causality, already at the classical level
(unless, as in \eq{eqnonlocAN}, we have  a gauge symmetry that allows us to gauge away the nonlocal term in the equations of motion).

However, quantum effective actions are in general nonlocal, as in eq.~(\ref{qed}), (\ref{SPolyakov}) or
(\ref{GammaRiegert}). Of course, this does not mean that they describe acausal physics. These nonlocal effective actions are just a way to express, with an action that can be used at tree level, the result of a  quantum computation in fundamental theories  which are local and  causal. Therefore, it is clear that their nonlocality has nothing to do with acausality. 
Simply, to reach the correct conclusions one must play QFT according to its rules. The variation of the quantum effective action does not give the classical equations of motion of the field. Rather, it provides   the  time evolution, or equivalently the equations of motion, obeyed by  the {\em vacuum expectation values} of the corresponding operators, as in \eq{Tmnvac}.
These equations of motion are obtained in a different way depending on whether   we consider  the in-in  or the in-out matrix elements. The  in-out expectation values  are obtained  using the  Feynman path  integral in \eq{Gamma}, and are indeed acausal. Of course, there is nothing wrong with it. The in-out matrix element are not observable quantities, but just auxiliary  objects which enter in intermediate steps in the computation of scattering amplitudes, and  the Feynman propagator, which is  acausal, enters everywhere in QFT computations.

The physical quantities, which can be interpreted as physical observables, are instead the in-in expectation values. For instance, $\langle 0_{\rm in}|\hat{g}_{\mu\nu} |0_{\rm in}\rangle$ can be interpreted as a semiclassical metric, while $\langle 0_{\rm out}|\hat{g}_{\mu\nu}|0_{\rm in}\rangle$ is not even a real quantity.
The equations of motion of the  in-in expectation values are obtained from
the Schwinger-Keldysh path integral, which automatically provides nonlocal {\em but causal} 
equations  \cite{Jordan:1986ug,Calzetta:1986ey}.
In practice, the equations of motion obtained from the Schwinger-Keldysh path integral turn out to be the same that one would obtain by treating formally the $\iBox$ operator in the variation, without specifying the Green's function, and replacing in the end $\iBox\ra\Box^{-1}_{\rm ret}$ in the equations of motion (see e.g.
\cite{Mukhanov:2007zz}).\footnote{In  the in-in formalism the equations of motions are more easily obtained using the tadpole method, i.e. writing a generic field $\phi$ as 
$\phi=\phi_{\rm cl}+\varphi$, where $\varphi$ are the quantum fluctuations over a classical configuration
$\phi_{\rm cl}$, and requiring that $\langle 0_{\rm in}|\varphi |0_{\rm in}\rangle=0$. See \cite{Boyanovsky:1994me,Collins:2012nq} for an instructive  computation, showing explicit how nonlocal but causal terms emerge in the in-in equations of motion.}

Thus nonlocal actions, interpreted as quantum effective actions, provide causal evolution equations for the in-in matrix elements.


\vspace*{-6mm}

\subsection{Degrees of freedom and ghosts}\label{sect:degrees}

\vspace*{-2mm}
Another subtle issue concerns the number of degrees of freedom described by a nonlocal theory such as (\ref{RR}). Let us at first treat it as we would do for a fundamental action. We write $\gmn=\emn+\hmn$ and  expand the quantum effective action to quadratic order over flat space.\footnote{The same treatment holds for the RT model, since at the level of the equations of motion linearized over flat space the RR and RT model are identical.} The corresponding flat-space action is~\cite{Maggiore:2014sia}
\be\label{Lquadr}
{\Gamma}^{(2)}_{\rm RR}=\int d^4x\, \[ \frac{1}{2}\hmn{\cal E}^{\mu\nu,\rho\sigma}\hrs
-\frac{1}{3}\, m^2\hmn P^{\mu\nu}P^{\rho\sigma}\hrs \], \,
\ee
where 
\be\label{defPmunu}
P^{\mu\nu}=\eMN-\frac{\paM\paN}{\Box}\, ,
\ee
where now $\Box$ is the flat-space d'Alembertian. We then add 
 the usual gauge fixing term of linearized massless gravity,
${\cal L}_{\rm gf}=-(\paN\bhmn ) (\parho\bar{h}^{\rho\mu})$, where $\bhmn=\hmn -(1/2)h\emn$.
Inverting the quadratic form we get the propagator  
$\tilde{D}^{\mu\nu \rho\s}(k)=-i\Delta^{\mu\nu \rho\s}(k)$,  where 
\bees
\Delta^{\mu\nu \rho\s}(k)&=&\frac{1}{2k^2}\, 
\( \eMR\eNS +\eMS\eNR-\eMN\eRS \) \nn\\
&&+ \frac{1}{6}\, \(\frac{1}{k^2}-\frac{1}{k^2-m^2}\)\eMN\eRS\, ,\label{4DeltakRT}
\ees
plus terms  proportional to $k^{\mu}k^{\nu}$, $k^{\rho}k^{\sigma}$ and
$k^{\mu}k^{\nu}k^{\rho}k^{\sigma}$, that give zero when contracted with a conserved energy-momentum tensor.
The  term in the second line in \eq{4DeltakRT} gives  an extra contribution to  $\tilde{T}_{\mu\nu}(-k)
\tilde{D}^{\mu\nu \rho\s}(k)\tilde{T}_{\rho\sigma}(k)$, equal to
\be\label{TT}
\frac{1}{6}\tilde{T}(-k)\[ -\frac{i}{k^2}+\frac{i}{k^2-m^2}\]
\tilde{T}(k)\, .
\ee
This term apparently describes the exchange of a healthy massless scalar plus a ghostlike massive scalar.  The presence of a ghost in the spectrum of the quantum theory would be fatal to the consistency of the model. However, once again, this conclusion comes from a confusion between the concepts of fundamental action and quantum effective action.

To begin, let us observe that it is important to distinguish between the effect of a ghost in the classical theory and its effect in the quantum theory. Let us consider first the classical theory. At linear order, the interaction between the metric perturbation and an external conserved energy-momentum tensor $\Tmn$ is given by
\be\label{SinthT}
S_{\rm int}=\int d^{4}x\,\hmn\TMN\, ,
\ee
where $\hmn$ is the solution  of the equations of motion derived from \eq{Lquadr}.  Solving them explitly and inserting the solution for $\hmn$ in \eq{SinthT} one finds~\cite{Maggiore:2013mea}
\be
S_{\rm int} =16\pi G\int \frac{d^4k}{(2\pi)^{4}}\, \tilde{T}_{\mu\nu}(-k)\Delta^{\mu\nu\rho\sigma}(k)
\tilde{T}_{\rho\sigma}(k)\, , 
\ee
with  $\Delta^{\mu\nu \rho\s}(k)$ given by \eq{4DeltakRT}. The quantity $\Delta^{\mu\nu \rho\s}(k)$ therefore plays the role of the propagator in the classical theory [and differs by a factor of $-i$ from the quantity usually called the propagator in the quantum theory, $\tilde{D}^{\mu\nu \rho\s}(k)=-i\Delta^{\mu\nu \rho\s}(k)$]. A `wrong' sign in the term proportional to $1/(k^2-m^2)$ in \eq{4DeltakRT} might then result in a classical instability. Whether this is acceptable or not must be studied on a case-by-case basis. For instance, taking $m={\cal O}(H_0)$, as we will do below, the instability will only develop on cosmological timescales. Therefore, it must be studied in the context of a FRW cosmology, where it will also compete with damping due to the Hubble friction. Whether this will result or not in a viable cosmological evolution, both at the level of background evolution and of cosmological perturbations, can only be deduced an explicit quantitative  study of the solutions of these cosmological equations. We will indeed see in Sect.~\ref{sect:cosmo} that the cosmological evolution obtained from this model is perfectly satisfying.

A different issues is the presence of a ghost in the spectrum of the quantum theory. After quantization a ghost carries negative energy, and induces vacuum decay through the associated production of ghosts and normal particles, which would be fatal to the consistency of the theory. However, here we must be aware of the fact that the spectrum of the quantum theory can be read from the free part of the {\em fundamental} action of the quantum theory. To apply blindly the same procedure to the quantum effective action is simply wrong. We  have already seen this in Sect.~\ref{sect:anominid} for the anomaly-induced effective action, where
the action (\ref{SPolyakov26sigmatot}) with $N>25$, or the action (\ref{anominducSeffD4}), naively seem to have a ghost, but in fact are perfectly healthy effective quantum actions, derived from fundamental QFTs that have  no ghost.
Another example that illustrates the sort of nonsense that one obtains if one tries to read the spectrum of the quantum theory from the quantum effective action $\Gamma$, consider for instance the one-loop effective action of QED, \eq{qed}. If we proceed blindly and quantize it as if it were a fundamental action, we would add to \eq{qed} a gauge fixing term ${\cal L}_{\rm gf}=-(1/2) (\pam A^{\mu})^2$ and invert the resulting quadratic form. We would then obtain, for  the propagator in the $m_e\ra 0$ limit, 
\be\label{propqed}
\tilde{D}^{\mu\nu}(k)=-i\frac{\eMN}{k^2}\, \[1-e^2(\mu)\beta_0\log\frac{k^2}{\mu^2}\]\, ,
\ee
plus terms proportional to $k^{\mu}k^{\nu}$ that cancel when contracted with a conserved current $j^{\mu}$.\footnote{Actually, the terms $k^{\mu}k^{\nu}$ can be made to vanish if we take also the gauge fixing as nonlocal, and given by $(-1/2) (\pam A^{\mu}) [1/e^2(\Box)] (\pan A^{\nu})$. The same could be done for the propagator in \eq{4DeltakRT}.} Using the identities
\be
\log\frac{k^2}{\mu^2}=\int_0^{\infty}dm^2\, \(\frac{1}{m^2-\mu^2}-\frac{1}{k^2+m^2}\)\, 
\ee
and
\be
\frac{m^2}{k^2(k^2+m^2)}=\frac{1}{k^2}-\frac{1}{k^2+m^2}
\ee
we see that the ``propagator" (\ref{propqed})  has the standard  pole of the electromagnetic field, proportional to $-i\eMN/k^2$ with a positive coefficient, 
 plus a continuous set of ghost-like poles proportional to $+i\eMN/(k^2+m^2)$, with $m$ an integration variable. We would then conclude that QED as a continuous spectrum of ghosts! Of course this is nonsense, and it is just an artifact of having applied to the quantum effective action a procedure that only makes sense for the fundamental action of a QFT. In fact, the proper interpretation of \eq{propqed} is that 
$\log (k^2/\mu^2)$ develops an imaginary part for $k^2<0$ (e.g. for $k_0\neq 0, \vk=0$, i.e. for a spatially uniform but time-varying electromagnetic field). This is due to the fact that, in the limit $m_e\ra 0$ in which we are working (or, more generally, for $-k^2>4m_e^2$), in such an external electromagnetic field there is a rate of creation of electron-positron pairs, and the imaginary part of the effective action describes the rate of pair creation  \cite{Dobado:1998mr}.

These general considerations show that  the spectrum of the theory cannot be read naively from the quantum  effective action. Thus, in particular, from the presence of a `ghost-like' pole obtained from the effective quantum action (\ref{Lquadr}), one cannot jump to the conclusion that the underlying fundamental theory has a ghost. In the next section we will 
 be more specific, and try to understand the origin of this `wrong-sign' pole in the RR  and RT theories.

\section{Localization of nonlocal theories}\label{sect:loc}

Nonlocal models can be formally written in a local form introducing auxiliary fields, as discussed in similar contexts in \cite{Nojiri:2007uq,Jhingan:2008ym,Koshelev:2008ie,Koivisto:2008dh,Koivisto:2009jn,Barvinsky:2011rk,Deser:2013uya}.
This reformulation is quite useful both for the numerical study of the equations of motion, and for understanding exactly why the  ghosts-like poles in \eq{4DeltakRT} do not correspond to states in the spectrum of the quantum theory. It is useful to first illustrate  the argument for the Polyakov  effective action, for which we know that it is the effective quantum action of a perfectly healthy fundamental theory.

\vspace{5mm}\noindent
{\em Localization of the Polyakov action}. In $D=2$ the Polyakov action becomes local when written in terms of the conformal factor. Let us however introduce a different localization procedure, that can be generalized to 4D. We start from \eq{SPolyakov}, 
\be\label{Ganom1}
\Gamma= c\int d^2x\sqrt{-g}\, R\iBox R\, ,
\ee
where we used the notation $c=-N/(96\pi)$. We now
introduce an auxiliary field $U$ defined by $U=-\iBox R$. At the level of the action, this can be implemented by introducing a Lagrange multiplier $\xi$, and writing
\be \label{SPolyakovloc1}
\Gamma=\int d^2x\sqrt{-g}\, \[-c RU  +\xi (\Box U+R)\]\, .
\ee
The variation with respect to $\xi$ gives
\be\label{BoxURD2}
\Box U=-R\, ,
\ee
so it enforces $U=-\iBox R$, while the variation with respect to $U$ gives $\Box\xi=cR$ and therefore
$\xi=c\iBox R=-cU$. This is an algebraic equation that can be put back in the action so that, after an integration by parts,  $\Gamma$ can be rewritten as~\cite{Antoniadis:2006wq}
\be\label{Ganom2}
\Gamma=c\int d^2x\sqrt{-g}\, \[\pa_a U\pa^a U-2UR\]\,.
\ee
The theories defined by \eqs{Ganom1}{Ganom2} are classically equivalent.  As a check, one can compute the energy-momentum tensor from \eq{Ganom2}, and verify that its classical trace is given by $T=4c\Box U=-4cR$. So \eq{Ganom2}, used as a classical action, correctly  reproduces the quantum trace anomaly (\ref{traceT})~\cite{Antoniadis:2006wq}. We can further manipulate  the action (\ref{Ganom2}) writing
$g_{ab}=e^{2\sigma}\eta_{ab}$. Using \eq{Rsigma2D} and introducing a new field $\varphi$ from
$U=2(\varphi+\sigma)$ to diagonalize the action, we get
\be\label{Ganom3}
\Gamma=4c\int d^2x\, \(\eta^{ab}\pa_a\varphi\pa_b\varphi-\eta^{ab}\pa_a\sigma\pa_b\sigma\)\, .
\ee
Taken litteraly, this action seems to suggest that in the theory there are two dynamical fields, $\varphi$ and $\sigma$. For $c>0$, $\varphi$ would be a ghost and $\sigma$ a healthy field, and viceversa if $c<0$ (in the Polyakov action (\ref{Ganom1}) $c=-N/(96\pi)<0$, but exactly the same computation could be performed with the action (\ref{SPolyakov26}), where 
$c=-(N-25)/(96\pi)$ can take both signs).
Of course, we know that this conclusion is wrong, since we know exactly the spectrum of the quantum theory at the fundamental level, which is made uniquely by the quanta of the conformal matter fields. As we mentioned, even taking into account the anomaly-induced effective action, still  $\sigma$ has no quanta in the physical spectrum, since they are eliminated by the physical-state condition  \cite{Polchinski:1989fn}. 
As for the auxiliary field $\varphi$, or equivalently $U$, there is no trace of its quanta in the physical spectrum. $U$ is an artificial field which has been introduced by the localization procedure, and there are no quanta associated with it.

This can also be understood purely classically, using the fact that, in $D=2$, the Polyakov action becomes local when written in terms of the conformal factor. Therefore, the classical evolution of the model is fully determined once we give the initial conditions on $\sigma$, i.e. $\sigma(t_i,x)$ and $\dot{\sigma}(t_i,x)$ at an initial time. Thus, once we localize the theory introducing $U$, the initial conditions on $U$ are not arbitrary. Rather, they are uniquely fixed by the condition that the classical evolution, in the formulation 
obtained from \eq{Ganom2}, must be equivalent to that in the original theory (\ref{SPolyakov}). In other words, $U$ is not the most general solution of \eq{BoxURD2}, which would be given by a particular solution of the inhomogeneous equation plus the most general solution of the associated homogeneous equation $\Box U=0$. Rather, it is just one specific solution, with given boundary conditions, such as $U=0$ when $R=0$ in \eq{BoxURD2}. Thus, if we are for instance in flat space, there are no arbitrary plane waves associated to $U$, whose coefficients $a_{\vk}$ and $a_{\vk}^*$ would be promoted to creation and annihilation operators in the quantum theory. In this sense, the situation is different with respect to the conformal mode $\sigma$: the conformal mode, at the quantum level, is a quantum field with its own  creation and annihilation operators, but the corresponding quantum states do no survive the imposition of the physical-state condition, and therefore do not belong to the physical Hilbert space. The $U$ field, instead,  is a classical auxiliary field and has not even creation and annihilation operators associated to it.

\vspace{5mm}\noindent
{\em Localization of the RR theory}. We next consider the RR model. To put the theory in a local form we introducing two auxiliary fields $U$ and $S$, defined by
\be\label{defU}
U=-\iBox R\, ,\qquad 
S=-\iBox U\, .
\ee
This can be implemented at the Lagrangian level by introducing two Lagrange multipliers $\xi_1,\xi_2$, and rewriting \eq{RR} as
\bees\label{S2}
\Gamma_{\rm RR}=\frac{\mplr^2}{2}\int d^4x \sqrt{-g}\, \[ R\( 1-\frac{m^2}{6} S\)-\xi_1(\Box U+R)-\xi_2 (\Box S+U)\]\nn
\, .
\ees
The equations of motion derived performing the variation of this action with respect to $\hmn$ is
\be
\Gmn=\frac{m^2}{6} K_{\mu\nu}+8\pi G\Tmn\, ,\label{Gmn}
\ee
where
\be\label{K}
K_{\mu\nu}=2S\Gmn-2\n_{\mu}\pan S 
+\gmn [
-2U+\parho S\paR U
 -(1/2) U^2]-(\pam S\pan U+\pan S\pam U)\, .
\ee
At the same time, the definitions (\ref{defU})  imply that $U$ and $S$ satisfy
\bees
\Box U&=&-R\, ,\label{BoxU}\\
\Box S &=&-U\, .\label{BoxS}
\ees
Using the equations of motion we can check  explicitly that $\n^{\mu}K_{\mu\nu}=0$, as it should, since the equations of motion  has been derived from a diff-invariant action.
Linearizing \eq{K} over flat space we get
\be\label{line1}
{\cal E}^{\mu\nu,\rho\sigma}\hrs
-\frac{2}{3}\, m^2 P^{\mu\nu}P^{\rho\sigma}
\hrs=-16\pi G\TMN\, ,
\ee
Let us we restrict to  the scalar sector, which is the most interesting for our purposes. We proceed as in GR, and use the diff-invariance of the nonlocal theory to fix the Newtonian gauge 
\be\label{Bardeen}
h_{00}=-2\Psi\, ,\qquad
h_{0i}=0\, ,\qquad h_{ij}=2\Phi\d_{ij}\, . 
\ee 
We also write the energy-momentum tensor in the scalar sector as 
\bees
T_{00}&=&\rho\, , \qquad T_{0i}=\pa_i\Sigma\, ,\\
T_{ij}&=&P\d_{ij}+[\pa_i\pa_j-(1/3)\d_{ij}\n^2]\Pi\, . 
\ees
A straightforward generalization of the standard computation performed in GR (see e.g. \cite{Jaccard:2012ut}) 
gives  four independent equations for the four scalar variables $\Phi,\Psi$, $U$ and $S$. For the Bardeen variables $\Phi$ and $\Psi$ we get~\cite{Maggiore:2014sia}\footnote{Compared to \cite{Maggiore:2014sia}, in \eq{Bardeen}
we have changed the sign in the definition of $\Psi$, in order to be consistent with the convention that we used  in \cite{Dirian:2014ara} when studying the cosmological perturbations of this model, compare with \eq{defPhiPsi} below.}
\bees
\n^2\[\Phi-(m^2/6) S\]&=&-4\pi G\rho\, ,\label{dof1}\\
\Phi+\Psi-(m^2/3) S&=& -8\pi G\Pi\label{dof2}\, .
\ees
Thus,  just as in GR, $\Phi$ and $\Psi$ remain non-radiative degrees of freedom, with a dynamics governed by a Poisson equation rather than by a Klein-Gordon equation. This should be contrasted with what happens when one linearizes massive gravity with a Fierz-Pauli mass term. In that case $\Phi$ becomes a radiative field that satisfies 
$(\Box-m^2)\Phi=0$ \cite{Deser:1966zzb,Alberte:2010it,Jaccard:2012ut}, and the corresponding jump in the number of radiative degrees of freedom of the linearized theory is just the vDVZ discontinuity. Furthermore, in local massive gravity with a mass term that does not satisfies  the Fierz-Pauli tuning, in the Lagrangian also  appears a term $(\Box\Phi)^2$ \cite{Jaccard:2012ut}, signaling the presence of a dynamical ghost. 

To linearize   \eq{BoxU} we first observe that, taking the trace of \eq{line1}, we get 
\be\label{R1Ph}
R^{(1)}- m^2P^{\mu\nu}\hmn=8\pi G (\rho-3P)\, ,
\ee
where 
\be\label{LinR1}
R^{(1)}= \pam\pan (\hMN-\eMN h)\, 
\ee 
is the linearized Ricci scalar. 
From \eq{defPmunu}, 
\be\label{PhmniBoxR}
P^{\mu\nu}\hmn=\frac{1}{\Box}( \Box h-\paM\paN\hmn)=-\frac{1}{\Box}R^{(1)}\, .
\ee
Therefore, \eq{R1Ph} can also be rewritten in the suggestive form
\be
\( 1+\frac{m^2}{\Box}\) R^{(1)}=8\pi G (\rho-3P)\, .
\ee
\Eq{PhmniBoxR} also implies that,  to linear order, 
\be\label{PmnU}
P^{\mu\nu}\hmn=U\, ,
\ee
and therefore \eq{R1Ph} can be rewritten as
\be
R^{(1)}=8\pi G (\rho-3P)+m^2U\, .
\ee
Inserting this into \eq{BoxU} we finally get
\be\label{dof3}
(\Box+m^2)U=-8\pi G (\rho-3P)\, ,
\ee
where, in all the linearized equations, $\Box=-\pa_0^2+\n^2$ is the flat-space d'Alembertian. Similarly the linearized equation for $S$ is just given by \eq{BoxS}, again with the flat-space d'Alembertian.

Thus, in the end, in the scalar sector we have two fields $\Phi$ and $\Psi$ which obey 
eqs.~(\ref{dof1}) and (\ref{dof2}) and are therefore non-radiative, just as, in GR. Furthermore, we have two fields $U$ and $S$ that satisfy Klein-Gordon equations with sources. In particular $U$ satisfies the massive KG equation (\ref{dof3}), so is clearly the field responsible for the ghost-like $1/(k^2-m^2)$ pole in
\eq{TT}, while $S$ satisfies a massless KG with source, and is the field responsible for the healthy $1/k^2$ pole in \eq{TT}. This analysis shows that  the potential source of problems is not one of the physical fields $\Phi$ and $\Psi$, but rather the auxiliary field $U$. However, at this point the solution of the potential problem becomes clear (see in particular the discussions in \cite{Koshelev:2008ie,Koivisto:2009jn,Barvinsky:2011rk,Deser:2013uya} in different nonlocal models, and in \cite{Maggiore:2013mea,Maggiore:2014sia,Foffa:2013sma} for the RR and RT models), and is in fact completely analogous 
to the situation that we have found for the Polyakov effective action.
In general, an equation such as $\Box U=-R$ is solved by $U=-\iBox R$, where
\be\label{defiBox}
\iBox R= U_{\rm hom}(x)-\int d^{4}x'\, \sqrt{-g(x')}\, G(x;x') R(x')\, ,
\ee
with $U_{\rm hom}(x)$ any solution of $\Box U_{\rm hom}=0$, and $G(x;x')$ a Green's function of the $\Box$ operator. The choice of the homogeneous solution is part of the definition of the $\iBox$ operator and therefore of the original nonlocal effective theory. In principle, the appropriate prescription would emerge once one knows the fundamental theory behind. In any case, there will be one prescription for what $\iBox$ means in the effective theory. This means that the auxiliary field $U$ is not {\em the most general} solution of $\Box U=-R$, which is  given by a solution of the inhomogeneous equation plus the most general solution of the associated homogeneous equation $\Box U=0$. Rather, it is just a single, specific, solution. In other words, the boundary conditions of the equation $\Box U=-R$ are fixed.  Whatever the choice made in the definition of $\iBox$, the corresponding homogeneous solution is fixed. For instance, in flat space this homogeneous solution is a superposition of plane waves, and  the coefficients $a_{\vk},a_{\vk}^*$ are fixed by the definition of $\iBox$ (e.g. at the value $a_{\vk}=a_{\vk}^*=0$ if the definition of $\iBox$ is such that $U_{\rm hom}=0$). They are not free parameters of the theory, and at the quantum level it makes no sense to promote them to annihilation and creation operators. There is no quantum degree of freedom associated to them.

\vspace{2mm}

To conclude this section, it is interesting to observe that the need of imposing boundary conditions on some classical fields, in order to recover the correct Hilbert state at the quantum level, is not specific to nonlocal effective actions. Indeed, GR itself can be formulated in such a way that requires the imposition of similar conditions~\cite{Jaccard:2012ut,Foffa:2013sma}. 
Indeed, let us consider GR linearized over flat space. To quadratic order, adding to the Einstein-Hilbert action the interaction term with a conserved energy-momentum tensor, we have
\be\label{Squadr}
S_{\rm EH}^{(2)}+S_{\rm int}=\int d^{4}x \,\[
\frac{1}{2}\hmn{\cal E}^{\mu\nu,\rho\sigma}\hrs +\frac{\kappa}{2}\, \hmn\TMN \]\, .
\ee
We decompose  the metric as
\be\label{decomphmn}
\hmn =\hmn^{\rm TT}+(\pam \eps_{\nu}+\pan \eps_{\mu}) +\frac{1}{3}\emn s\, ,
\ee
where $\hmn^{\rm TT}$ is transverse and traceless,
\be
\paM \hmn^{\rm TT}=0\, ,\qquad
\eMN \hmn^{\rm TT}=0\, . 
\ee
Thus, the 10 components of $\hmn$ are split into the 5 components of the TT tensor $\hmn^{\rm TT}$, the four components of $\eps_{\mu}$, and the scalar $s$.
Under a linearized diffeomorphism
$\hmn\ra\hmn -(\pam\xin+\pan\xim)$, the four-vector $\eps_{\mu}$ transforms as $\eps_{\mu}\ra\eps_{\mu}-\xi_{\mu}$, while $\hmn^{\rm TT}$ and $s$ are gauge invariant. We similarly decompose $\Tmn$. 
Plugging \eq{decomphmn} into \eq{Squadr} $\eps_{\mu}$ cancels (as it is obvious from the fact that \eq{Squadr} is invariant under linearized diffeomorphisms and $\eps_{\mu}$ is a pure gauge mode), and we get
\be\label{SEH2nl}
S_{\rm EH}^{(2)}+S_{\rm int}=\int d^{4}x \,\frac{1}{2} \[\hmn^{\rm TT}\Box (h^{\mu\nu })^{\rm TT}
-\frac{2}{3}\, s\Box s\]
+\frac{\kappa}{2}\[ \hmn^{\rm TT}(\TMN)^{\rm TT}+\frac{1}{3}sT\]
\, .
\ee
The equations of motion derived from $S_{\rm EH}^{(2)}+S_{\rm int}$ are
\be\label{Boxs}
\Box\hmn^{\rm TT}=-\frac{\kappa}{2}\Tmn^{\rm TT}\, ,\qquad
\Box s=+\frac{\kappa}{4}T\, .
\ee
This result seems to suggest that in ordinary massless GR we have six propagating degrees of freedom: the five components of the transverse-traceless tensor $\hmn^{\rm TT}$,  plus the scalar $s$. Note that  $\hmn^{\rm TT}$ and  $s$ are gauge invariant, so they cannot be  gauged away. Furthermore, from \eq{SEH2nl}
the scalar $s$ seems a ghost! 

Of course, we know that in GR only the two components with helicities $\pm 2$ are true propagating degrees of freedom. In fact, the resolution of this apparent puzzle is that the variables $\hmn^{\rm TT}$ and $s$ are nonlocal functions of the original metric. Indeed, inverting \eq{decomphmn}, one finds
\bees
s&=&P^{\mu\nu}\hmn\, ,\label{sPhmn}\\
\hmn^{\rm TT}&=&\hmn -\frac{1}{3}P_{\mu\nu}h
-\frac{1}{\Box}(\pam\paR\hnr+\pan\paR\hmr)+
\frac{1}{3}\,\emn \frac{1}{\Box}\paR\paS\hrs \nn\\
&&+\frac{2}{3}\frac{1}{\Box^2}\pam\pan\paR\paS\hrs\, ,\label{defhath}
\ees
where  $P^{\mu\nu}$ is the nonlocal operator (\ref{defPmunu}). Observe that the nonlocality is not just in space but also in time. Therefore,  giving initial conditions on a given time slice for the metric is not the same as providing the initial conditions on  $\hmn^{\rm TT}$ and $s$, and the proper counting of dynamical degrees of freedom gets mixed up. If we want to study GR in terms of the variables $\hmn^{\rm TT}$ and $s$, which are nonlocal functions of the original variables $\hmn$, we can do it, but we have to be careful that the number of independent initial conditions that we impose to evolve the system must 
remains the same as in the standard Hamiltonian formulation of GR. This means in particular that the initial conditions on $s$ and on the components of $\hmn^{\rm TT}$  with helicities $0,\pm 1$ cannot be freely chosen, and in particular the solution of the homogeneous equations $\Box s=0$ associated to the equation $\Box s=(\kappa/4)T$ is not arbitrary. It is fixed, e.g. by the condition that $s=0$ when $T=0$. Just as for the auxiliary field $U$ discussed above, there are no quanta associated to $s$ (nor to the components of $\hmn^{\rm TT}$  with helicities $0,\pm 1$), just as in the standard $3+1$ decomposition of the metric there are no quanta associated to the Bardeen potentials $\Phi$ and $\Psi$.

The similarity between the absence of quanta for the field $U$ in the localization procedure of the RR model, and the absence of quanta for $s$ in GR, is in fact more than an analogy. 
Comparing \eqs{PmnU}{sPhmn} we  see that, at the level of the linearized theory, $U$ reduces just to $s$ in the $m=0$ limit. The boundary condition that eliminates the quanta of $U$ in the RR theory  therefore just reduces to the boundary condition that eliminates the quanta of $s$ in GR.

The bottomline of this discussion is that the `wrong-sign' pole in \eq{TT} is not due to a ghost in the quantum spectrum of the underlying fundamental theory. It is simply due to an auxiliary field that enters the dynamics at the classical level, but has no associated quanta in the physical spectrum of the theory.
A different question is whether this auxiliary field might induce instabilities in the classical evolution. Since we will take $m$  of order of the Hubble parameter today, $H_0$, any such instability would only develop on cosmological timescale, so it must be studied on a FRW background, which we will do in the next section.

The above analysis was performed for the RR model. For the RT model the details of the localization procedure  are technically different~\cite{Maggiore:2013mea,Kehagias:2014sda}. In that case we define again $U=-\iBox R$, and we also introduce
$S_{\mu\nu}=-U\gmn=\gmn \iBox R$. We then compute  $S^T_{\mu\nu}$ using \eq{splitSmn}. Thus, \eq{RT} is localized in terms of an auxiliary scalar field $U$ and the auxiliary four-vector field $S_{\mu}$ that enters through \eq{splitSmn}, obeying the coupled system
\bees
\Gmn +\frac{m^2}{6}\, \(2U\gmn + \n_{\mu}S_{\nu}+\n_{\nu}S_{\mu}\)&=&8\pi G\,\Tmn\, ,
\label{v2loc1}\\
\Box U&=&-R\, ,\\
(\d^{\mu}_{\nu}\Box +\n^{\mu}\n_{\nu})S_{\mu}&=&-2\pan U\, , \label{panU}
\ees
where the latter equation is obtained by taking the divergence of \eq{splitSmn}. We see that, at the full nonlinear level, the RT model is different from the RR model. However, linearizing over flat space they become the same. In fact in this case, using  \eq{PhmniBoxR},  to linear order we have
\be
\Smn\equiv\gmn\iBox R\simeq -\emn P^{\rho\sigma}\hrs\, .
\ee
In flat space the extraction of the transverse part can be easily performed using
\eq{extractSmnT}, without the need of introducing auxiliary fields. This gives, again to linear order, 
$\Smn^T=-P_{\mu\nu}P^{\rho\sigma}\hrs$.
Using the fact that, to linear order, 
$\Gmn^{(1)} =-(1/2){\cal E}_{\mu\nu,\rho\sigma}\hRS$,
we see that
the linearization of \eq{RT} over flat space gives the same equation as \eq{line1}.
Thus, the RR and RT model coincide at linear order over flat space, but not on a general background (nor at linear order over a non-trivial background, such as FRW).

It should also be stressed that the RR and RT models are not theories of massive gravity. The graviton remains massless in these theories. Observe also, from \eq{4DeltakRT}, that when we linearize over flat space
the limit $m\ra 0$ of the propagator is smooth, and there is no vDVZ discontinuity, contrary to what happens in massive gravity. 
The continuity with GR has also been explicitly verified for the 
\Sch solution~\cite{Kehagias:2014sda}.\footnote{See app.~B of \cite{Dirian:2016puz} for the discussion of a related  issue on the comparison with Lunar Laser Ranging, raised in \cite{Barreira:2014kra}.}

\section{Cosmological consequences}\label{sect:cosmo}

We can now explore the cosmological consequences of the RT and RR models, as well as of some of their extensions that we will present below, beginning with the background evolution, and then moving to cosmological perturbation theory and to the comparison with cosmological data.

\subsection{Background evolution and self-acceleration}\label{sect:backg}

We begin with the  background evolution (we closely follow the original discussions in \cite{Maggiore:2013mea,Foffa:2013vma} for the RT model and \cite{Maggiore:2014sia} for the RR model). It is convenient to use the localization procedure discussed in Sect.~\ref{sect:loc}, so we deal with a set of coupled differential equations, rather than with the original integro-differential equations.

\subsubsection{The RT model} Let us begin with the RT model. In FRW, at the level of background evolution, for symmetry reasons the spatial component $S_i$ of the auxiliary field $S_{\mu}$ vanish, and the only variables are $U(t)$ and $S_0(t)$, together with the scale factor $a(t)$.
\Eqst{v2loc1}{panU} then become
\bees
H^2-\frac{m^2}{9}(U-\dot{S}_0)&=&\frac{8\pi G}{3}\rho\,
\label{loc1} \\
\ddot{U}+3H\dot{U}&=&6\dot{H}+12H^2\, ,\label{loc2}\\
\ddot{S}_0+3H\dot{S}_0-3H^2S_0&=&\dot{U}\, .\label{loc3}
\ees
We supplement these equations with the initial conditions
\be\label{initcond}
U(t_*)=\dot{U}(t_*)=S_0(t_*)=\dot{S}_0(t_*)=0\, ,
\ee
at some time $t_*$ deep in the radiation dominated (RD) phase. We will come back below to how the results depend on this choice. Observe that we do not include a cosmological constant term. Indeed, our aim is first of all to see if  the nonlocal term produces a self-accelerated solution, without the need of a cosmological constant.

\begin{figure}[t]
\begin{center}
\includegraphics[width=0.45\columnwidth]{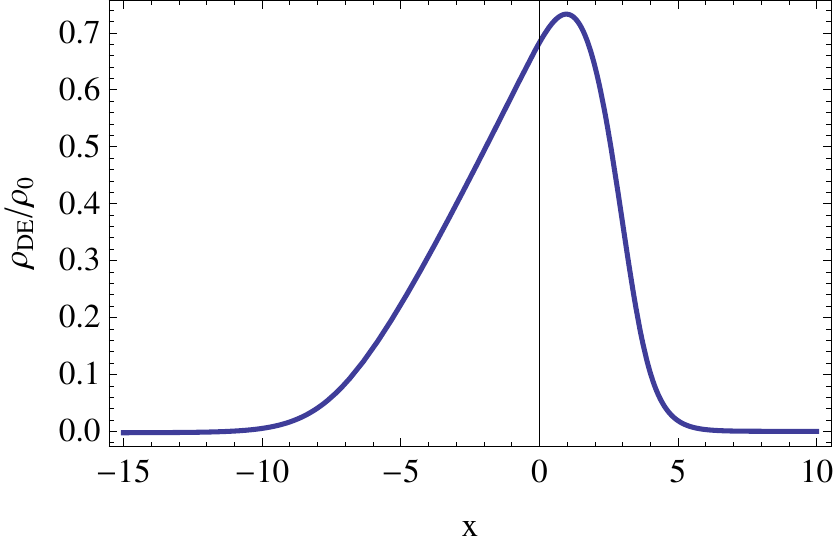}\hspace*{2mm}
\includegraphics[width=0.45\columnwidth]{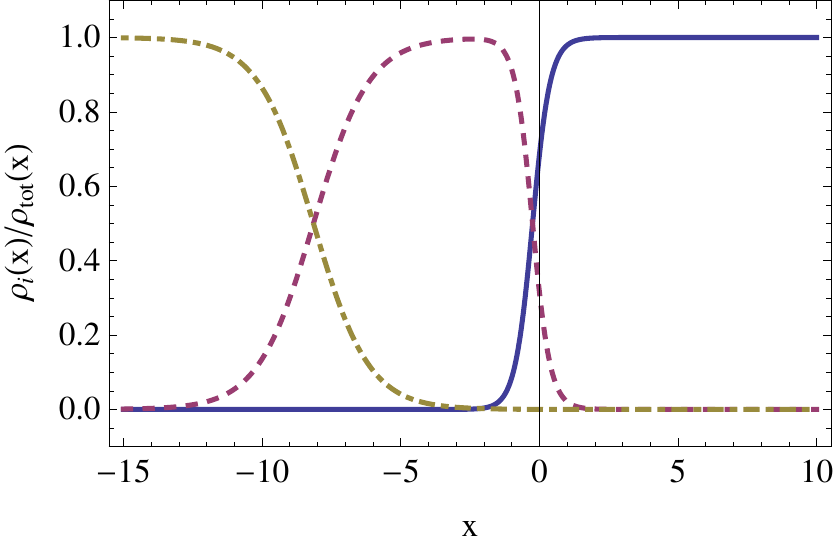}
\caption{\label{fig:DE} Left panel:  the function $\rde(x)/\rho_0$, against $x=\ln a$, for the RT model (from \cite{Maggiore:2013mea}).
Right panel: the energy fractions  $\Omega_i=\rho_i(x)/\rho_{c}(x)$ for  $i=$~radiation  (green, dot-dashed) matter (red, dashed) and dark energy (blue solid line)  (from \cite{Foffa:2013vma}).  }
\end{center}
\end{figure}

\begin{figure}[t]
\begin{center}
\includegraphics[width=0.42\columnwidth]{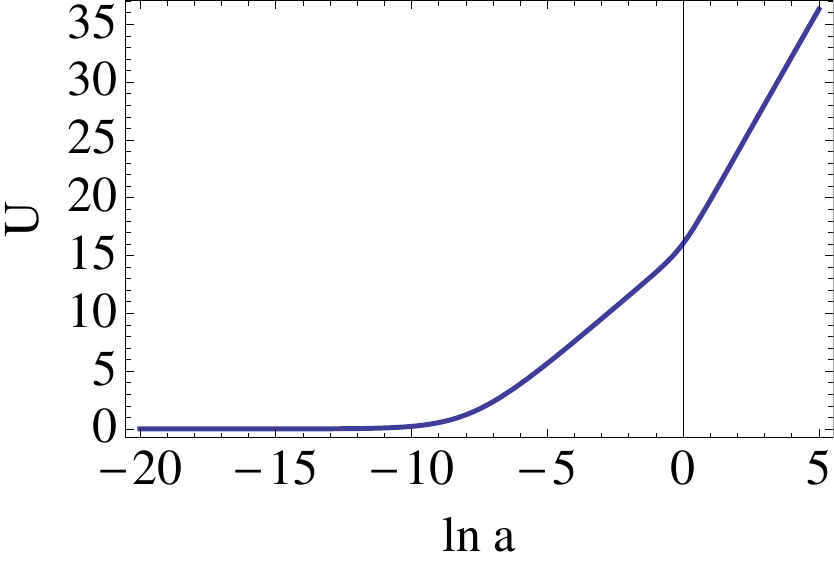}\hspace*{5mm}
\includegraphics[width=0.45\columnwidth]{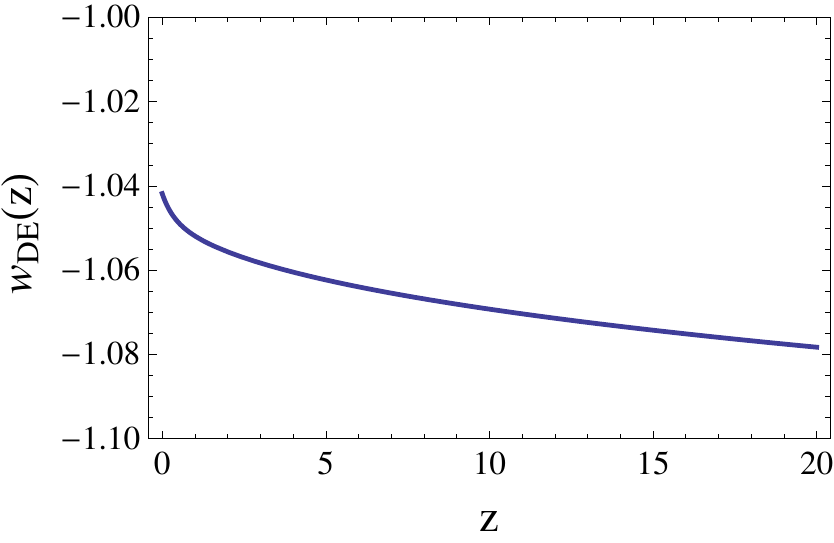}
\caption{Left: the background evolution of the auxiliary field $U$, for the RT model. Right: $w_{\rm DE}$ as a function of the redshift $z$, for the RT model (from \cite{Foffa:2013vma}). \label{fig:fitDeltaw}
}
\end{center}
\end{figure}

It is convenient to pass to dimensionless variables,  using
$x\equiv \ln a(t)$ instead of $t$ to parametrize the temporal evolution. We denote $df/dx=f'$, and we define 
$Y=U-\dot{S}_0$, $h=H/H_0$, 
$\Omega_i (t)=\rho_i(t)/\rho_c(t)$ (where $i$ labels radiation, matter and dark energy), and  $\Omega_i\equiv \Omega_i (t_0)$, where $t_0$ is the present value of cosmic time.
Then the Friedmann equation reads
\be
h^2(x)=\Omega_M e^{-3x}+\Omega_R e^{-4x}+\g Y(x)
\, ,\label{hLCDM}\\
\ee
where $\g\equiv  m^2/(9H_0^2)$. This shows that  there is an effective DE density
\be
\rde(t)=\rho_0\g Y(x)\, , 
\ee
where $\rho_0=3H_0^2/(8\pi G)$. We can trade $S_0$  for $Y$, and rearrange the equations so that $U$ and $Y$ satisfy the coupled system of equations
\bees
&&\hspace*{-5mm}Y''+(3-\zeta)Y'-3(1+\zeta)Y=3U'-3(1+\zeta)U\, ,\label{sy1}\\
&&\hspace*{-5mm}U''+(3+\zeta)U'=6(2+\zeta)\label{sy3}\, ,\\
&&\hspace*{-5mm}\zeta(x)\equiv\frac{h'}{h}=-\, \,
\frac{3\Omega_M e^{-3x}+4\Omega_R e^{-4x}
-\g Y' }{2(\Omega_M e^{-3x}+\Omega_R e^{-4x}+\g Y)}\label{syz}\, .
\ees
The result of the numerical integration is shown in Fig.~\ref{fig:DE}. In terms of the variable $x=\ln a$, radiation-matter equilibrium is at $x=x_{\rm eq}\simeq -8.1$, while the present epoch corresponds to $x=0$. From the left panel of Fig.~\ref{fig:DE} we see that the effective DE vanishes in  RD. This is a consequence of the fact that, in RD, $R=0$, together with our choice of boundary conditions  $U(t_*)=\dot{U}(t_*)=0$ at some initial value $t_*$ deep in RD. As a consequence, $\iBox R$ remains zero in an exact RD phase, and only begins to grow when it starts to feel the effect of non-relativistic matter.
The evolution of the auxiliary field $U=-\iBox R$ is shown in the left panel of Fig.~\ref{fig:fitDeltaw}. We see however that, as we enter in the matter-dominated (MD) phase, the effective DE density start to grow, until it eventually dominates, as we see from the right panel of Fig.~\ref{fig:DE}.
The numerical value of $\ode$ today can be fixed at any desired value, by choosing the parameter $m$ of the nonlocal model (just as in $\Lambda$CDM one can chose $\ola$ by fixing the value of the cosmological constant). In Fig.~\ref{fig:DE} $m$ has been chosen so that, today, $\ode\simeq 0.68$, i.e. $\oma\simeq 0.32$. This is obtained by setting $\gamma \simeq 0.050$, which corresponds to $m\simeq 0.67 H_0$. Of course, the exact value of $\oma$, and therefore of $m$, will eventually be fixed by Bayesian parameter estimation within the model itself, as we will discuss below. 

We also define, as usual, the effective equation-of-state parameter of dark energy, $w_{\rm DE}$, from\footnote{The same expression for $w_{\rm DE}(x)$ can  be obtained defining an effective DE pressure 
$p_{\rm DE}$ from the trace of the $(i,j)$ component of 
the modified Einstein equation (\ref{v2loc1}), and defining   $w_{\rm DE}(x)$ from 
$p_{\rm DE}=w_{\rm DE}\rho_{\rm DE}$. The equivalence of the two definitions is a consequence of the fact that, because of the extraction of the transverse part in the RT model (and because of the general covariance of the action for the RR model), energy-momentum conservation is automatically ensured.}
\be\label{defwDE}
\dot{\rho}_{\rm DE}+3(1+w_{\rm DE})H\rho_{\rm DE}=0\, .
\ee
Once fixed $m$ so to obtain the required value of $\oma$, $\rde(x)$ is fixed, and therefore we get a pure prediction for the evolution of $w_{\rm DE}$ with time. The right panel of
Fig.~\ref{fig:fitDeltaw} shows the result, plotted as a function of redshift $z$. We observe that $w_{\rm DE}(z)$ is on the phantom side, i.e. $w_{\rm DE}(z)<-1$. This is a general consequence of \eq{defwDE}, together with the fact that, in the RT model, $\rho_{\rm DE}>0$, $\dot{\rho}_{\rm DE}>0$, and $H>0$, so 
$(1+w_{\rm DE})$ must be negative. 
Near the present epoch we can compare the numerical evolution with  the widely used fitting function~\cite{Chevallier:2000qy,Linder:2002et}
\be\label{fitCP}
w_{\rm DE}(a)= w_0+(1-a) w_a\, ,
\ee  
(where $a=e^x$), and we get  $w_0 \simeq -1.04$,  $w_a \simeq -0.02$.
These results  are quite interesting, because they show that, at the level of background evolution, the nonlocal term generates an effective DE, which produces a self-accelerating solution with 
$w_{\rm DE}$ close to $-1$.

It is interesting to observe that, in terms  of the field $U=-\iBox R$, \eq{RT} can be replaced by the system of equations
\bees\label{RTU}
\Gmn +\frac{m^2}{3}\( U\gmn\)^{\rm T}&=&8\pi G\,\Tmn\, ,\\
\Box U=-R\, .\label{boxURagain}
\ees
We now observe that, under a shift $U(x)\ra U(x)+u_0$, where $u_0$ is a constant, \eq{boxURagain} is unchanged, while  $(u_0\gmn)^T=u_0\gmn$, since $\n^{\mu}\gmn=0$. Then \eq{RTU} becomes
\be\label{u0cosmconst}
\Gmn +\frac{m^2}{3}\( U\gmn\)^{\rm T}=8\pi G\,\[\Tmn-\frac{m^2u_0}{24\pi G}\gmn\]\, .
\ee
We see that in principle one could chose $u_0$ so to cancel any vacuum energy term in $\Tmn$. 
In particular, given that $m\simeq H_0$, one can cancel a constant positive vacuum energy $T_{00}=\rvac={\cal O}(\mplr^4)$ by choosing a negative value of $u_0$ such that $-u_0={\cal O}(\mplr^2/H_0^2)\sim 10^{120}$ (viceversa, choosing a positive value of $u_0$ amounts to introducing a positive cosmological constant).
This observation is interesting, but unfortunately by itself is not a solution of the cosmological constant problem. We are simply trading the large value of the vacuum energy into a large value of the shift parameter in the transformation $U(x)\ra U(x)+u_0$, and the question is now why the shifted field should have an initial condition $U(t_*)=0$, or anyhow $U(t_*)={\cal O}(1)$, rather than an astronomically large initial value.

The next point to be discussed is how the cosmological background evolution depends on the choice of initial conditions (\ref{initcond}). To this purpose, let us consider first \eq{sy3}. In any given epoch, such as RD, MD, or e.g. an earlier inflationary de~Sitter (dS) phase, the parameter $\zeta$ has an approximately constant value $\zeta_0$, with  $\zeta_0=0$ in dS, $\zeta_0=-2$ in RD and $\zeta_0=-3/2$ in MD. In the approximation of constant $\zeta$ \eq{sy3} can be integrated analytically, and has the solution~\cite{Maggiore:2013mea} 
 \be \label{pertU}
U(x)=\frac{6(2+\zeta_0)}{3+\zeta_0}x+u_0
+u_1 e^{-(3+\zeta_0)x}\, ,
\ee
where the coefficients $u_0,u_1$ parametrize the general solution of the homogeneous equation $U''+(3+\zeta_0)U=0$. The constant $u_0$ corresponds to the reintroduction of a cosmological constant, as we have seen above. We will come back to its effect in Sect.~\ref{sect:exten}. The other solution of the homogeneous equation, proportional to $e^{-(3+\zeta_0)x}$, is instead a decaying mode, in all cosmological phases. Thus, 
the solution with initial conditions $U(t_*)=\dot{U}(t_*)=0$ has a marginally stable direction, corresponding to the possibility of reintroducing a cosmological constant, and  a stable direction, i.e. is an attractor in the $u_1$ direction. Perturbing the initial conditions is equivalent to introducing a non-vanishing value of $u_0$ and $u_1$. We see that the introduction of $u_0$ will in general lead to differences in the cosmological evolution, which we will explore below, while $u_1$ corresponds to an irrelevant direction. In any case, it is reassuring that there is no growing mode in the solution of the homogeneous equation. Consider now \eq{sy1}. Plugging \eq{pertU} into \eq{sy1} and solving for $Y(x)$ 
we get \cite{Maggiore:2013mea}
\bees
Y(x)&=&-\frac{2(2+\zeta_0)\zeta_0}{(3+\zeta_0)(1+\zeta_0)}
+\frac{6(2+\zeta_0)}{3+\zeta_0} x+u_0
-\frac{6(2+\zeta_0) u_1 }{2\zeta_0^2+3\zeta_0-3}e^{-(3+\zeta_0)x} 
\nn\\
&& +a_1 e^{\a_{+}x}+ a_2 e^{\a_{-}x}\, ,\label{pertY}
\ees
where 
\be\label{apm}
\a_{\pm}=\frac{1}{2}\[-3+\zeta_0\pm \sqrt{21+6\zeta_0+\zeta_0^2}\]\, .
\ee
In particular, in dS there is a growing mode with  $\alpha_+=(-3+\sqrt{21})/2\simeq 0.79$. In RD both modes are decaying, and the mode that decays more slowly is the one with 
$\alpha_+=(-5+\sqrt{13})/2\simeq -0.70$ while in MD again both modes are decaying, and 
$\a_{+}=(-9+\sqrt{57})/4\simeq -0.36$. Thus, if we start the evolution in RD,  in the space $\{u_0,u_1,a_1,a_2\}$ that parametrizes the initial conditions of the auxiliary fields, there is one marginally stable direction and three stable directions. However, if we start from an early inflationary era, there is a growing mode corresponding to the $a_1$ direction. Then $Y$ will grow during dS (exponentially in $x$, so as a power of the scale factor), but will then decrease again during RD and MD. We will study the resulting evolution in Sect.~\ref{sect:exten}, where we will see that even in this case a potentially viable background evolution emerges. In any case, it is important that in RD and MD there is no growing mode, otherwise the evolution would necessarily  eventually lead us far from an acceptable FRW solution. This is indeed what happens
in the model (\ref{GmnT}), where the homogeneous solutions associated to an auxiliary field are unstable both in RD and in MD (see app.~A of \cite{Foffa:2013vma}), and is the reason why we have discarded that model.

\subsubsection{The RR model}
Similar results are obtained for the RR model. Specializing to a FRW background, and  using the dimensionless field $W(t)=H^2(t)S(t)$ instead of $S(t)$, 
\eqst{Gmn}{BoxS} become
\bees
&&h^2(x)=\Omega_M e^{-3x}+\Omega_R e^{-4x}+\g Y\label{syh}\\
&&U''+(3+\zeta) U'=6(2+\zeta)\, ,\label{syU}\\
&&W''+3(1-\zeta) W'-2(\zeta'+3\zeta-\zeta^2)W= U\, ,\label{syW}
\ees
where again $\gamma= m^2/(9H_0^2)$, $\zeta=h'/h$ and
\be\label{defYRR}
Y\equiv \frac{1}{2}W'(6-U') +W (3-6\zeta+\zeta U')+\frac{1}{4}U^2\, .
\ee
From this form of the equations we see that  there is again an effective dark energy density, given by $\rde=\rho_0\gamma Y$. 

\begin{figure}[t]
\centering
\includegraphics[width=0.45\columnwidth]{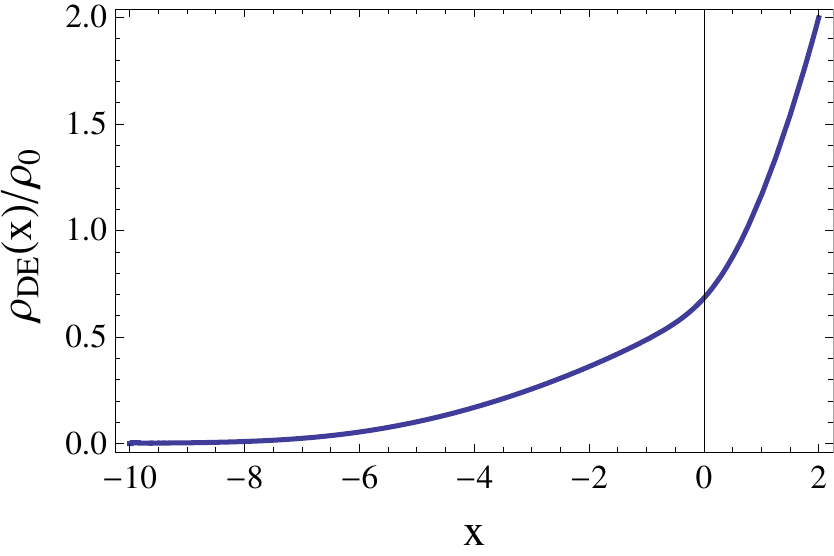}
\includegraphics[width=0.47\columnwidth]{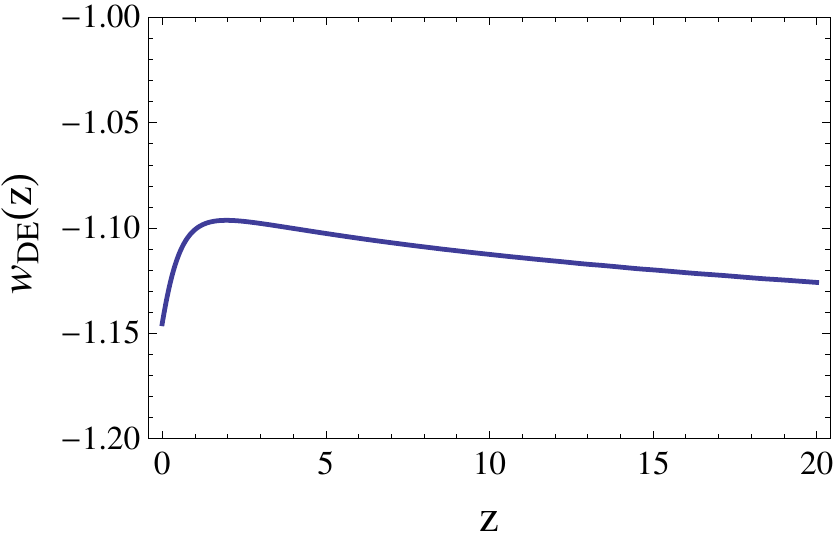}
\includegraphics[width=0.47\columnwidth]{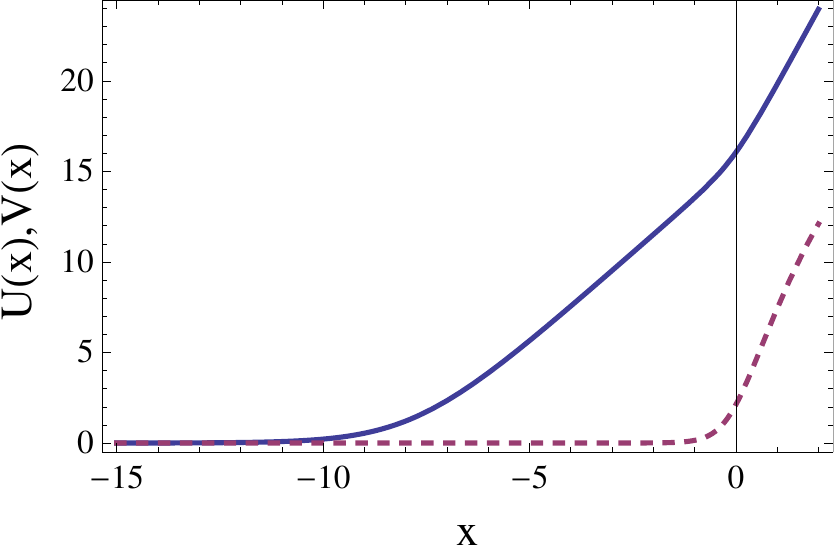}
\caption{\label{fig:rhoDE_RR} Upper left panel: the function $\rde(x)/\rho_0$  against $x=\ln a$, for the RR model.
Upper right panel:   the function $w_{\rm DE}(z)$. Lower panel: the background evolution of the auxiliary fields $U$  (blue solid line) and $V$ (red dashed line).
From \cite{Dirian:2014ara}.
}
\end{figure}

To actually perform the numerical integration of these equations, and also to study the perturbations, it can be more convenient to use a variable $V(t)=H_0^2S(t)$ instead of $W(t)=H^2(t)S(t)$. Then 
\eqst{syh}{syW} are replaced by
\bees
&&h^2(x)=\frac{\Omega_M e^{-3x}+\Omega_R e^{-4x}+(\g/4) U^2}
{1+\g[-3V'-3V+(1/2)V'U']}\,,\label{syh2}\\
&&U''+(3+\zeta) U'=6(2+\zeta)\, ,\label{syU2}\\
&&V''+(3+\zeta) V'=h^{-2} U\label{syV}\, .
\ees
In \eqs{syU2}{syV} appears $\zeta=h'/h$. In turn, $h'$ can be computed explicitly from
\eq{syh2}. The resulting expression contains $V''$ and $U''$, which can  be eliminated using again
\eqs{syU2}{syV}. This gives
\be\label{zetaRR}
\zeta= \frac{1}{2(1-3 \gamma V)}
\[ h^{-2}\Omega'+3\g\(h^{-2}U+U'V'-4V'\)\]\, ,
\ee
where $\Omega(x)=\Omega_M e^{-3x}+\Omega_R e^{-4x}$. Then \eqs{syU2}{syV}, with $h^2$ given by \eq{syh2} and $\zeta$ given by \eq{zetaRR}, provide a closed set of second order equations for $V$ and $U$, whose numerical integration is straightforward.

The result of the numerical integration is shown in Fig.~\ref{fig:rhoDE_RR}.  Similarly to 
\eq{initcond} for the RT model, we set initial  conditions $U=U'=V=V'=0$ at some initial time $x_{\rm in}$ deep in RD (we will see in Sect.~\ref{sect:effectinfl} how the results depend on this choice).
In this case  we get $w_0 \simeq -1.14$, 
$w_a \simeq 0.08$~\cite{Maggiore:2014sia}, so the RR model differs more from $\Lambda$CDM, compared to the RT model, at the level of background evolution. In the RR model, to obtain for instance a value $\oma=0.32$, i.e. $\ode=0.68$, we must fix $m\simeq 0.28 H_0$.

The dependence on the initial conditions can be studied as before. The equation for $U$ is the same as in the RT model, so the  homogeneous solution for $U$ is again  $u_0+u_1 e^{-(3+\zeta_0)x}$. The homogeneous equation for $V$ is the same as that for $U$, so  similarly the homogeneous solution for $V$ is 
$v_0+v_1 e^{-(3+\zeta_0)x}$.
In the early Universe we have $-2\leq \zeta_0\leq 0$ and all these terms are either constant or exponentially decreasing, which means that the 
solutions for both $U$ and $V$ are stable  in MD, RD, as well as in a previous  inflationary stage. From this point of view the RR model differs from the RT model which, as we have seen, has a growing mode during a dS phase. Note also the the constant $u_0$ now no longer has the simple interpretation of a cosmological constant term since, contrary to \eq{panU},  \eq{BoxS} is not invariant under $U\ra U+u_0$.

\subsection{Cosmological perturbations}\label{sect:cosmpert}

In order to asses the viability of these models, the next step is the study of their cosmological perturbations. This has been done in \cite{Dirian:2014ara}. Let us considering first the scalar perturbations.  We work 
in the  Newtonian gauge, and write the metric perturbations as
\be\label{defPhiPsi}
ds^2 =  -(1+2 \Psi) dt^2 + a^2(t) (1 + 2 \Phi) \delta_{ij} dx^i dx^j\, .
\ee
We then add the perturbations of the auxiliary fields, see below,
we  linearize the equations of motion and go in momentum space. We denote by $k$ the comoving momenta, and  define
\be
\kappa \equiv k/k_{\rm eq}\, , 
\ee
where
$k_{\rm eq}=a_{\rm eq} H_{\rm eq}$ is the wavenumber of the mode that enters the horizon at matter-radiation equilibrium. To illustrate our numerical results, we   
use as reference values $\kappa = 0.1,1,5$.  The mode with $\kappa=5$ entered  inside the horizon already during RD, while 
the mode  $\kappa=1$ reentered at matter-radiation equality. In contrast,  the mode with $\kappa=0.1$  was  outside the horizon during RD and most of MD, and re-entered at $z\simeq 1.5$. Overall, these three values of $k$ illustrate well  the $k$ dependence of the results, covering the range of $k$ relevant   to the regime of linear structure formation.

We summarize here the results for the RT and RR models, referring the reader to \cite{Dirian:2014ara} for details and for the (rather long) explicit expression of the perturbation equations.

\subsubsection{RT model}

In the RT model we expand the auxiliary fields as
\be
U(t,\vx)=\bar{U}(t)+\d U(t,\vx)\, ,\qquad S_{\mu}(t,\vx)=\bar{S}_{\mu}(t) +\d S_{\mu}(t,\vx)\, .
\ee
In FRW,  the background value
$\bar{S}_{i}$ vanishes because there is no preferred spatial direction, but of course its perturbation $\d S_i$ is a dynamical variable. As with any vector, we can decompose it into a transverse and longitudinal part,
$\d S_i=\d S_i^{\rm T}+\pa_i (\d S)$
where $\pa_i(\d S_i^{\rm T})=0$. Since we  restrict here to scalar perturbations, we only retain $\d S$, and write $\d S_i=\pa_i (\d S)$. Thus in this model the scalar  perturbations  are given by $\Psi,\Phi,\d U,\d S_0$ and $\d S$, see also  \cite{Kehagias:2014sda,Nesseris:2014mea}.

\begin{figure}[t]
\begin{center}
\includegraphics[width=0.45\columnwidth]{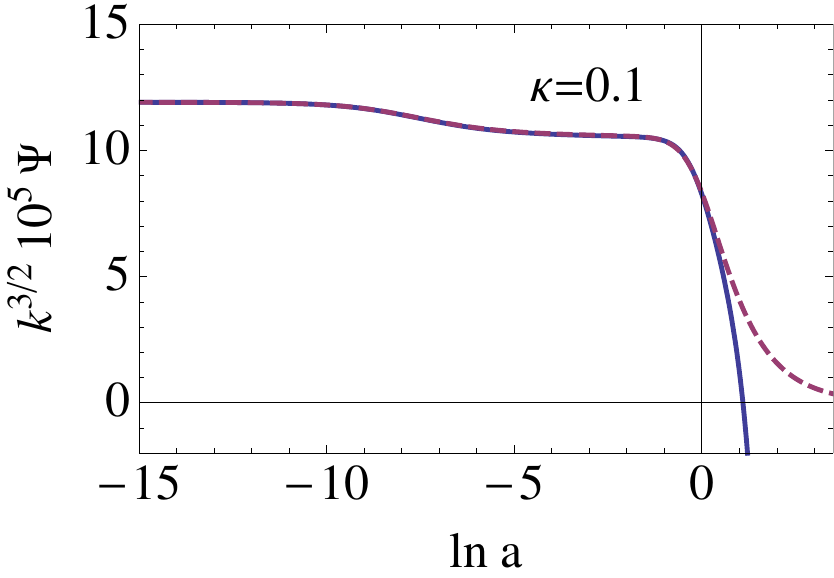}
\includegraphics[width=0.45\columnwidth]{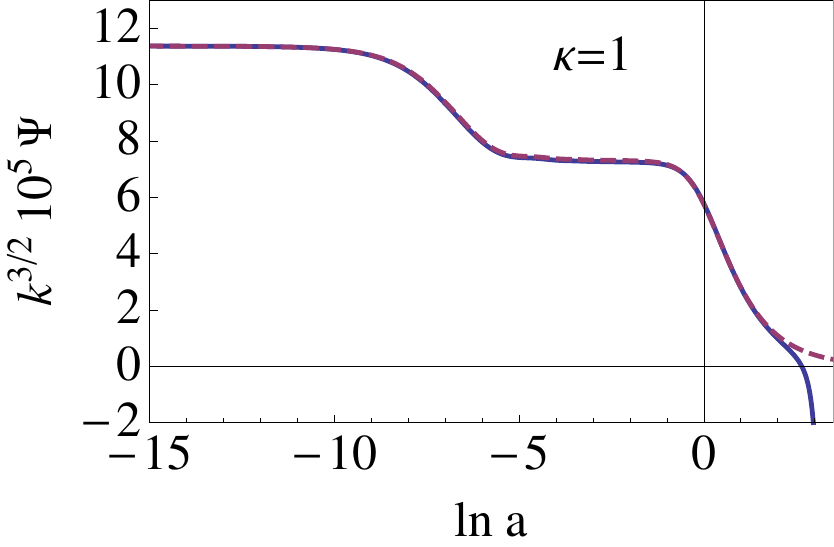}
\includegraphics[width=0.45\columnwidth]{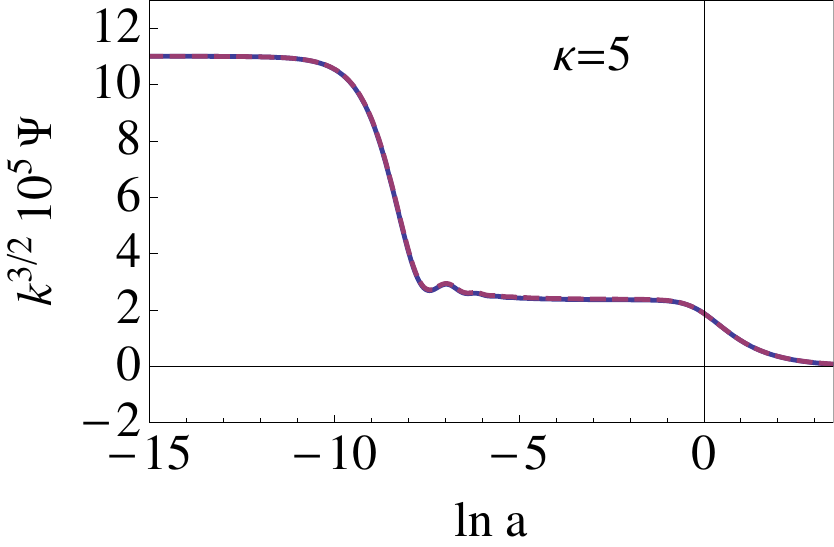}
\includegraphics[width=0.45\columnwidth]{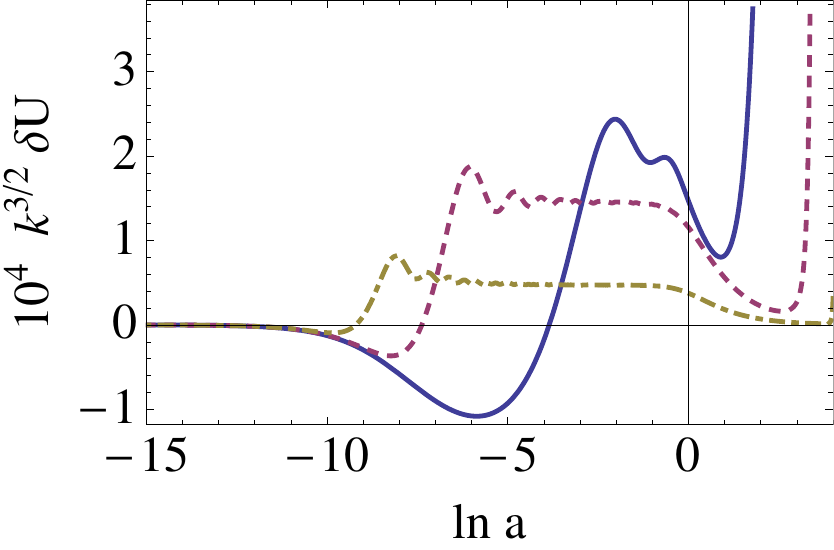}
\caption{\label{fig:P123RT} $k^{3/2}\Psi(a;k)$  in the RT model (blue solid line) and in $\Lambda$CDM (purple dashed line), as a function of $x=\ln a(t)$, for  
$\kappa = 0.1$ (left upper panel), $\kappa=1$ (right upper panel), $\kappa=5$ (lower left panel).
Observe that the quantity that we plot is $k^{3/2}\Psi(a;k)$ multiplied by a factor $10^5$.
Lower right panel: the evolution of the perturbations $\d U$ for $\kappa=0.1$ (blue solid line),
$\kappa=1$ (purple, dashed) and $\kappa=5$ (green, dot-dashed).}
\end{center}
\end{figure}

\begin{figure}[t]
\begin{center}
\includegraphics[width=0.45\columnwidth]{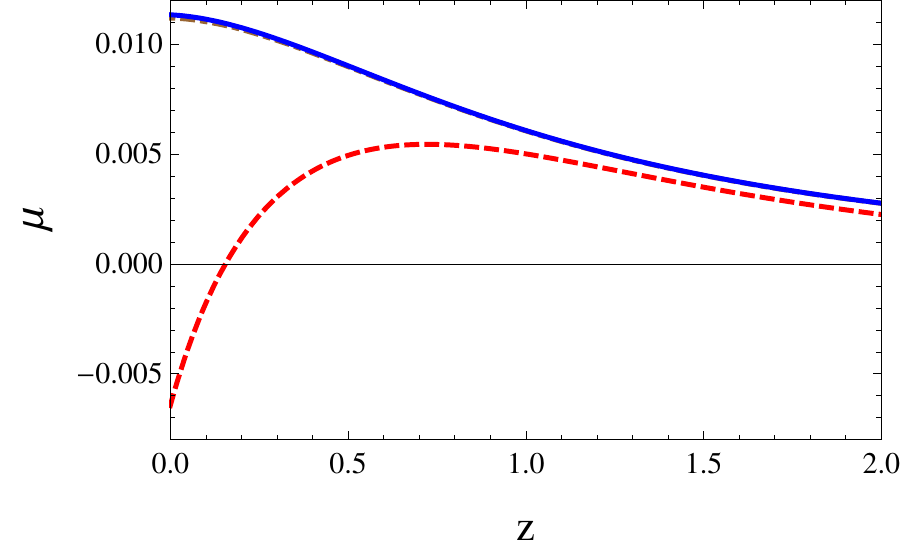}
\includegraphics[width=0.45\columnwidth]{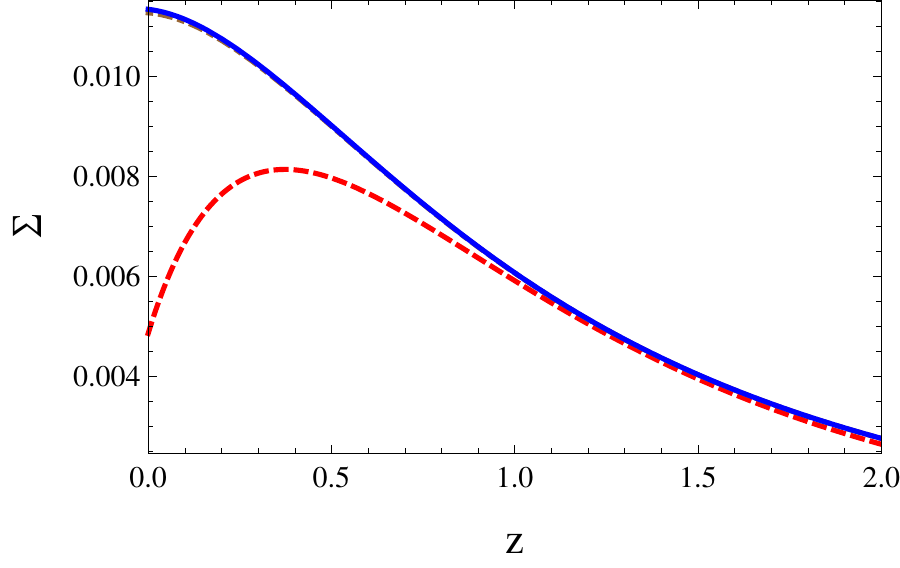}
\includegraphics[width=0.45\columnwidth]{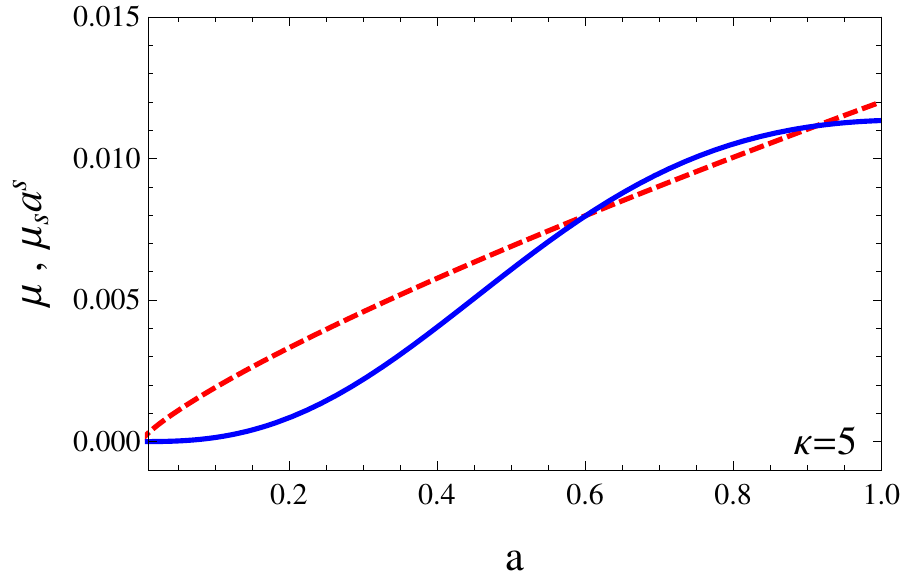}
\caption{\label{fig:muSigmaRT} Upper left panel: $\mu(z;k)$, as a function of the redshift $z$, for  
$\kappa = 0.1$ (red dashed) $\kappa=1$ (brown dot-dashed)
and $\kappa =5$ (blue solid line), for the RT model.  The curves for $\kappa = 1$ and  $\kappa=5$ are almost indistinguishable on this scale.
Upper right panel: the same for $\Sigma(z;k)$.
Lower panel: $\mu(a;k)$, as a function of the scale factor $a$, for  
$\kappa =5$ (blue solid line), for the RT model, compared to the function $\mu(a)=\mu_s a^s$ with $\mu_s=0.012$ and $s=0.8$ (red dashed). 
}
\end{center}
\end{figure}

Fig.~\ref{fig:P123RT} shows the time evolution of the Fourier modes of the Bardeen variable $\Psi_k$ for our three reference values of $\kappa$ (blue solid line) and compare with the corresponding result in 
$\Lambda$CDM (purple dashed line).\footnote{Figs.~\ref{fig:P123RT} and \ref{fig:muSigmaRT} have been obtained by Dirian {\em et al.} in the work leading to \cite{Dirian:2014ara} although there, for reasons of space, we  only published the corresponding figures relative to the RR model. Note also that the quantity plotted as $\Psi$ in fig.~6 of \cite{Dirian:2014ara} was actually $-\Psi$.} 
As customary, we actually plot $k^{3/2}\Psi_k$, whose square gives the variance of the field per unit logarithmic interval of momentum, according to
\be
\langle\Psi^2(\vx)\rangle=\int\frac{d^3k}{(2\pi)^3}\langle |\Psi_{\vk}|^2\rangle=\frac{1}{2\pi^2}\int_{0}^{\infty}\frac{dk}{k}\, \langle |k^{3/2}\Psi_k|^2\rangle\, ,\label{4Phi2power}
\ee
where the bracket is the ensemble average over the initial conditions, that we take to be the standard adiabatic initial conditions. Note also that, if start the evolution choosing real  initial conditions on $\Psi_k$, it remains real along the evolution.

We see from fig.~\ref{fig:P123RT} that, up to the present time $x=0$, the evolution of the perturbations is well-behaved, and very close to that of $\Lambda$CDM, even if in the cosmological future the perturbations will enter the nonlinear regime much earlier than for $\Lambda$CDM. In particular, the perturbation of the `would-be' ghost field $U$, up to the present time, are small, with $k^{3/2}U_k\sim 10^{-4}$.
Observe that in the cosmological future the perturbation becomes non-linear, both for $\Psi_k$ and for $\d U_k$, with the nonlinearity kicking in earlier for the lower-momentum modes.\footnote{Nevertheless, even the longest observable wavelength, which can be observed through their effect on the CMB, remain well linear up to the present epoch. We will see indeed from a full Boltzmann code analysis in Sect.~\ref{sect:Baye} that the nonlocal models fit very well the CMB data.}
 This can be understood as follows. Any classical instability possibly induced by the nonlocal term will only develop on a timescale $t$ such that $mt$ is (much) larger than one. However, we have seen that, to reproduce the typical observed value of $\oma$, $m$ is of order $H_0$, and in fact numerically smaller, with $m\simeq 0.28 H_0$ for the RT model (see sect.~\ref{sect:Baye} for accurate Bayesian parameter estimation). Thus,  instabilities induced by the nonlocal term, if present, only develop on a timescale larger or equal than to a few times  $H_0$, and therefore in the cosmological future.

Beside following the cosmological evolution for the fundamental perturbation variables, such as $\Psi_k(x)$ (recall that $x\equiv \ln a(t)$ is our time-evolution variable, not to be confused with a spatial variable!), the  behavior of the perturbations can also be conveniently described by some indicators of deviations from $\Lambda$CDM. 
Two useful quantities are the functions $\mu(x;k)$  and $\Sigma(x;k)$, defined by
\bees
\Psi&=&[1+\mu(x;k)]\Psi_{\rm GR}\, ,\label{defmu}\\
\Psi-\Phi&=&[1+\Sigma(x;k)] (\Psi-\Phi)_{\rm GR}\label{defSigma}\, ,
\ees
where the subscript `GR' denotes the same quantities computed in GR, assuming a $\Lambda$CDM model with the same value of $\oma$ as the modified gravity model. The advantage of using $\Psi$ and $\Phi-\Psi$ as independent combinations is that the former enters in  motion of non-relativistic particles, while the latter determines the light propagation. The numerical results for the RT model are shown in upper panels of Fig.~\ref{fig:muSigmaRT}. We see that, in this model, the deviations from $\Lambda$CDM are very tiny, of order of $1\%$ at most, over the relevant wavenumbers and redshifts. In the forecast for experiments, $\mu(x;k)$ is often approximated as a function independent of $k$, with a power-like dependence on the scale factor,
\be\label{muas}
\mu(a)=\mu_s a^s\, . 
\ee
For the RT model we find that the scale-independent approximation is good, in the range of momenta relevant for structure formation, but the functional form (\ref{muas}) only catches the gross features of the $a$-dependence. The lower panel of Fig.~\ref{fig:muSigmaRT} compares the function $\mu(a,k)$ computed numerically for $\kappa=5$, with the function 
(\ref{muas}), setting $\mu_s=0.012$ and $s=0.8$.


Another useful indicator of deviations from GR is the effective Newton's constant, which is defined so that the Poisson equation for the Bardeen variable $\Phi$ takes the same for as in GR, with Newton's constant $G$ replaced by a function $G_{\rm eff}(x;k)$. In the RT model, for modes inside the horizon, 
\cite{Nesseris:2014mea,Dirian:2014ara},
\be\label{GeffGRTlargek}
\frac{G_{\rm eff}}{G}=1+{\cal O}\(\frac{1}{\hat{k}^2}\)\, ,
\ee
where $\hat{k}=k/(aH)$.
This gives  again the information that, for the RT model, deviations from $\Lambda$CDM in structure formations are quite tiny. We will see  in more detail in Sect.~\ref{sect:Baye} how the predictions of the model compare with that of $\Lambda$CDM for CMB,SNae, BAO and structure formation data.

\subsubsection{RR model}
In the RR model, in the study of perturbations we find convenient to use $U$ and $V=H_0^2S$ (rather than $W=H^2(t)S$). In the scalar sector
we   expand the metric as in \eq{defPhiPsi} and the auxiliary fields as
$U(t,\vx)=\bar{U}(t)+\d U(t,\vx)$, $ V=\bar{V}(t) +\d V(t,\vx)$.
Thus, in this model the scalar  perturbations  are described by $\Psi,\Phi,\d U$ and $\d V$.

\begin{figure}[t]
\centering
\includegraphics[width=0.45\columnwidth]{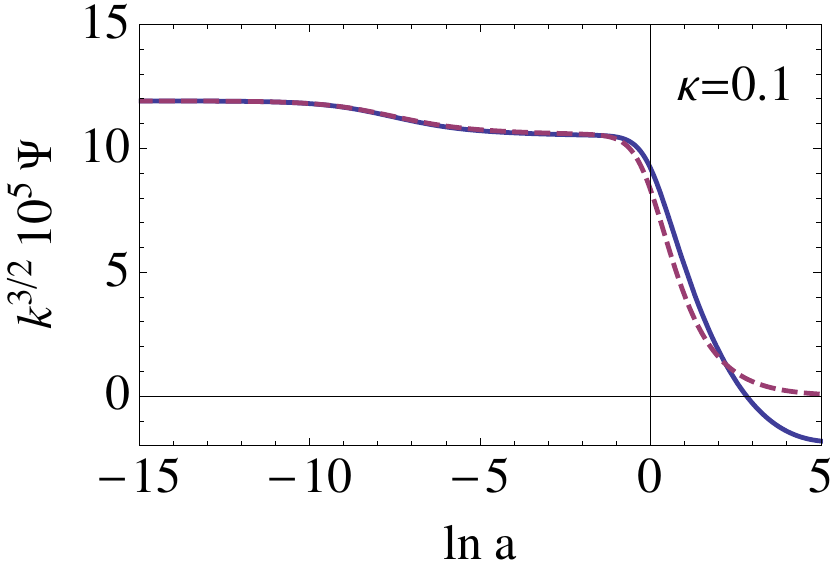}
\includegraphics[width=0.45\columnwidth]{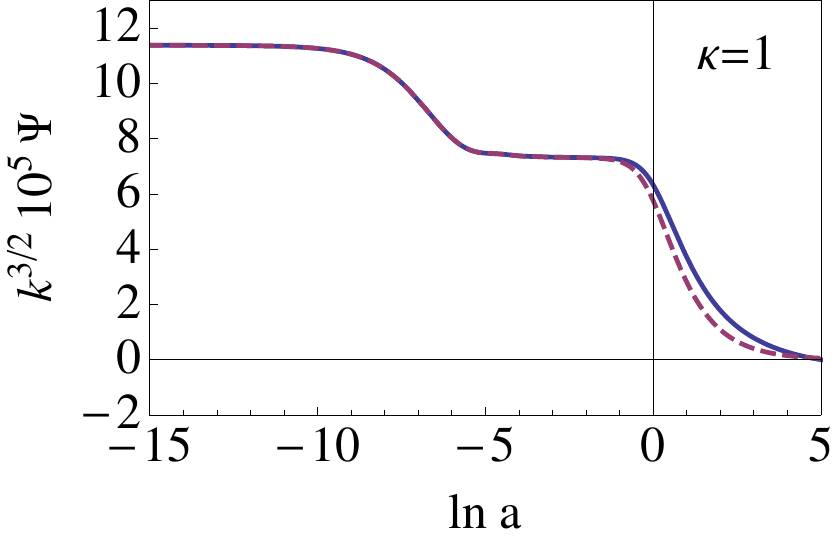}
\includegraphics[width=0.45\columnwidth]{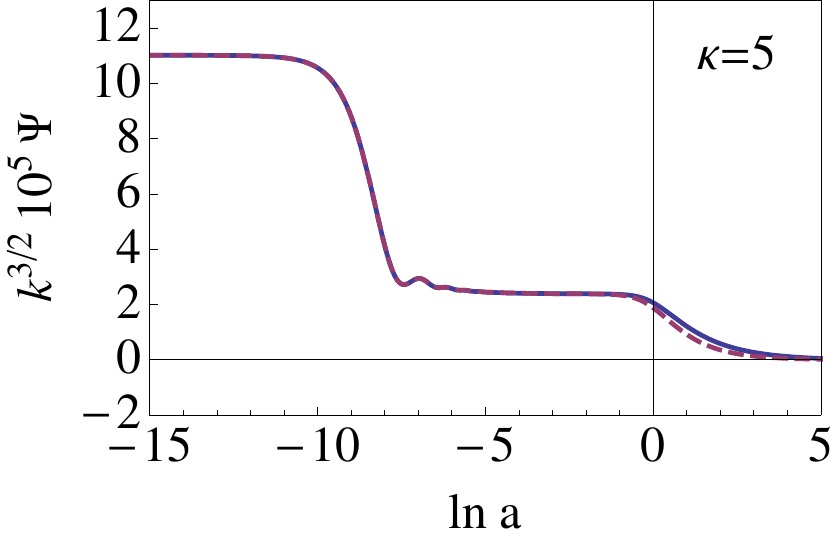}
\includegraphics[width=0.45\columnwidth]{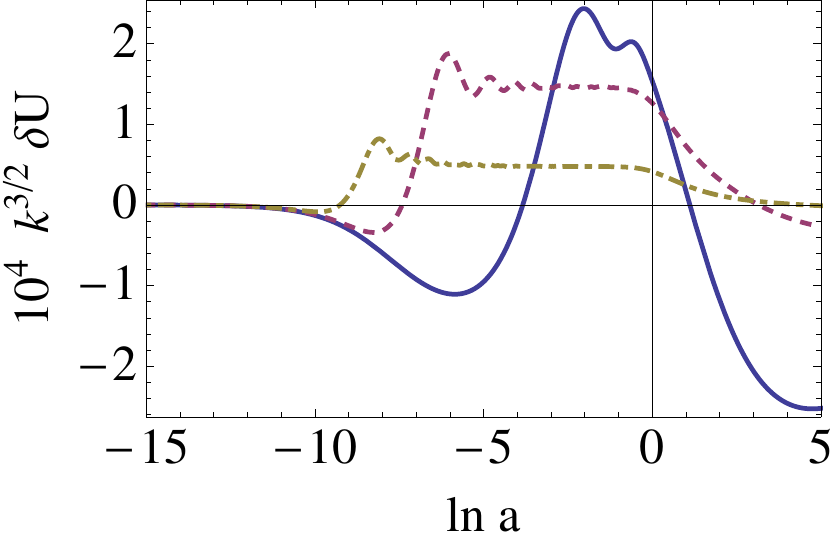}
\caption{\label{fig:F1F2} $k^{3/2}\Psi(a;k)$  from the RR model (blue solid line) and from $\Lambda$CDM (purple dashed line), as a function of $x=\ln a(t)$, for  
$\kappa = 0.1$ (left upper panel), $\kappa=1$ (right upper panel), $\kappa=5$ (left lower panel).
Observe that, on the vertical axis, we plot $10^5k^{3/2}\Psi(a;k)$. Adapted from \cite{Dirian:2014ara}.
Lower right panel: the evolution of the perturbations $\d U$ for $\kappa=0.1$ (blue solid line),
$\kappa=1$ (purple, dashed) and $\kappa=5$ (green, dot-dashed).
}
\end{figure}

The results for the evolution of $\Psi$ are shown in Fig.~\ref{fig:F1F2}. We see that again the perturbations are well-behaved, and very close to $\Lambda$CDM. 
Compared to the RT model, the deviations from $\Lambda$CDM are somewhat larger, up to the present epoch. However, contrary to the RT model, they also stay relatively close to $\Lambda$CDM even in the cosmological future.

\begin{figure}[th]
\centering
\includegraphics[width=0.45\columnwidth]{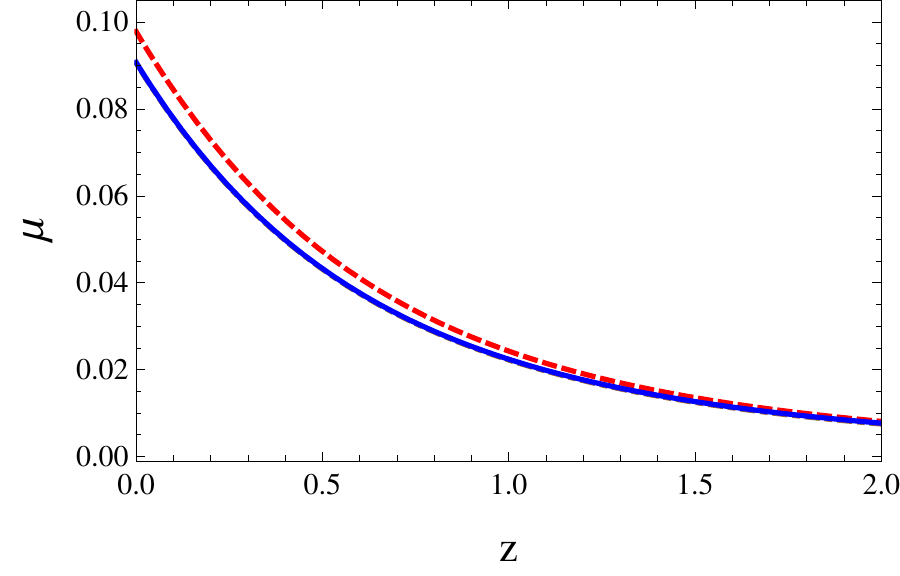}
\includegraphics[width=0.45\columnwidth]{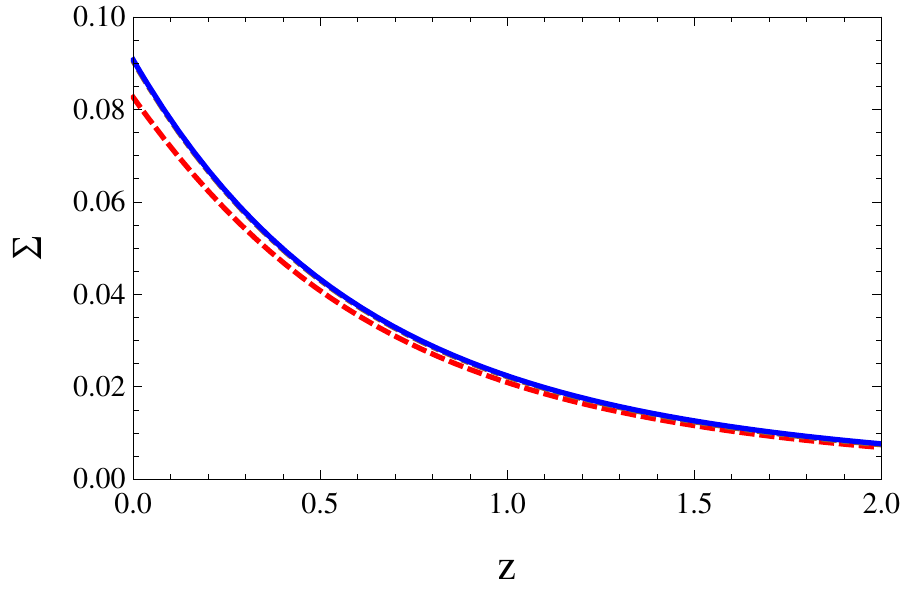}
\includegraphics[width=0.45\columnwidth]{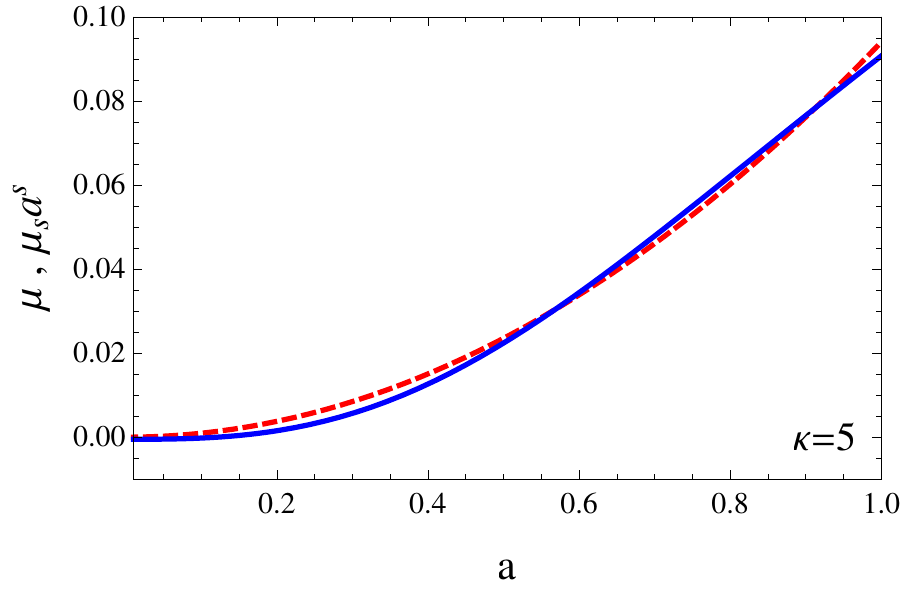}
\includegraphics[width=0.45\columnwidth]{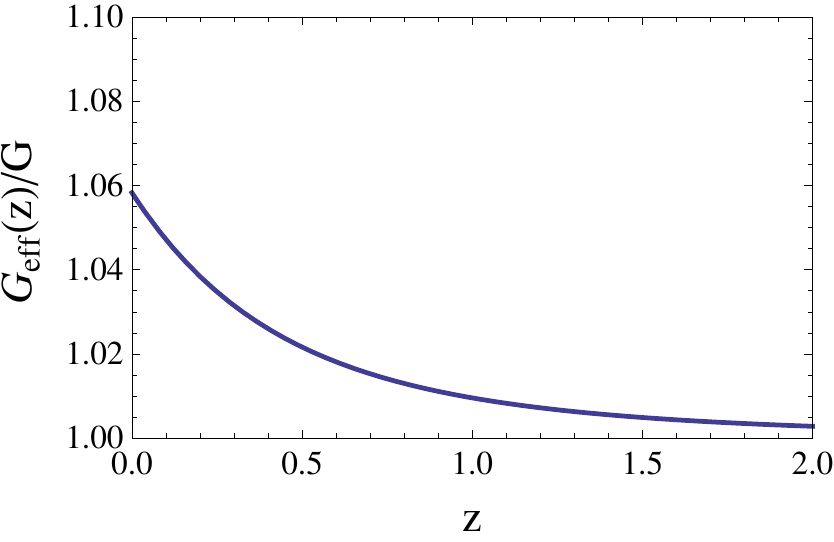}
\caption{\label{fig:N3N3b} Upper left panel: $\mu(z;k)$, as a function of the redshift $z$ for the RR model, for   $\kappa = 0.1$ (red dashed) $\kappa=1$ (brown dot-dashed)
and $\kappa =5$ (blue solid line).  The curves for $\kappa = 1$ and  $\kappa=5$ are almost indistinguishable on this scale.
Upper right panel: the same for $\Sigma(z;k)$. 
Lower left panel: $\mu(a;k)$, as a function of the scale factor $a$, for  
$\kappa =5$ (blue solid line), for the RT model, compared to the function $\mu(a)=\mu_s a^s$ with 
$\mu_s=0.094 $ and $s=2$ (red dashed).  Lower right panel:
$G_{\rm eff}/G$ as a function of the redshift $z$, for sub-horizon modes, for the RR model.
From \cite{Dirian:2014ara}.
}
\end{figure}

The functions $\mu$ and $\Sigma$ are shown as functions of the redshift in the 
upper panels of Fig.~\ref{fig:N3N3b}, for our three reference value of the wavenumber.  
At a redshift such as $z=0.5$, typical for the comparison with structure formation data,  they are of order $5\%$, so again larger than in the RT model.
For the RR model $\mu$, as a function of the scale factor, is well reproduced by \eq{muas}, with
\be
\mu_s=0.094\, ,\qquad  s=2\, , 
\ee
see the lower panel of Fig.~\ref{fig:N3N3b}. By comparison, 
the forecast for {\sc Euclid} on the error $\sigma(\mu_s)$ over the parameter $\mu_s$, for fixed cosmological parameters, is $\sigma(\mu_s)=0.0046$ for $s=1$ and 
$\sigma(\mu_s)=0.014$ for $s=3$~\cite{Song:2010fg}. Thus (barring the effect of degeneracies with other cosmological parameters), we expect that the accuracy of
{\sc Euclid} should be sufficient to test the
prediction for  $\mu_s$ from the RR model, and  possibly also for the RT model.

Finally, the effective Newton's constant in the RR model, for sub-horizon scales, is given by
\be\label{Geff1suk2}
\frac{G_{\rm eff}(x;k)}{G} =  \frac{1}{1-3\gamma\bar{V}(x)}\, \[ 1 +{\cal O}\(\frac{1}{\hat{k}^2}\)\]\, .
\ee
Thus in the sub-horizon limit, $G_{\rm eff}(x;k)$ becomes independent of $k$. However, contrary to the RT model, it retains a time dependence. The behavior of $G_{\rm eff}$ as a function of the redshift is shown in the lower right panel of Fig.~\ref{fig:N3N3b}.

Nonlinear structure formation has also been studied, for the RR model, using $N$-body simulations~\cite{Barreira:2014kra}. The result is that, in the high-mass tail of the distribution,  massive dark matter haloes are slightly more abundant, by about $10\%$ at $M\sim 10^{14}\msun/h_0$. The halo density profile is also spatially  more concentrated, by about $8\%$ over a range of masses.\footnote{The result of \cite{Barreira:2014kra} were obtained using, for the RR model, the value of the cosmological parameters obtained in $\Lambda$CDM, before parameter estimation for the RR models was performed in 
\cite{Dirian:2014bma,Dirian:2016puz}, see below. It would be interesting to repeat the analysis using the best-fit parameters of the RR model, and to extend it also to the RT model.}

Tensor perturbations have also been studied in \cite{Dirian:2016puz,Cusin:2015rex}, for both the RR and RT models, and again their evolution is well behaved, and very close to that in $\Lambda$CDM.

\subsection{Bayesian parameter estimation and comparison with $\Lambda$CDM}\label{sect:Baye}

The results of the previous sections show that the RR and RT nonlocal models give a viable cosmology at the background level, with an accelerated expansion obtained  without the need of a cosmological constant. Furthermore, their cosmological perturbations are well-behaved and in the right ballpark for being consistent with the data, while still sufficiently different from $\Lambda$CDM to raise the hope that the models might be distinguishable with present or near-future observations. We can therefore go one step forward, and implement the cosmological perturbations in a Boltzmann code, and perform Bayesian parameter estimation. We can then  compute the relevant chi-squares or Bayes factor, to see if these  models can `defy' 
$\Lambda$CDM, from the point of view of fitting the data. We should stress that this is a level of comparison with the data, and with $\Lambda$CDM, that none of the other infrared modifications of GR widely studied in the last few years has ever reached. The relevant analysis has been performed in \cite{Dirian:2014bma}, using the {\em Planck} 2013 data then available, together with supernovae and BAO data, and updated and extended in
\cite{Dirian:2016puz}, using the {\em Planck} 2015 data. 

In particular, in \cite{Dirian:2016puz}
we  tested the nonlocal models against the {\em Planck} 2015 TT, TE, EE and lensing data from Cosmic Microwave Background (CMB),  isotropic and anisotropic Baryonic Acoustic Oscillations (BAO)  data, JLA supernovae,  $H_0$ measurements and growth rate data, implementing the perturbation equations in a modified CLASS \cite{Class} code.  As independent cosmological parameters we take the Hubble parameter today
$H_0 = 100 h \, \rm{km} \, \rm{s}^{-1} \rm{Mpc}^{-1}$, the physical baryon and cold dark matter density fractions today $\omega_b = \Omega_b h^2$ and $\omega_c = \Omega_c h^2$, respectively, the amplitude  $A_s$ and  the spectral tilt $n_s$ of primordial scalar perturbations  and  the reionization optical depth  $\tau_{\rm re}$, so we have a $6$-dimensional parameter space.  For  the neutrino masses  we use the same values as in  the \textit{Planck} 2015 baseline analysis \cite{Planck_2015_CP}, i.e. two massless and a massive neutrinos, with
$\sum_{\nu} m_{\nu}=0.06$~eV, and we fix the effective number of neutrino species to 
$N_{\rm eff}=3.046$. 

Observe that, in the spatially flat case that we  consider, in $\Lambda$CDM  the dark energy density fraction $\ola$ can be taken as a derived parameter, fixed in terms of the other parameters by the flatness condition. Similarly, in the nonlocal models $m^2$ can be taken as a derived parameter,  fixed again  by the flatness condition. Thus, not only the nonlocal models have the same number of parameters as $\Lambda$CDM, but in fact the independent parameters  can be chosen so that  are exactly the same in the nonlocal models and in $\Lambda$CDM.

The results are shown in Table~\ref{partable}. On the left table we combine the {\em Planck } CMB data with JLA supernovae and with a rather complete set of BAO data, described in \cite{Dirian:2016puz}. On the right table we also add a relatively large value of $H_0$, of the type suggested by local measurement. The most recent analysis of local measurements, which appeared after  \cite{Dirian:2016puz} was finished, gives
$H_0 = 73.02 \pm 1.79$ \cite{Riess:2016jrr}. In the last row we give the difference of $\chi^2$, with respect to the model that has the lowest $\chi^2$.  Let us recall that, according to the standard Akaike or Bayesian information criteria,  in the comparison between two models with the same number of parameters, a difference $|\Delta \chi^2| \leq  2$ implies  statistically equivalence between the two models compared, while $|\Delta \chi^2| \gtrsim 2$ suggests ``weak evidence'', and $|\Delta \chi^2|\gtrsim 6$ indicates ``strong evidence''.\footnote{The comparison of the $\chi^2$ is not genuinely Bayesian.  A more accurate method for comparing models, which is fully Bayesian, is based on Bayes factors. We checked in \cite{Dirian:2016puz} that the results  obtained from the computation of the Bayes factor are  in full agreement with that obtained from the comparison of the $\chi^2$.}

\begin{table}[t]
\centering
\resizebox{12cm}{!}{
\begin{tabular}{|l||c|c|c||c|c|c|} 
 \hline 
\multicolumn{1}{|l||}{ } & \multicolumn{3}{|c||}{BAO+Planck+JLA} & \multicolumn{3}{|c|}{BAO+Planck+JLA+$(\rm{H_0}=73.8$)} \\ \hline
 \hline 
 Param & $\Lambda$CDM  & RT &RR & $\Lambda$CDM  & RT &RR\\ \hline 
$100~\omega_{b }$  & $2.228_{-0.015}^{+0.014}$ &$2.223_{-0.014}^{+0.014}$ &$2.213_{-0.014}^{+0.014}$

& $2.233_{-0.014}^{+0.014}$ &$2.226_{-0.014}^{+0.014}$ &$2.217_{-0.014}^{+0.014}$ \\

$\omega_c$ & $0.119_{-0.0011}^{+0.0011}$  & $0.1197_{-0.00096}^{+0.0011}$ &  $0.121_{-0.001}^{+0.001}$ 

&  $0.1185_{-0.0011}^{+0.00097}$& $0.1194_{-0.001}^{+0.001}$ &$0.1207_{-0.00097}^{+0.00096}$\\ 

\red{$H_0$ } & $\red{67.67}_{-0.5}^{+0.47}$ &$\red{68.76}_{-0.51}^{+0.46}$ &$\red{70.44}_{-0.56}^{+0.56}$

&  $\red{ 67.93}_{-0.43}^{+0.48}$ & $\red{68.91}_{-0.5}^{+0.49}$ & $\red{70.65}_{-0.54}^{+0.52}$\\ 

$\ln (10^{10}A_{s })$ & $3.066_{-0.026}^{+0.019}$&$3.056_{-0.023}^{+0.021}$ & $3.027_{-0.023}^{+0.027}$ 

& $3.077_{-0.019}^{+0.026}$&  $3.061_{-0.022}^{+0.026}$ & $3.031_{-0.022}^{+0.018}$\\ 

$n_{s}$  & $0.9656_{-0.0043}^{+0.0041}$& $0.9637_{-0.0041}^{+0.0039}$ &  $0.9601_{-0.0039}^{+0.004}$ 

& $0.9671_{-0.0041}^{+0.0041}$ &$0.9645_{-0.0041}^{+0.004}$ &$0.9611_{-0.004}^{+0.0038}$\\ 

$\tau_{\rm re}$  & $0.06678_{-0.013}^{+0.011}$ &  $0.0611_{-0.013}^{+0.011}$& $0.04516_{-0.012}^{+0.014}$ 

& $0.07275_{-0.01}^{+0.014}$ &  $0.0641_{-0.012}^{+0.013}$& $0.04791_{-0.011}^{+0.01}$\\  \hline

$z_{\rm re}$  & $8.893_{-1.2}^{+1.1}$ & $8.359_{-1.2}^{+1.2}$ & $6.707_{-1.2}^{+1.7}$

& $9.435_{-0.85}^{+1.3}$ & $8.636_{-1.1}^{+1.3}$ &  $7.02_{-1.2}^{+1.1}$\\ 

$\sigma_8$  &$0.817_{-0.0095}^{+0.0076}$ & $0.8283_{-0.0093}^{+0.0085}$ & $0.8443_{-0.0099}^{+0.01}$

& $0.8197_{-0.0075}^{+0.0096}$ &$0.8298_{-0.0086}^{+0.0095}$ & $0.8456_{-0.0088}^{+0.0081}$ \\  

$\gamma$ & $-$ & $5.15(4) \times 10^{-2}$ & $9.21(7) \times 10^{-3}$ 

& $-$ & $5.17(4) \times 10^{-2}$ & $9.24(7) \times 10^{-3}$\\

\hline 
$\chi^2_{\rm min}$  &$13631.0$  &$13631.6$ &$13637.0$

&$13637.5$  &$13636.1$ &$13638.9$\\

\red{$\Delta \chi^2_{\rm min}$}  &\red{$0$}  &\red{$0.6$} &\red{$6.0$}

&\red{$1.4$}  &\red{$0$} &\red{$2.8$}\\
\hline
\end{tabular}}
\caption{\label{partable} Parameter tables for $\Lambda$CDM and the nonlocal models. Beside the six parameters that we have chosen as our fundamental parameters, we give also the values of the derived quantities $z_{\rm re}$ (the redshift to reionization) and $\sigma_8$ (the variance of the linear matter power spectrum in a radius of $8$ Mpc today) For the RR and RT models, among the derived parameters, we also give $\gamma=m^2/(9H_0^2)$. From \cite{Dirian:2016puz}.}
\end{table}

Thus,  for the case BAO+Planck+JLA, $\Lambda$CDM and the RT model are statistically equivalent, while the RR model is on the border of being strongly disfavored.  Among the various parameter, a particularly interesting result concerns $H_0$, which in the nonlocal models is predicted to be higher than in $\Lambda$CDM. Thus,  adding a high prior on $H_0$, of the type suggested by local measurements, goes in the direction of favoring the nonlocal models, as we see from the right table. In this case $\Lambda$CDM and the RT model are still statistically equivalent, although now with a slight preference for the RT model, while the RR model becomes only slightly disfavored with respect to the RR model, 
$\chi^2_{\rm RR}-\chi^2_{\rm RT}\simeq 2.8$,
and statistically equivalent to 
$\Lambda$CDM, $\chi^2_{\rm RR}-\chi^2_{\Lambda{\rm CDM}}\simeq 1.4$.

\begin{figure}[t]
\centering
\includegraphics[width=0.45\columnwidth]{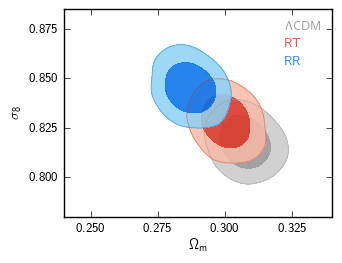}
\includegraphics[width=0.5\columnwidth]{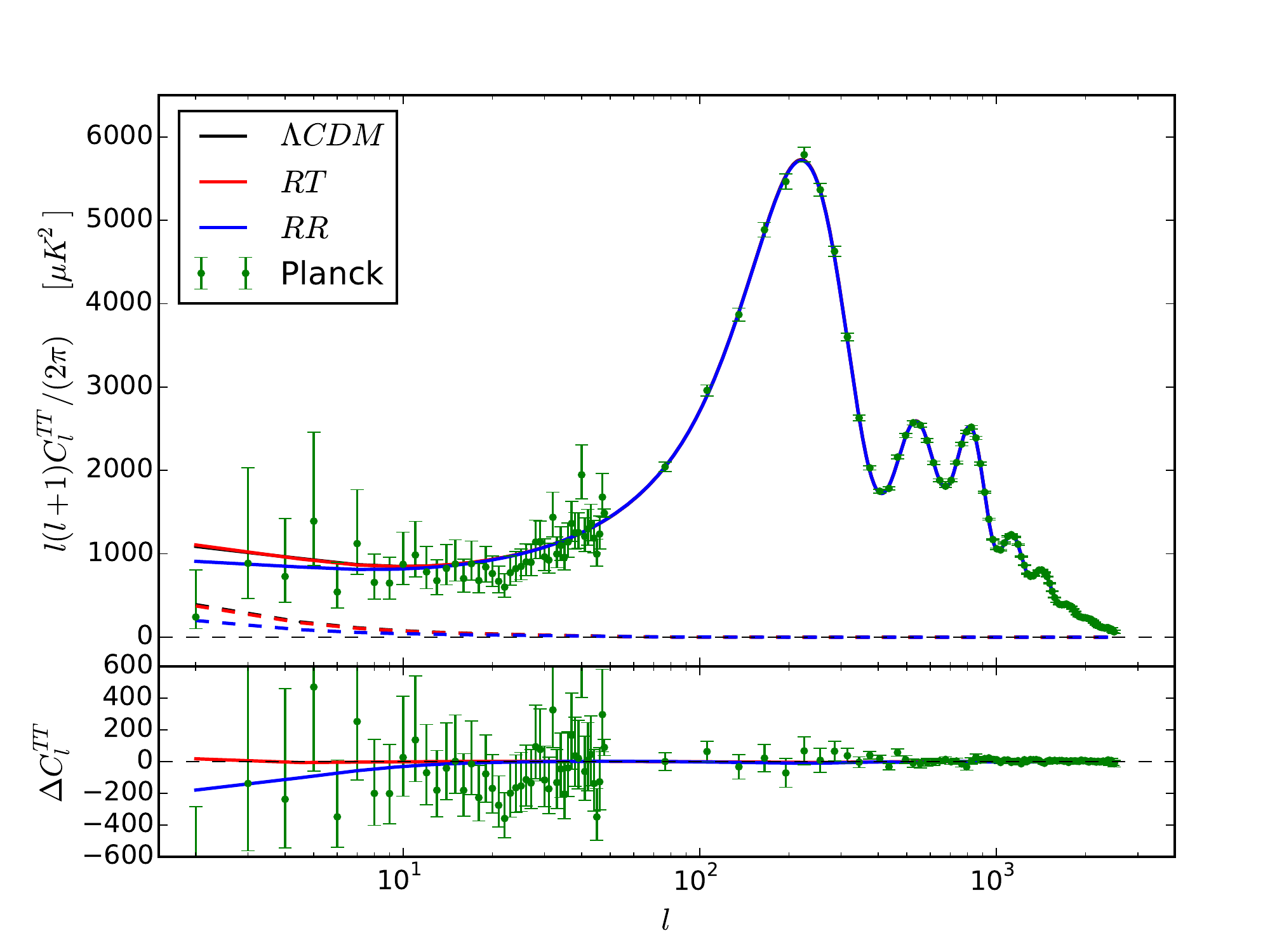}
\caption{\label{s8_Om}Left panel: $\sigma_8 - \Omega_m$ contour plot for Planck+BAO+JLA+$(H_0=70.6)$. Right panel: Upper plot: temperature power spectrum (thick), and the separate contribution from the late ISW contributions (dashed), for $\Lambda$CDM (black), RT (red) and RR (blue), using the best fit values of the parameters determined from BAO+JLA+Planck.
The black and red lines are indistinguishable on this scale. The lower plot shows the residuals for $\Lambda$CDM and difference of RT (red) and RR (blue) with respect to $\Lambda$CDM. Data points are from Planck 2015 \cite{Planck_2015_CP} (green bars). Error bars correspond to $\pm 1 \sigma$ uncertainty.
From \cite{Dirian:2016puz}.}
\end{figure}

In Table~\ref{partable} we also give the derived values of $\gamma=m^2/(9H_0^2)$ for the nonlocal models. These central values for $\gamma$ correspond to 
\bees
&&m/H_0 \simeq 0.288\, \qquad ({\rm RR\, \, model})\, ,\hspace{8mm}\\
&&m/H_0 \simeq 0.680\, \qquad ({\rm RT\,\, model})\, .
\ees
From the values in the Table, in the case BAO+Planck+JLA,
we  find, for the total matter fraction $\oma=(\omega_c+\omega_b)/h_0^2$, the mean values $\oma=\{0.308,0.300,0.288\}$ for $\Lambda$CDM, the RT and RR models, respectively, and    $h_0^2\oma=\{0.141,0.142,0.143\}$, which is practically constant over the three models. Using
BAO+Planck+JLA+$(\rm{H_0}=73.8$) these numbers change little, and become
$\oma=\{0.305,0.298,0.286\}$ for $\Lambda$CDM, the RT and the RR model, see \cite{Dirian:2016puz}  for full  details, and plots of one and two-dimensional likelihoods. In particular, the left panel of
Fig.~\ref{s8_Om} shows the two-dimensional likelihood in the plane $(\oma,\sigma_8)$. We see that the nonlocal models predict slightly higher values of $\sigma_8$ and slightly lower values of $\oma$.
The fit to the CMB temperature power spectrum, obtained with the data in Table~\ref{partable}, is shown in
the right  panel of
Fig.~\ref{s8_Om}.\footnote{We should also stress that  the analysis in \cite{Dirian:2014bma,Dirian:2016puz} has been performed using, for  the sum of the neutrino masses,  the value of the \textit{Planck} baseline analysis \cite{Planck_2015_CP}, 
$\sum_{\nu} m_{\nu}=0.06$~eV, which is the smallest value consistent with neutrino oscillations. Increasing the neutrino masses  lowers $H_0$. In $\Lambda$CDM this would increase the tension with local measurements, which is the main reason for choosing them in this way in the \textit{Planck} baseline analysis.
However, we have seen that the non-local models, and particularly the RR model, predict higher values of $H_0$, so they can accommodate larger neutrino masses without entering in tension with local measurements. A larger prior on neutrino masses would therefore favor the nonlocal models over $\Lambda$CDM.  This possibility is currently being investigated  \cite{DirainBarreira:inprep}.}

\subsection{Extensions of the minimal models}\label{sect:exten}

The RR and RT models, as discussed above, are a sort of `minimal models', that allow us to begin to explore, in a simple and predictive setting, the effect of nonlocal terms. However, even if the general philosophy of the approach should turn out to be correct,  it is quite possible that the actual model that describes Nature will be more complicated. A richer phenomenology can indeed be obtained with some well-motivated extensions of these models, as we discuss in this section.

\subsubsection{ Effect of a previous inflationary era}\label{sect:effectinfl}

The minimal models studied above are characterized by the fact that the initial conditions for the auxiliary fields and their derivatives are set to zero during RD. As we have discussed in Sect.~\ref{sect:loc}, the choice of initial conditions on the auxiliary fields is part of the definition of the model, and different initial conditions define different nonlocal models. In principle, the correct prescription should come from the fundamental theory. 
We now consider the effect of more general initial conditions, in particular of the type that could be naturally generated by a previous phase of inflation.\footnote{We are assuming here that the effective nonlocal theory given by the RR or RT model is still valid at the large energy scales corresponding to primordial inflation. Whether this is the case can only be ascertained once one has a understood the mechanism that generates these nonlocal effective theories from a fundamental theory.}
 
\vspace{2mm}\noindent
{\em RT model.} 
We consider first the effect of $u_0$ in the  RT model~\cite{Foffa:2013vma}.
From  \eq{pertU} we see that  the most general initial condition of $U$ amounts to a generic choice of the parameters $u_0$ and $u_1$, at some given initial time. The parameter $u_1$ is associated to a decaying mode, so the solution obtained with a nonzero value of $u_1$ is quickly attracted toward that with $u_1=0$. However, $u_0$ is a constant mode.  We have seen in \eq{u0cosmconst} that, in the RT model, the introduction of $u_0$ corresponds to adding back a cosmological constant term. From \eq{u0cosmconst} we find that the corresponding value of the energy fraction associated to a cosmological constant, $\ola$, is given by $\ola=\gamma u_0$. 
In the case $u_0=0$, for the RT model,  $\gamma\simeq 5\times 10^{-2}$, see Table~\ref{partable}. Then the effect of a non-vanishing $u_0$ will be small as long as $|u_0|\ll 20$. However larger values of $u_0$ can be naturally generated by a previous inflationary era.
 Indeed, we see from \eq{pertU} that in a deSitter-like inflationary phase, where $\zeta_0\simeq 0$, if we start the evolution  at an initial time $t_i$ at beginning an inflationary era  and set $U(t_i)=\dot{U}(t_i)=0$, we get, during inflation
\be\label{U4xxi}
U(x)=4(x-x_i)+\frac{4}{3}\( e^{-3(x-x_i)}-1\)\, ,
\ee
where $x_i=x(t_i)$. At the end of inflation, $x=x_f$, we therefore have
\be\label{Uxf}
U(x_f)\simeq 4\Delta N\, ,
\ee
where $\Delta N=x_f-x_i\gg 1$. 
Consider next the auxiliary field $Y(x)$.  If we choose the initial conditions at the beginning of inflation so that the growing mode is not excited, i.e. $a_1=0$ in \eq{pertY}, at the end of inflation we also have
$Y(x_f)\simeq 4\Delta N$. These values for  $U(x_f)$  and $Y(x_f)$ can be taken as initial conditions for the subsequent evolution during RD. The corresponding results where shown in~\cite{Foffa:2013vma}. This choice of $a_1$ is however a form of tuning of the initial conditions on $Y$. Here we consider the most generic situation in which $a_1\neq 0$. 
In this case during inflation $Y$ will grow to a value of order $\exp\{ 0.79\Delta N$\}, where $\Delta N$ is the number of efolds and $\alpha_+\simeq 0.79$ in a deSitter-like inflation. It will then decrease as $\exp\{-0.70 x\}$ during the subsequent RD phase, see \eq{apm}.

Despite the  growth during inflation (exponential in $x$, so power-like in the scale factor $a$), the DE  density associated to $Y$, $\rho_{\rm DE}=\gamma Y\rho_0$, is still totally negligible in the inflationary phase,  because $\rho_0={\cal O}({\rm meV}^4)$ is utterly negligible compared to the energy density during inflation. Thus, this growth of $Y$   does not affect the dynamics at the inflationary epoch, nor in the subsequent RD era. Nevertheless, this large initial value at the end of inflation can produce a different behavior of $Y$ near the present epoch, when the effective DE term $\gamma Y(x)$ becomes important.\footnote{Two caveats are however necessary here. First, as already mentioned, we are assuming that the nonlocal models are valid in the early inflationary phase. Second, we 
are assuming that the large value of $Y$ generated during inflation is still preserved by reheating.  
During reheating  the energy density of the inflaton field is transferred to the radiation field. Since $\gamma Y$ is just the DE energy density, it is  in principle possible that 
even the energy density  associated to $Y$ is transferred to the radiation field, just as the inflaton energy density. In this case  the evolution could resume at the beginning of RD with a  small initial value of $Y$. Since, during RD, $Y$ only has decaying modes,  the solution would then be quickly attracted back to that obtained setting $Y(x_*)=0$ at some $x_*$ in RD. }

\begin{figure}[t]
\includegraphics[width=0.45\columnwidth]{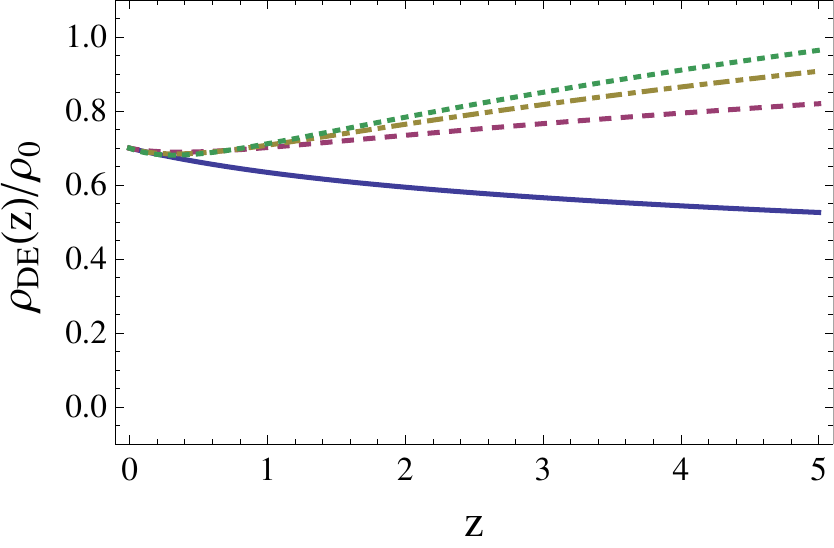}\hspace{2mm}
\includegraphics[width=0.45\columnwidth]{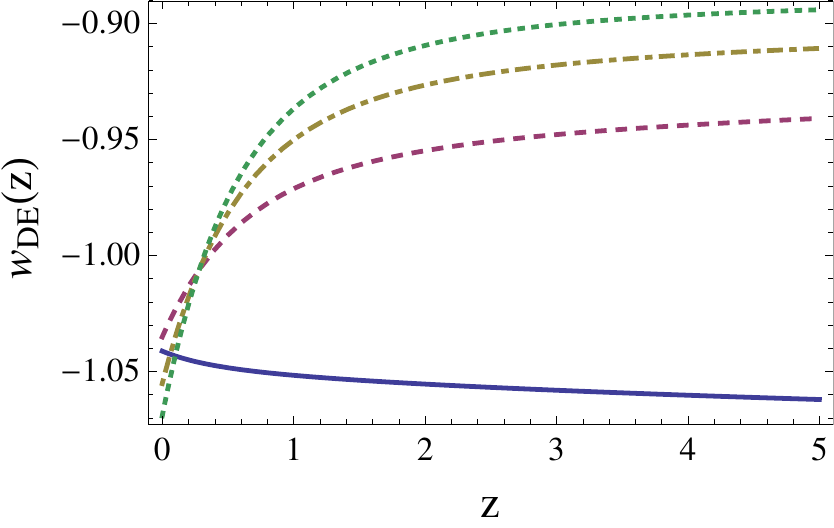}
\caption{Left panel:  $\rho_{\rm DE}/\rho_0$ for the RT model,  shown against the redshift $z$, for the initial conditions  on $U$ and $Y$ corresponding to the minimal model (i.e. $U=U'=Y=Y'=0$ at an initial value $x_{\rm in}=-15$ in RD, blue solid line), and for the initial values  of $U,U',Y,Y'$ given by an inflationary phase with $M=10^{3}$~GeV (red dashed), $M=10^{10}$~GeV (brown, dot-dashed) and
$M=10^{16}$~GeV (green, dotted).  In each case we adjust $\gamma$ so to maintain  fixed $\oma= 0.30$,  
which gives $\gamma=5.16\times 10^{-2}$ for the minimal model, and
$\gamma\simeq\{2.72\times 10^{-3},1.04\times 10^{-3},3.76\times 10^{-4}\}$ for
$M=\{10^3,10^{10},10^{16}\}$~GeV, respectively. 
Right panel: the corresponding results for $w_{\rm DE}$.
\label{fig:nonvanishing_u0andY0}}
\end{figure}

To be more quantitative let us recall that, if inflation takes place at a scale $M\equiv(\rho_{\rm infl})^{1/4}$, the minimum number of efolds required to solve the flatness and horizon problems is given by
\be\label{DeltaNmin}
\Delta N\simeq 64-\log\frac{10^{16}\, {\rm GeV}}{M}\, .
\ee
The inflationary scale  $M$ can range from a maximum value of order $O(10^{16})$~GeV (otherwise, for larger values  the effect of GWs produced during inflation would have already been detected in the CMB temperature anisotropies) to a minimum  value around  1~TeV, in order not to spoil the predictions of the standard big-bang scenario.
Assuming instantaneous reheating, the value of the scale factor $a_*$ at which inflation ends and RD begins is given by $\rho_{\rm infl}=\rho_{R,0}/a_*^4$, where $\rho_{R,0}$ is the present value of the radiation energy density, and as usual we have set the present value $a_0=1$. Plugging in the numerical values, for $x_*=\log a_*$ we find
\be
x_*\simeq -65.9+\log\frac{10^{16}\, {\rm GeV}}{M}\, .
\ee
Recall also that RD  ends and MD starts at $x=x_{\rm eq}\simeq -8.1$. 
Thus, assuming that the number of efolds $\Delta N$ is the minimum necessary to solve the horizon and flatness problems, during RD  (i.e. for $x_*<x<x_{\rm eq}$) we have 
\be\label{Ydix}
\log [Y(x)]\simeq 0.79\Delta N -0.70 (x-x_*)\, ,
\ee
where we used the fact that, during RD, $Y(x)\propto e^{-0.70 x}$, see \eq{apm}.
In Fig.~\ref{fig:nonvanishing_u0andY0} we show the result for $\rde$ and $w_{\rm DE}$ obtained starting the evolution from a value $x_{\rm in}=-15$ deep in RD, setting as initial conditions $U(x_{\rm in})=4\Delta N$, $U'(x_{\rm in}) =0$, and with $Y(x_{\rm in})=\exp\{  0.79\Delta N -0.70 (x_{\rm in}-x_*)\}$ and
$Y'(x_{\rm in})=-0.70 Y(x_{\rm in})$, as determined by \eq{Ydix}. We show the result  for three different values of the inflationary scale $M$, and also show again, as a reference curve, the result for  the minimal RT model. We see that the results, already for the background evolution, are quantitatively different from the minimal case. Comparing with the observational limits of $w_{\rm DE}(z)$ from Fig.~5 of the {\em Planck} DE paper~\cite{Ade:2015rim} we see that the predictions of these non-minimal nonlocal models for $w_{\rm DE}(z)$ are still consistent with the observational bounds, so even these models are observationally viable, at least at the level of background evolution. Observe that now, in the past,  $w_{\rm DE}(z)$ is no longer phantom, since 
$\rde(x)=\gamma Y(x)$ now starts from a large initial value and, at the beginning, it decreases.  Then,  $w_{\rm DE}(z)$ crosses the phantom divide at $z\simeq 0.35$ (for $M=10^{3}$~GeV), and 
$z\simeq 0.32$ (for $M=10^{10}$ and $M=10^{16}$~GeV). It is quite interesting to observe that, in the RT  model, an early inflationary phase leaves an imprint on the equation of state of dark energy today, so that one could in principle infer the inflationary scale from a measurement of the function $w_{\rm DE}(z)$.

\vspace{2mm}\noindent
{\em RR model.} The situation in the RR model is  different, because now the homogeneous solutions associated to the auxiliary fields $U$ and $V$ in \eqs{syU2}{syV}  only have constant or decreasing modes, in all cosmological epochs. In a deSitter epoch, setting $\zeta(x)=\zeta_0=0$, the solution of \eq{syU2} is still given by \eq{U4xxi}, so again at the end of inflation $U(x_f)\simeq 4\Delta N$. 
Neglecting the second term in \eq{U4xxi}, we
can set $U(x)\simeq 4(x-x_i)$ on the right-hand side of \eq{syV}. Taking into account that during a deSitter inflationary phase $h(x)$ is constant, at a value $h_{\rm dS}=H_{\rm dS}/H_0$, the equation for $V(x)$ becomes
\be
V''+3V'=\frac{4(x-x_i)}{h^2_{\rm dS}}\, .
\ee
 If we start the evolution  at an initial time $x_i$ at beginning an inflationary era  with initial conditions $V(x_i)=V'(x_i)=0$ we get, during inflation,
\be
V(x)=\frac{2}{27h^2_{\rm dS} }\[ 9(x-x_i)^2 -6(x-x_i)+2 \(1-e^{-3(x-x_i)}\)\]\, .
\ee
Then, at the end of inflation, $V(x_f)\simeq 2(\Delta N)^2/(3h^2_{\rm dS})$. This value is totally negligible, since even for an inflationary scale as small as $M=1$~TeV, $h^2_{\rm dS}\sim 10^{15}$. Thus,  
as initial conditions for the subsequent evolution in RD, we can take 
$U(x_{\rm in})=4\Delta N$ and $V(x_{\rm in})=0$, at a value $x_{\rm in}$ deep in RD.  Of course, one could take an initial value $V(x_{\rm in})={\cal O}(1)$, but this would not really affect the result. The point is that, for $V$, inflation does not generate a very large value at the beginning of RD.

\begin{figure}[t]
\centering
\includegraphics[width=0.45\columnwidth]{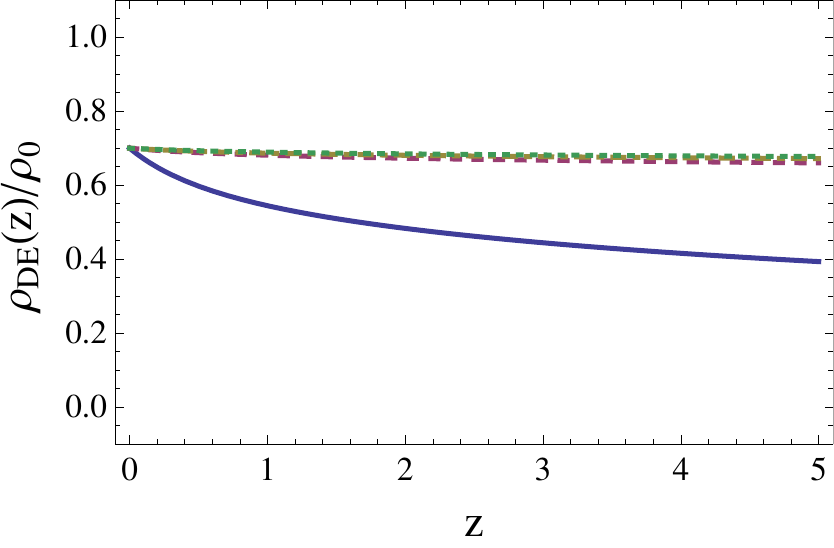}\hspace{2mm}
\includegraphics[width=0.45\columnwidth]{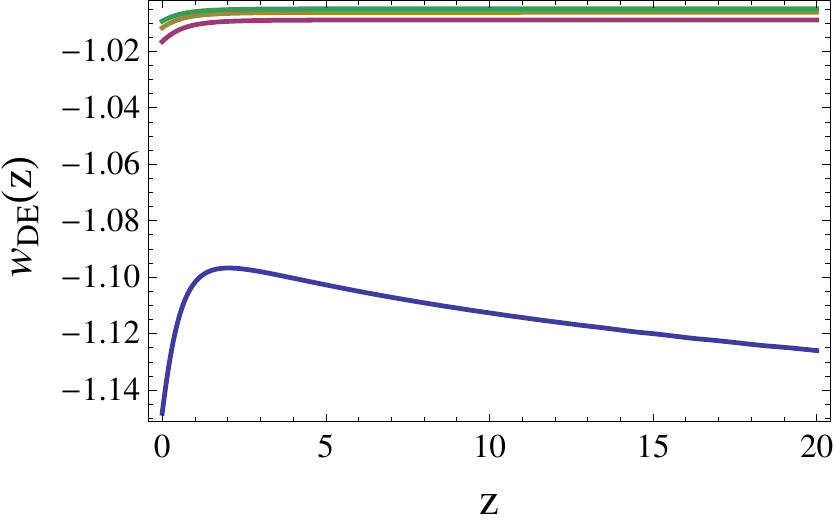}
\includegraphics[width=0.45\columnwidth]{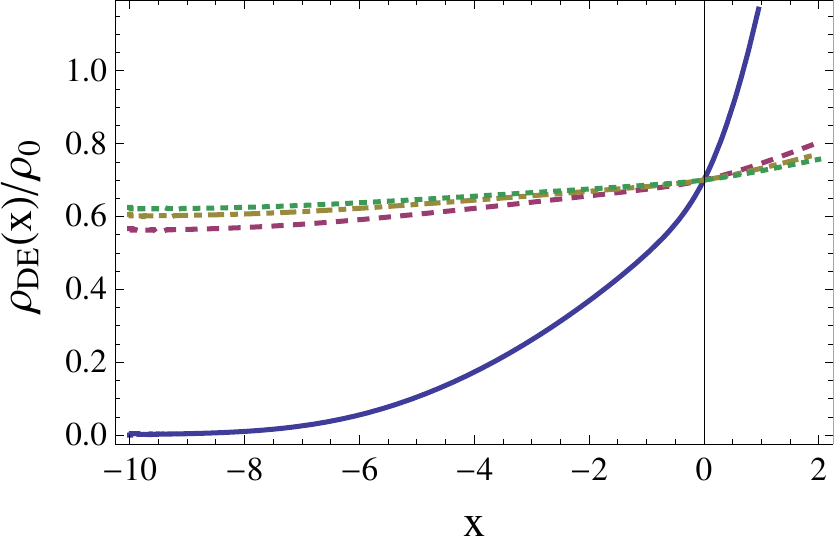}
\caption{Upper left panel:  $\rho_{\rm DE}(z)/\rho_0$ for the RR model, for the initial conditions  on $U$ and $Y$ corresponding to the minimal model (i.e. $U=U'=V=V'=0$ at an initial value $x_{\rm in}=-15$ in RD, blue solid line), and for the initial values  of $U$ given by an inflationary phase with $M=10^{3}$~GeV (red dashed), $M=10^{10}$~GeV (brown, dot-dashed) and
$M=10^{16}$~GeV (green, dotted).  In each case we adjust $\gamma$ so to maintain  fixed $\oma= 0.30$, which gives $\gamma=9.12\times 10^{-3}$ for the minimal model, and
$\gamma\simeq\{1.18\times 10^{-4},5.87\times 10^{-5},3.73\times 10^{-5}\}$ for
$M=\{10^3,10^{10},10^{16}\}$~GeV, respectively. 
Upper right panel: the corresponding results for $w_{\rm DE}$.
Lower panel: the function $\rde(x)/\rho_0$  against $x=\ln a$
\label{fig:nonvanishing_u0_RR}}
\end{figure}

The result is shown in Fig.~\ref{fig:nonvanishing_u0_RR} where, again, we express $\Delta N$ in terms of  the inflationary scale using \eq{DeltaNmin}.
We see that the RR model with a large initial value of $u_0$ gets  closer and closer to  $\Lambda$CDM. We find that $w_{\rm DE}(z=0)$ ranges from the value $-1.017$ for $M=10^{3}$~GeV to the value 
$-1.009$ for $M=10^{16}$~GeV, so the deviation with respect to $\Lambda$CDM are of order $(1-2)\%$.
Observe also that  in the cosmological future $\rde (x)$ continues to grow, although slowly, see the lower panel in
Fig.~\ref{fig:nonvanishing_u0_RR}. 

From the point of view of the comparison with observations, a sensible strategy is therefore to start from the minimal  RR model, since it predicts the largest deviations from $\Lambda$CDM and therefore 
can be more easily falsified (or verified). Indeed, already the next generation of experiments such as {\sc Euclid} should be able to discriminate  clearly the minimal RR model from $\Lambda$CDM. However, one must keep in mind that the non-minimal model with a large value of $u_0$ is at least as well motivated physically as the `minimal' model,  but  more difficult to distinguish
from  $\Lambda$CDM.

The RR model with a large value of $u_0$ is also conceptually interesting because it gives an example of a dynamical DE model that effectively generates a dark energy that, at least up to the present epoch, behaves almost like a cosmological constant, without however relying on a vacuum energy term, and therefore without suffering from the lack of technical naturalness associated to vacuum energy. Observe that these nonlocal models do not solve the coincidence problem, since in any case we must choose $m$ of order $H_0$, just as in $\Lambda$CDM we must choose the cosmological constant $\Lambda$ of order $H_0^2$. However, depending on the physical origin of the nonlocal term, the mass parameter $m$ might not suffer from the problem of large radiative corrections that renders the cosmological constant technically unnatural.

\begin{figure}[t]
\centering
\includegraphics[width=0.45\columnwidth]{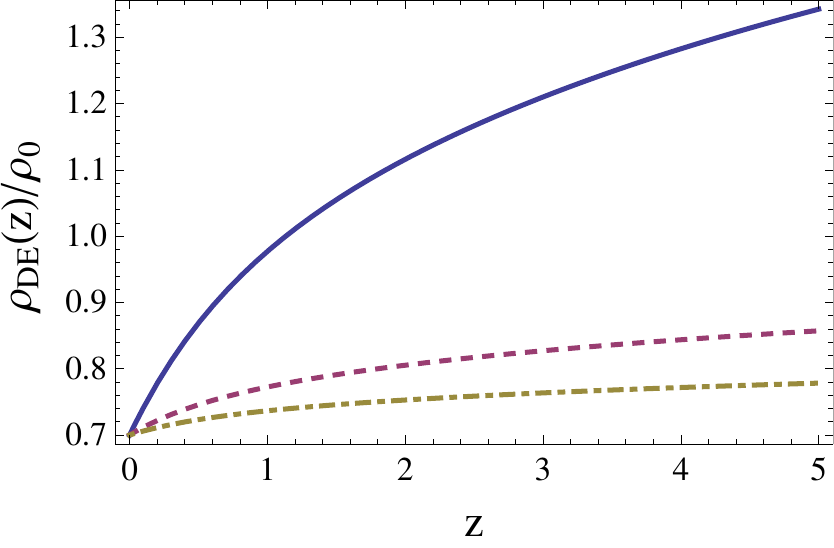}\hspace{2mm}
\includegraphics[width=0.45\columnwidth]{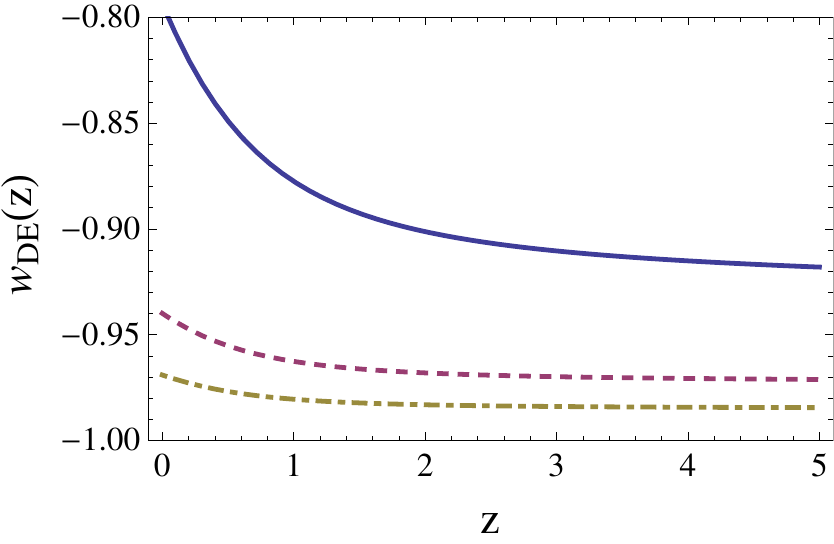}
\caption{Left panel:  $\rho_{\rm DE}(z)/\rho_0$ for the RR model, for the initial conditions  $U=u_0$, $U'=V=V'=0$ at an initial value $x_{\rm in}=-15$ in RD, with $u_0=-30$ (blue solid line), $-60$ (red dashed) and $-100$ (brown, dot-dashed). The corresponding values of $m/H_0$ are 
$\{0.42,0.12,0.06\}$.
These values 
corresponds to regime of the `path~B' solutions of \cite{Nersisyan:2016hjh}.  Right panel: the corresponding function $w_{\rm DE}(z)$.
\label{fig:pathB}}
\end{figure}

Observe also that, just as in $\Lambda$CDM, the inflationary sector is a priori distinct from the sector that provides acceleration at the present epoch. Thus, one can in principle supplement the nonlocal models with any inflationary sector at high energy, adding an inflaton field with the desired inflaton potential, just as one does for $\Lambda$CDM. However in these nonlocal models, and particularly in the RR model, there is a very natural choice, which is to connect them to Starobinski inflation, since in a model where is already present a nonlocal term proportional to  $R\Box^{-2}R$ is quite natural to also admit  a local $R^2$ term. As first  suggest in  ~\cite{Maggiore:2015rma,Cusin:2016nzi}, one can then  consider a model of the form
\be\label{6RRStar2}
S=\frac{\mplr^2}{2}\int d^4x\, \sqrt{-g}\, 
\[ R+\frac{1}{6M_{\rm S}^2} R\( 1- \frac{\Lambda_{\rm S}^4}{\Box^2}\) R\]\, ,
\ee
where $M_{\rm S}\simeq 10^{13}$~GeV is the mass scale of the Starobinski model and $\Lambda_{\rm S}^4=M_{\rm S}^2m^2$. As discussed in \cite{Cusin:2016nzi}, at early times the non-local term is irrelevant and we recover the standard inflationary evolution, while at late times the local $R^2$ term becomes irrelevant and we recover the evolution of the non-local models, although with initial conditions on the auxiliary fields determined by the inflationary evolution.

A general study of  the effect of the  initial conditions on the auxiliary fields in the RR model has been recently performed in~\cite{Nersisyan:2016hjh}. In particular, it has been observed that there is a critical value $\bar{u}_0\simeq -14.82+0.67\log\gamma $. For initial conditions $u_0>\bar{u}_0$ the evolution is of the type  that we have discussed above (denoted as `path~A' in \cite{Nersisyan:2016hjh}).  For
$u<\bar{u}_0$ a qualitatively different solution (`path~B') appears. On this second branch, after the RD and MD epoch, there is again a DE dominated era, where however $w_{\rm DE}$ gets close to $-1$ but still remaining in the non-phantom region $w_{\rm DE}>-1$ (and, in the cosmological future, approaches asymptotically an unusual  phase with $w_{\rm DE}=1/3$, $\ode\ra-\infty$ and $\oma\ra+\infty$, see Fig.~4 of \cite{Nersisyan:2016hjh}).  In Fig.~\ref{fig:pathB} we show the evolution in the recent epoch for such a solution, for three different values of $u_0=-30,-60,-100$.  As we see from \eq{defYRR},
the DE density in this case starts  in RD from a non-vanishing value 
$\rde(x_{\rm in})/\rho_0=(\gamma/4)u_0^2$. For instance, for $u_0=-60$, requiring $\oma=0.3$ fixes $\gamma\simeq 0.00157$, so 
$\rde(x_{\rm in})/\rho_0\simeq 1.4$. It then decreases smoothly up to the present epoch, where 
$\rde(x=0)/\rho_0\simeq 0.7$,
resulting in a non-phantom behavior for $w_{\rm DE}(z)$.\footnote{In the RT model the situation is different. Indeed, in \cite{Foffa:2013vma} it was found that   
cosmological  solutions such that, today, $\rde(x=0)/\rho_0$ is positive and equal to, say,  $0.7$, 
only exist for $u_0$ larger than a critical value $\bar{u}_0\,\simeq\, -12$. Thus,  again `path~A' solutions only exists only for $u_0$ larger than a critical value, but below this critical value there are no viable `path~B' solutions. The reason can be traced to the fact  that in the RT model a non-vanishing initial value of $u_0$ corresponds to $\rde(x_{\rm in})/\rho_0=\gamma u_0$, linear in $u_0$, while in the RR model corresponds to $\rde(x_{\rm in})/\rho_0=(\gamma/4)u_0^2$. Thus, a negative value of $u_0$ in the RT model implies a negative initial value of $\rde(x_{\rm in})/\rho_0$, resulting in a qualitatively different evolution. In particular, for $u_0$ negative and sufficiently large, it becomes impossible to obtain $\rde(x=0)/\rho_0$ positive and equal to $0.7$ by the present epoch.}

For sufficiently large values values of $-u_0$, this second branch is still cosmologically viable (while we see from the figure that, e.g., $u_0=-30$ gives a value of $w_{\rm DE}(0)$ too far from $-1$ to be observationally viable), and has been compared to JLA supernovae in~\cite{Nersisyan:2016hjh}.  Observe however that a previous inflationary phase would rather generate the initial conditions corresponding to `path A' solutions.

\subsection{Exploring the landscape of nonlocal models}
 
The study of nonlocal infrared modifications of GR is a relatively recent research direction, and one needs some orientation as to which  nonlocal models might be viable and which  are not. At the present stage, the main reason for exploring variants of the models presented is not just to come out with one more nonlocal  model that fits the data. Indeed, with the RT and RR models, both in their minimal and non-minimal forms discussed above, we already have a fair number of models to test against the data. Rather, our main motivation at present is  that identifying features of the nonlocal models that are viable might shed light on the underlying mechanism that generates their specific form of  nonlocality from a fundamental local theory.

A first useful hint comes from the fact, remarked in Sect.~\ref{sect:effmodGR}, that at the level of models defined by equations of motions such as  eqs.~(\ref{GmnT}) or (\ref{RT}), models where $\iBox$ acts on a tensor such as $\Gmn$ or $\Rmn$ are not cosmologically viable, while models involving $\iBox R$, such as the RT model, are viable. A similar analysis can be performed for models defined directly at the level of the action. At quadratic order in the curvature, 
a basis for the curvature-square terms is given by  $\Rmnrs^2$, $\Rmn^2$ and $R^2$. Actually, for cosmological application it is convenient to trade the Riemann tensor $\Rmnrs$ for the Weyl tensor $C_{\rho\sigma\mu\nu}$. A natural generalization of the nonlocal action (\ref{RR}) is then given by
\be\label{actionTotal}
S_{\rm NL}=\frac{\mplr^2}{2}\int d^4 x \sqrt{-g}\,
\left[R-\mu_1 R\frac{1}{\Box^2}R-\mu_2 C^{\mu\nu\rho\sigma}\frac{1}{\Box^2}C_{\mu\nu\rho\sigma}-\mu_3\RMN\frac{1}{\Box^2}\Rmn
\right]\,,
\ee
where $\mu_1$, $\mu_2$ and $\mu_3$ are  parameters with dimension of squared mass. This extended model has been studied in \cite{Cusin:2015rex}, where it has been found that
the term $\RMN\Box^{-2}\Rmn$ is ruled out since it gives  instabilities in the cosmological evolution at the background level. The Weyl-square term instead does not contribute to the background evolution, since the Weyl tensor vanishes in FRW, and it also has well-behaved scalar perturbations. However, its tensor perturbations are unstable \cite{Cusin:2015rex}, which again rules out this term. 

These results indicate that models in which the nonlocality involves $\iBox$ applied on the Ricci scalar, such as the RR and RT model, play a special role. 
This is particularly interesting since, as we saw in \eq{m2s2}, a term
$R\Box^{-2} R$ has a specific physical meaning, i.e. it corresponds to a
diff-invariant mass term for the conformal mode. The same holds for the RT model, since at linearized order over Minkowski it is the same as the RR model. This provides an interesting direction of investigation for understanding the physical origin of these nonlocal models, that we will pursue further in Sect.~\ref{sect:toward}.

One can then further explore the landscape of nonlocal models, focusing on extensions of the RR model. Indeed, already the RT model can be considered as a nonlinear extension of the RR model, since the two models become the same when linearized over Minkowski. An action for the RT model would probably include further nonlinear terms beside $R\Box^{-2}R$, such as higher powers of the curvature associated to higher powers of $\iBox$. We have seen in Sect.~\ref{sect:Baye} that the RT model appears to be the one that fits best the data, so it might be interesting to explore other physically-motivated nonlinear extensions of the RR model. In particular, in \cite{Cusin:2016nzi} we have explored 
two  possibilities, that could be a sign of an underlying conformal symmetry, and that we briefly discuss next.

\vspace{2mm}\noindent
{\em The $\Delta_4$ model}.
A first  option is to consider the model whose effective quantum action is
\be\label{actDelta4}
\Gamma_{\Delta_4}=\frac{\mplr^2}{2}\int d^4x\, \sqrt{-g}\, \[ R-\frac{m^2}{6} R\frac{1}{\DP} R\]\, .
\ee
where $\DP$ is the Paneitz operator (\ref{defDP}).
This operator depends only on the conformal structure of the metric,  and we have seen that it  appears  in the nonlocal anomaly-induced effective action in four dimensions. In FRW the model can again be localized using two auxiliary fields $U$ and $V$, so that the full system of equations reads~\cite{Cusin:2016nzi}
\bees
&&h^2(x)=\frac{\Omega(x) +(\gamma/4) U^2}{ 1+\gamma[ -3V'-3 V +(1/2)V'(U'+2U) ]}\, ,\label{fun1}\\
&&U''+(5+\zeta)U'+(6+2\zeta)U=6(2+\zeta)\, ,\label{fun3}\\
&&V''+(1+\zeta)V'=h^{-2}U\, ,\label{fun2}
\ees
where as usual $\Omega(x)=\Omega_M e^{-3x}+\Omega_R e^{-4x}$. The effective DE density can then be read from $\rde(x)/\rho_0=h^2(x)-\Omega(x)$. 
In the `minimal' model with initial conditions  $U(x_{\rm in})=U'(x_{\rm in})=V(x_{\rm in})=V'(x_{\rm in})=0$ at some value $x_{\rm in}$ deep in RD, we find that the evolution leads to $w_{\rm DE}(z=0)\simeq -1.36$, too far away from $-1$ to be consistent with the observations. Also, contrary to the RR model,  there is no constant homogeneous solution for $U$ in RD  and MD, because of a presence of a term proportional to $U$ in \eq{fun3}. Rather, the homogeneous solutions 
are $U=e^{\a_{\pm} x}$  with $\a_{+}=-2$ and $\a_{-} =-(3+\zeta_0)$, which are both negative in all three eras, and indeed whenever 
$\zeta_0>-3$, which is always the case in the early Universe. Therefore, there is no `non-minimal' model in this case. No large value for $U$ or $V$ is generated during inflation, and in any case even a large initial value at the end of inflation would  decrease exponentially in RD, quickly approaching the solution of the minimal model. Therefore, this model is not cosmologically viable. 

\begin{figure}[t]
\centering
\includegraphics[width=0.48\columnwidth]{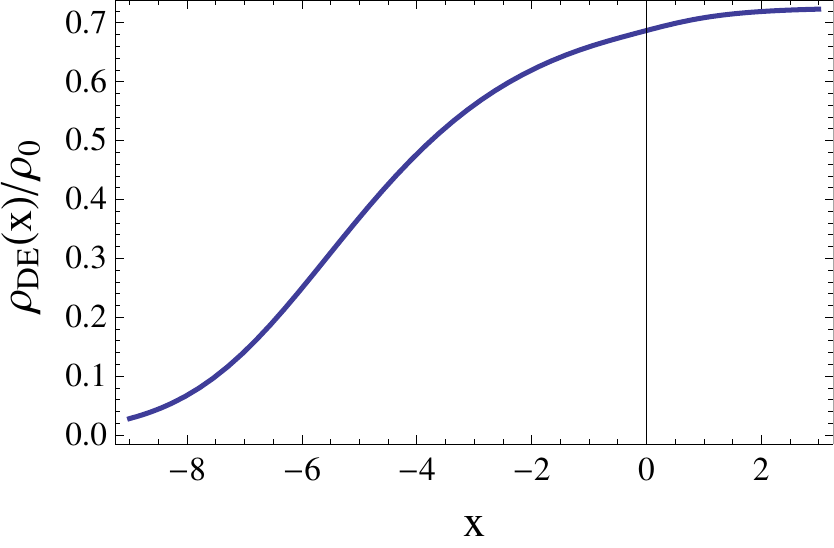}\hspace{2mm}
\includegraphics[width=0.48\columnwidth]{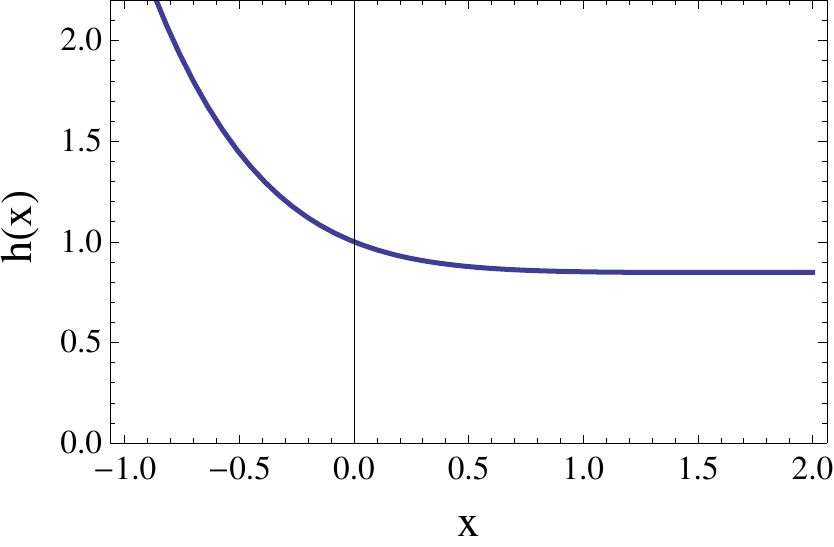}
\includegraphics[width=0.48\columnwidth]{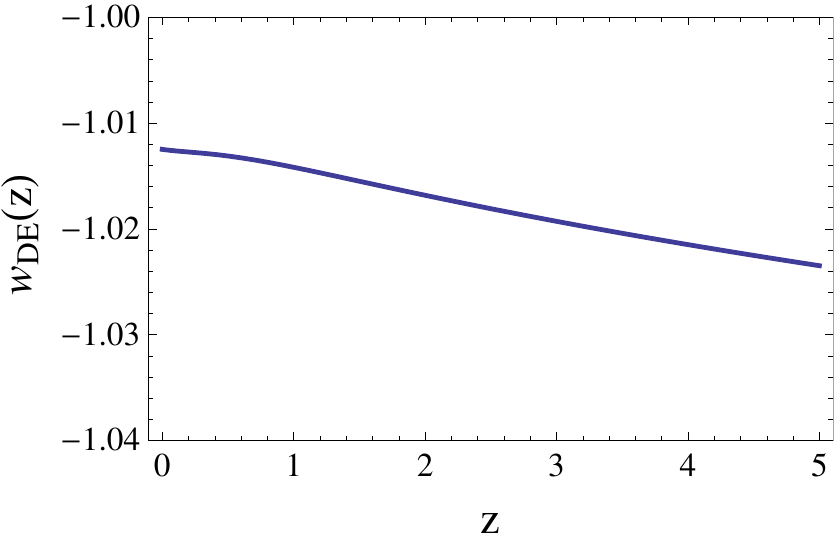}
\caption{\label{Lfig:rho} 
Upper left panel: the dark energy density $\rde(x)/\rho_0$ for the conformal RR model.
Upper right panel:  the Hubble parameter $h(x)$. 
Bottom panel: $w_{\rm DE}(z)$.
From~\cite{Cusin:2016nzi}.
}
\end{figure}

\vspace{2mm}\noindent
{\em The conformal RR model}. Another natural modification related to conformal symmetry would be to replace the $\Box$ operator in the RR model (or in the RT model) by the `conformal d'Alembertian'
$[-\Box +(1/6)R]$~\cite{Cusin:2016nzi}, which again  only depends on the conformal structure of space-time. We will call it the `conformal RR model'. More generally, one can also study the model~\cite{Mitsou:2015yfa}
\be\label{xiRR}
\Gamma_{\rm \xi RR}=\frac{\mplr^2}{2}\int d^4x\, \sqrt{-g}\, \[ R-\frac{m^2}{6} R\frac{1}{(-\Box+  \xi R)^2} R\]\, ,
\ee
with $\xi$ generic, although only  $\xi=1/6$ is related to conformal invariance.
Its study is a straightforward repetition of 
the analysis for the RR model. We can localize it by introducing two fields $U=(-\Box +\xi R)^{-1}R$ and $S=(-\Box +\xi R)^{-1}U$, and then \eqst{syh2}{syV} become
\bees
&&h^2(x)=\frac{\Omega(x) +(\gamma/4) U^2}{ 1+\gamma[ -3(V-\xi UV)'-3 (V-\xi UV) +(1/2)V'U' ]}\, ,\label{Lfun1}\\
&&U''+(3+\zeta)U'+6\xi (2+\zeta) U=6(2+\zeta)\, ,\label{Lfun2}\\
&&V''+(3+\zeta)V'+6\xi (2+\zeta)V=h^{-2}U\, ,\label{Lfun3}
\ees
where again
$\zeta\equiv h'/h$.
This models has some novel features compared to the $\xi=0$ case~\cite{Mitsou:2015yfa}. Indeed, as we see from Fig.~\ref{Lfig:rho}, the DE density goes asymptotically to a constant, and correspondingly also the Hubble parameter becomes constant, so the evolution approaches that of $\Lambda$CDM. This can also be easily undestood analytically, observing that in a regime of constant (and non-vanishing) $R$, the operator $(-\Box+\xi R)^{-1}$ acting on $R$ reduces to $(\xi R)^{-1}$. Then the nonlocal term in the action (\ref{xiRR}) reduces to a cosmological constant $\Lambda=m^2/(12\xi^2)$, leading  to 
a de~Sitter era with $H^2=\Lambda/3= m^2/(6\xi)^2$, i.e. $H=m/(6\xi)$.  Similarly, from \eq{Lfun2} we see  that, asymptotically, $U\ra 1/\xi$. Note that this solution only exists for $\xi\neq 0$. In particular, for the conformal RR model we have  $\xi=1/6$, so asymptotically $H\ra m$ and  $h\ra 3\gamma^{1/2}$, in full agreement with the numerical result in Fig.~\ref{Lfig:rho}. 

As we see from the bottom panel in Fig.~\ref{Lfig:rho},  for the physically more relevant case $\xi=1/6$, $w_{\rm DE}(z)$ is very close to $-1$,  for all redshifts of interest. Therefore, similarly to the non-minimal RR model discussed in Sect.~\ref{sect:effectinfl},
the conformal RR model is phenomenologically viable but more difficult to distinguish from $\Lambda$CDM, compared to the minimal RR model with $\xi=0$.

\section{Toward a fundamental understanding}\label{sect:toward}

The next question is how one could hope to derive the required form of the nonlocalities,  from a fundamental local QFT. This is still largely work in progress, and we just  mention here some relevant considerations, following 
refs.~\cite{Maggiore:2015rma,Maggiore:2016fbn}.

\subsection{Perturbative quantum loops}

The first idea that might come to mind is whether perturbative loop corrections can generate the required nonlocality. 
We have indeed seen that, among several other terms, the expansion in \eq{expan} also produces a term of the form 
$\mu^4R\Box^{-2}R$, where $\mu$ is the mass of the relevant matter field (scalar, fermion or vector) running in the loops. One could then try to argue~\cite{Codello:2015pga} that the previous terms in the expansion,  
such as $R\log(-\Box/\mu^2)R$ or  $\mu^2 R\Box^{-1}R$, do not produce self-acceleration in the present cosmological epoch, and just retain the $\mu^4R\Box^{-2}R$ in the
hope to effectively reproduce the RR model. Unfortunately, it is easy to see that this idea does not work. Indeed, as we have seen in detail in Sect.~\ref{sect:Qeffact},
to obtain a nonlocal contribution we must be in the regime in which the particle is light with respect to the relevant scale, $|\Box/\mu^2|\gg 1$. In the cosmological context the typical curvature scale is given by the Hubble parameter, so at a given time $t$ a particle of mass $\mu$ gives a nonlocal contribution only if $\mu^2\lsim H^2(t)$. In the opposite limit $\mu^2\gg H^2(t)$ it rather gives the local contribution (\ref{dec}). Thus, to produce a nonlocal contribution at the present cosmological epoch, we need $\mu^2\lsim H_0^2$. Then, retaining only the Einstein-Hilbert term and the 
$\mu^4R\Box^{-2}R$ term, we get an effective action of the form
\be\label{formfactmu}
\Gamma=\int d^{4}x \sqrt{-g}\, \bigg[\frac{\mplr^2}{2}R - R \,\frac{\mu^4}{\Box^2} R\bigg]\, ,
\ee
apart from a coefficient $\delta={\cal O}(1)$ that we have reabsorbed in $\mu^4$. Comparing with \eq{RR} we see that we indeed get the RR model, but with a value of the mass scale $m$ given by 
\be\label{mmumplr}
m\sim \frac{\mu^2}{\mplr}\, .
\ee
Since $\mu\,\lsim\, H_0$, for $m$ we get the ridiculously small value $m\,\lsim\, H_0 (H_0/\mplr)\sim  10^{-60} H_0$. To obtain a value of $m$ of order $H_0$ we  should rather use in \eq{mmumplr} a value 
$\mu\sim (H_0\mplr)^{1/2}$, which is of the order of the meV (such as a neutrino!). However, in this case $\mu\gg H_0$, and for such a particle at the present epoch we are in the regime (\ref{dec}) where the form factors are local.
Therefore we cannot obtain the RR model with a value $m\sim H_0$, as would be required for obtaining an interesting cosmological model. The essence of the problem is that, with perturbative loop corrections, the term $R\Box^{-2}R$ in \eq{formfactmu} is unavoidably suppressed, with respect to the Einstein-Hilbert term, by a factor proportional $1/\mplr^2$.\footnote{It has been pointed out in \cite{Nersisyan:2016hjh} that such a small value of $m^2$ could be compensated using a nonminimal model with a large value of $|u_0|$. This would however lead to a model indistinguishable from $\Lambda$CDM. Furthermore, with $m/H_0\sim 10^{-60}$, the required value of $u_0$ would be huge. For instance, in the RT model $\ola=\gamma u_0$. Since $\gamma\sim (m/H_0)^2$, this would require $u_0\sim 10^{120}$. In the RR model, where the effective DE is quadratic in $u_0$, this would still require $u_0\sim 10^{60}$. Observe  that one should also tune the matter content so that the term $\mu^4/\Box^2$ in $k_W(-\Box/\mu^2)$ vanishes, since we have seen that this term induces unacceptable instabilities in the tensor sector.}

\subsection{Dynamical mass generation for the conformal mode}\label{sect:dynmassgen}

The above considerations suggest to look for some non-perturbative mechanism that might generate dynamically the mass scale $m$ \cite{Maggiore:2015rma}. An interesting hint, that follows from the exploration of the landscape of nonlocal models presented above, is that the models that are phenomenologically viable are only those, such as the RR and RT model, that have an interpretation in terms of a mass term for the conformal mode, as we saw in \eq{m2s2}. Thus, a mechanism that would generate dynamically a mass for the conformal mode would automatically give the RR model, or one of its nonlinear extensions such as the RT model or the conformal RR model. Dynamical mass generation requires non-perturbative physics, in which case it emerges as a very natural consequence, as we know from experience with several two-dimensional models, as well as from QCD. As we discussed, an effective  mass term for the gluon, given by the gauge-invariant but nonlocal expression (\ref{Fmn2}), is naturally generated in QCD. The question is therefore whether some sector of gravity can become non-perturbative in the IR, in particular in spacetimes of cosmological relevance such as deSitter. Indeed, it is well know that in deSitter space large IR fluctuations can  develop. This is true already in the purely gravitational sector, since the graviton propagator grows without bound at large distances, and in fact the fastest growing  term comes from the conformal mode~\cite{Antoniadis:1986sb}, although the whole subject of IR effects in deSitter is quite controversial (see e.g. \cite{Woodard:2015kqa} for a recent discussion and references).

Another promising direction for obtaining strong IR effects is given by the quantum dynamics of the conformal factor, which includes the effect of the anomaly-induced effective action. Indeed,  the term 
$\sigma\bar{\Delta}_4\sigma$ in \eq{anominducSeffD4} can induce long-range correlations, and possibly a phase transition reminiscent of the BKT phase transition in two dimensions~\cite{Antoniadis:1996pb}.
Further work is needed to put this picture on  firmer ground.

\vspace{5mm}\noindent
{\bf Acknowledgments.}
I am grateful to Giulia Cusin, Yves Dirian, Stefano Foffa, Maud Jaccard, Alex Kehagias, Nima Khosravi, Martin Kunz,  Michele Mancarella, Ermis Mitsou and Valeria Pettorino
for collaboration on different aspects of this project over the last few years. I also thank Ilya Shapiro for useful discussions.
This work is supported by the Fonds National Suisse and by the SwissMAP National Center for Competence in Research.

\bibliographystyle{utphys}						
\bibliography{myrefs_massive}

\end{document}